\documentclass[aps,rmp,reprint,amsmath,amssymb,graphicx,longbibliography,nofootinbib]{revtex4-1}

\usepackage{smartdiagram}
\usepackage{color}
\usesmartdiagramlibrary{additions}
\usepackage{graphicx}
\usepackage{cancel}

\usepackage{bm}

\usepackage{amsmath}
\usepackage{mathrsfs}
\usepackage{calrsfs}
\usepackage{multirow}
\DeclareMathAlphabet{\pazocal}{OMS}{zplm}{m}{n}

\usepackage{graphicx}
\usepackage{dcolumn}
\usepackage{bm}
\usepackage{color} 
\usepackage{CJK}
\usepackage{dsfont}
\usepackage[extension=xxx]{hyperref}

\newcommand {\fabs}[1] {\left| #1 \right|}

\newcommand {\fabsq}[1] {\left| #1 \right|^2}
\newcommand {\la}{\langle}
\newcommand {\ra}{\rangle}

\newcommand{\cL}{{\cal{L}}}

\newcommand{\ket}[1]{\ensuremath{|#1\rangle}}

\newcommand{\braket}[2]{\langle#1|#2\rangle}
\newcommand{\ketbra}[2]{|#1\rangle\langle#2|}

\newcommand{\beq}{\begin{equation}}
\newcommand{\eeq}{\end{equation}}
\newcommand{\beqa}{\begin{eqnarray}}
\newcommand{\eeqa}{\end{eqnarray}}
\def\ra{\rangle}
\def\la{\langle}\def\beq{\begin{equation}}

\definecolor{mygreen}{rgb}{0.12, 0.7, 0.17}

\begin{document}
\title{Shortcuts to adiabaticity: concepts, methods, and applications  
}

\author{D. Gu\'ery-Odelin}
\affiliation{Laboratoire de Collisions Agr\'egats R\'eactivit\'e, CNRS UMR 5589, IRSAMC, Universit\'e de Toulouse (UPS), 118 Route de Narbonne, 31062 Toulouse CEDEX 4, France}

\author{A. Ruschhaupt}
\affiliation{Department of Physics, University College Cork, Cork, Ireland}

\author{E. Torrontegui}
\affiliation{Instituto de F\'\i sica Fundamental IFF-CSIC, Calle Serrano 113b, 28006 Madrid, Spain}

\author{A. Kiely}
\affiliation{Department of Physics, University College Cork, Cork, Ireland}

\author{S. Mart\'\i nez-Garaot}
\affiliation{Departamento de Qu\'{\i}mica F\'{\i}sica, UPV/EHU, Apdo.
644, 48080 Bilbao, Spain}

\author{J. G. Muga}
\affiliation{Departamento de Qu\'{\i}mica F\'{\i}sica, UPV/EHU, Apdo.
644, 48080 Bilbao, Spain}

\date{\today}
\begin{abstract}
Shortcuts to adiabaticity (STA) are fast  routes to the final results of slow, adiabatic
changes of the controlling parameters of a system. The shortcuts are designed by a set of 
analytical and numerical methods suitable for different systems and conditions.   
A motivation to apply STA methods to quantum systems is to manipulate 
them on timescales shorter than decoherence times. Thus shortcuts to adiabaticity  have become instrumental in 
preparing and driving internal and motional states in atomic, molecular, and solid-state physics. Applications range from  information transfer and processing based on gates or analog paradigms, 
to interferometry and metrology. The multiplicity of STA paths for the controlling parameters    
may be used to enhance robustness versus noise and perturbations, or to optimize 
relevant variables.    
Since adiabaticity is a widespread phenomenon, STA methods  also extended beyond the quantum world, 
to  optical devices,
classical mechanical systems, and statistical physics. 
Shortcuts to adiabaticity  combine well with other concepts and techniques, in particular with optimal control theory, 
and pose fundamental scientific and engineering questions such as finding speed limits, 
quantifying the third law, or determining  process energy costs and efficiencies.   
We review concepts, methods and applications of shortcuts to adiabaticity  and outline promising prospects, 
as well as open questions and challenges
ahead.  
\end{abstract}
\pacs{37.10.Gh, 37.10.Vz, 03.75.Be}

\maketitle

\tableofcontents{}

\section{Introduction\label{intro}}
\subsection{Overview of shortcuts to adiabaticity} 

\begin{figure}[b]
\begin{center}
\includegraphics[height=4.cm,angle=0]{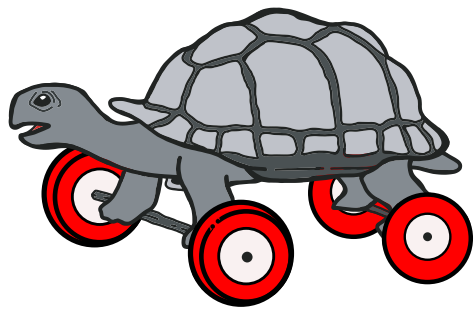}
\end{center}
\caption{\label{turtle}
(Color online) A turtle on wheels is a good metaphor for Shortcuts to Adiabaticity. The image is inspired with her permission by the artist work of Andree Richmond.}
\end{figure}

\begin{figure}[t]
\begin{center}
\includegraphics[height=5cm,angle=0]{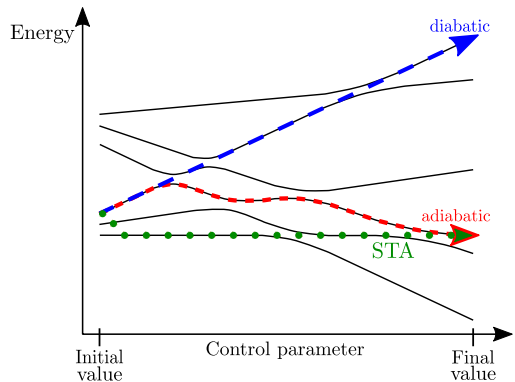}
\end{center}
\caption{\label{STAscheme}
(Color online) Schematic example of adiabatic, diabatic, and STA processes. The system is initially in the third level. In the adiabatic path ({\color{red} - - - } red short dashed) the system evolves always along the third level. In a diabatic evolution ({\color{blue}${\boldsymbol{---}}$} blue long dashed) the system gets excited by jumping across avoided crossings. Along an STA path ({\color{mygreen}$\boldsymbol{\cdot\cdot\cdot}$} green dots) the system does not always travel along the third level but it arrives at the third level in a shorter time (the time dimension is not explicitly shown in the figure).}
\end{figure}

``Shortcuts to adiabaticity'' (STA) are fast routes to the final results of slow, adiabatic  
changes of the controlling parameters of a system.
Adiabatic processes are here broadly defined as
those for which the slow changes of the controls leave 
some dynamical properties invariant, the ``adiabatic invariants'', such as the quantum number in quantum systems,
or phase-space areas in classical systems. Figure \ref{turtle} portrays informally the central idea of shortcuts to adiabaticity. Figure \ref{STAscheme} also illustrates the concept pictorially by comparing  diabatic, adiabatic, and STA 
paths in a discrete system. 

The shortcuts rely on specific time dependences of the 
control parameters, and/or  on the addition of auxiliary time-dependent couplings or interactions
with respect to some reference Hamiltonian or, more generally, Liouvillian or transition-rate matrix.   
STA methods were first applied in simple quantum systems: two- and three-level systems, or a particle in a time-dependent harmonic
oscillator. 
They have since come to encompass a much broader domain since slow processes are quite common as a simple way to prepare the state of a system  or 
to change conditions avoiding excitations 
in a wide spectrum of areas,  from    
atomic, molecular, and optical physics, solid state,  or chemistry, to  classical mechanical systems
and engineering. In parallel to such a large scope, different methods   
have been developed and applied.

The description of shortcuts given above
needs some caveats and clarifications.
In  many quantum mechanical applications,
the final state in an STA 
process  reproduces the set of adiabatic probabilities 
to find the system in the eigenstates of the final Hamiltonian. 
The mapping of probabilities from initial to final 
settings, without final excitation but allowing for excitations on route, 
holds for any possible initial state in some state-independent STA protocols and simple
enough systems. These state-independent shortcuts are ideal for maximal robustness and 
to reduce dependence on temperature, so as to avoid  costly and time-consuming
cooling to the ground state.
However, except for idealized models, state independence 
only holds approximately and for a domain of parameters, for example within the 
harmonic approximation in a trapped ion.    
In other applications state independence is not really necessary or it may be 
difficult or impossible to achieve, as in chaotic systems.
Yet  
shortcuts may still be found  for a chosen specific state, typically the ground state, or for a state subspace.  

A further caveat is that the scope of STA methods
has in practice outstripped the original, pure aim in many applications. For example,  these 
methods 
can be extended with minimal changes to drive general transitions, regardless of whether the initial
and final states can be
connected adiabatically, such as transitions where the initial state is an eigenstate of the initial Hamiltonian whereas the final state is not an eigenstate of the final Hamiltonian. This broader perspective, merges with inverse engineering methods of the Hamiltonian to achieve arbitrary transitions or unitary transformations.\\

{\it Motivations.} There are different  motivations for the speedup that depend on the setting. In optics, time is often substituted by length to quantify the rate of change,
so  the shortcuts imply  shorter, 
more compact optical devices. In mechanical engineering, 
we look for fast and safe protocols, say of robotic cranes,  to enhance productivity. 
In microscopic quantum systems, slowness
often implies decoherence, the accumulation of errors and perturbations, or even the escape of the system from its confinement. The shortcuts  provide a useful toolbox  to avoid or mitigate these problems and thus to 
develop  quantum technologies.  Moreover, with shorter process times experiments can be repeated more often to increase  
signal-to-noise ratios.   

A generic  and important feature of STA apart from the speed achieved is 
that there are typically many alternative  routes for the control parameters,
and this flexibility can be used to 
optimize physically relevant variables, for example to minimize transient energy excitations and/or energy
consumptions, or to maximize robustness against perturbations.\\  
 
{\it Shortcuts to adiabaticity in quantum technologies.} Indeed, shortcuts have been mostly developed and applied for quantum systems, and much of the current interest in STA is rooted on the quest  for quantum technologies.    
STA combine  well with the two main paradigms of quantum information processing:  

- In  the gate-based 
paradigm, shortcuts to adiabaticity  contribute to improve and speed up gates or state preparations, and to
perform elementary operations like moving atoms leaving them unexcited.  
Shortcuts to adiabaticity may be applied to all physical platforms, such as trapped ions, cold atoms, Nitrogen-Vacancy centers, superconducting circuits, quantum dots, or atoms in cavities.       

Apart from fighting decoherence,  implementing  ``scalable'' architectures towards larger quantum systems 
is a second major challenge to develop quantum information processing. Scalability benefits
 from STA  in two ways:
directly, through the design of operations explicitly intended to achieve it (see an example in Sec. \ref{trapped}),
 as well as indirectly,
since a mitigated decoherence reduces  the need for highly demanding,
qubit-consuming error-correction codes.

- The second main paradigm of quantum information processing is  
adiabatic computing or quantum annealing, which  may be accelerated or made possible by changing initial and/or  final Hamiltonians,   modifying  the interpolation path between initial
and final Hamiltonians, 
or adding auxiliary terms to the transient Hamiltonian.
For adiabatic quantum computing and other applications with complex systems,
progress is being made to find effective STA without using difficult-to-find information
on spectra and eigenstates.
The more recent paradigms of topological quantum information   processing or   measured based quantum computation  may also
benefit from STA.

%
Shortcuts to adiabaticity are also having an impact or are expected to be  useful in 
quantum technologies other than quantum information processing such as interferometry and metrology, communications, 
and micro-engines or refrigerators in quantum thermodynamics.\\  

{\it Fundamental questions.} Finally, apart from being a practical aid to process design, STA  also played a role and
will continue to be instrumental in clarifying 
fundamental concepts  such as 
quantum-classical relations, quantum speed limits and trade off relations between timing, energy, robustness, entropy or information, 
the third law of thermodynamics, and the proper characterization of energy costs
of processes.  

%

\subsection{About this review}

{ This review} provides a broad overview of concepts, methods, and applications of shortcuts to adiabaticity. 
A strong motivation for writing it is the breadth of systems and areas where STA are used. The hope is that this work will assist in stimulating discussions and information transfer between different domains. There are already examples that demonstrate the benefits of such an interaction. Analogies found between systems as disparate as individual ions in Paul traps and mechanical cranes or optical waveguide devices and atomic internal states have proved fruitful. The link between Lewis-Riesenfeld invariants and Lax pairs is another unexpected example. 

The development of STA methods and applications has been quite rapid since \textcite{Chen2010_063002}, where 
the expression ``shortcuts to adiabaticity'' was coined.  
Precedents before 2010 exist where  STA concepts had been applied 
\cite{Unanyan1997,Emmanouilidou2000,Couvert2008,Masuda2010,Muga2009,Schmiedl2009,Motzoi2009,Rezek2009,Salamon2009}.  
In the post-2010 period, the early work of Demirplak and Rice \cite{Demirplak2003,Demirplak2005,Demirplak2008},
and  \textcite{Berry2009} on ``counterdiabatic (CD) driving'' has been particularly influential, 
as well as methodologies such as ``inverse engineering'', ``invariants'', ``scaling laws'', ``fast-forward'', or ``local adiabatic'' methods.          
To add further flexibility to this already rich scenario of approaches, some methods provide in general a multiplicity of control protocols. 
Moreover, STA methods relate synergistically to or overlap partially with other control methods.  
In particular STA blend well with  
Optimal Control Theory (OCT), decoherence-free subspaces (DFS) \cite{Wu2017_042104},
linear response theory \cite{Acconcia2015}, or perturbative and variational schemes (see Sec. \ref{ssec:robustness} and Sec. \ref{ssec:variational}).
There is in summary a dense network of different approaches and STA protocols available, frequently hybridized. 

Some of the basic STA techniques  rely on specific formalisms (invariants and scaling, CD driving, or fast-forward)  that, 
within specific domains, can be related to each other
 and be made potentially equivalent because of underlying common structures. 
For example, a particular protocol to speed up the transport of a particle by defining a trap path, 
may be  found as the result of ``invariant-based engineering'', 
``fast-forward'', or ``local-CD'' approaches.  This convergence for  specific operations and methods might be misleading 
because it does not extend to
all systems and circumstances. No single all-inclusive  theory exists for all STA, and each of the major methodologies  
carries different limitations, different construction recipes, and different natural application domains. 
A widespread misconception is to identify ``STA''  with one particular approach, often CD driving, and even with one particular protocol 
within one approach.  We shall indeed pay due attention to 
unifying concepts and connections among the STA formalisms, which are of course useful and worth stressing,     
but at the same time it is hard to overemphasize    
that the diversity of approaches is a powerful asset of the shortcuts to adiabaticity  that explains their  versatility. 
  
New hybrid, or approximate methods are being
created as we write and more will be devised to adapt to diverse needs and systems.  
The review is also intended to map and characterize the options, help users to navigate among them, and encourage  
the invention of novel more efficient or goal-adapted approaches.  A number of techniques that fit naturally into the definition of STA but have not
been usually tagged as  STA  will be also mentioned  
to promote transfer of ideas.  Examples of this are  the Derivative Removal of Adiabatic Gate (DRAG) and        Weak AnHarmonicity With
Average Hamiltonian (WAHWAH) approaches
to implement fast pulses free from spurious transitions in superconducting qubits 
\cite{Motzoi2009,Schutjens2013}.\\

{\it Terminology.}  
Unsurprisingly, the various  approaches to STA have been described and used with inconsistent
terminologies by different groups and communities. Indeed the rapid
growth of STA-related work  and its extension across many disciplines has brought up 
different uses for the same words (``superadiabaticity'' and ``fast forward'' are  clear examples of polysemic terms),
and  different expressions for the same concept or method (for example ``transitionless quantum driving'' and  
``counterdiabatic approach'').  
This review is also intended to clarify some commonly found uses or expressions, making explicit our preference
when the polysemy may lead to confusion.\\

{\it Scope and related reviews.}
We intend the review to be didactical for the non-initiated but also comprehensive so that the experts in the different subfields
may find a good starting point for exploring other STA-related areas. 
Several recent reviews on partial aspects or 
overlapping topics are useful companions: 
\textcite{Torrontegui2013_117} is the previous
most comprehensive review on the subject, but 
the number of new applications in previously unexplored fields, experiments, and theoretical results since its publication 
clearly surpasses the work reviewed there. We shall mostly pay attention to the 
work done  after \textcite{Torrontegui2013_117}, but for presenting the key concepts some overlap is allowed.       
Work on the counterdiabatic method has been reviewed recently in \textcite{Kolodrubetz2017} with emphasis on geometric aspects and classical-quantum 
relations.  
For the approaches to speed up adiabatic computing see \textcite{Albash2018,Takahashi2019}.
For a review on STA to  control quantum critical dynamics see \textcite{delCampo2015}. 
See also \textcite{Menchon-Enrich2016} about spatial adiabatic passage and \textcite{Vitanov2017} about Stimulated 
Raman adiabatic passage (STIRAP), a technique that has been often sped up with STA.  
Finally, \textcite{Masuda2016_51} review the counterdiabatic and fast-forward methods focusing on applications to polyatomic molecules and Bose-Einstein condensates. Also, a Focus issue 
of New Journal of Physics on Shortcuts to Adiabaticity  provides a landscape of  current 
tendencies \cite{delCampo2019}.\\   

{\it Structure.}
The review is organized into ``Methods" of STA in Sec. \ref{methods} and ``Applications" from Sec. \ref{qt} to \ref{sec:cms}. We address applications of STA to quantum science and technology in Section \ref{qt}. This includes the main physical platforms and the quantum system is mainly considered as a closed quantum system. The connections between STA and quantum thermodynamic concepts, a topic between closed and open quantum systems, are reviewed in Sec.  \ref{ec}. STA and open quantum systems is the topic of Sec. \ref{sec:open}. Sections \ref{od} and \ref{sec:cms} deal with results and peculiarities of two fields that also offer a promising arena for practical applications: optics in Section \ref{od} and classical systems in Section \ref{sec:cms}.

%
%

We have tried to keep acronyms to a minimum but some terms (among the most prominent: shortcuts to adiabaticity, STA;  or counterdiabatic, CD)
appear so many times that the use of the acronym is well justified. Other acronyms are already of widespread use, such as 
OCT for Optical Control Theory.  
To facilitate reading, the full list of acronyms is provided in Appendix A. 
As for notation, an effort towards consistency has been made, so the  
symbols often differ from the ones in the original papers. 
Beware of possible multiple uses 
of some symbols (constants in particular). The context and text should help to avoid any confusion.   
\section{Methods\label{methods}}
\subsection{Overview of inverse engineering approaches\label{ria}}
We begin with an overview of ``inverse engineering''. This expression refers to inferring the time variation of the control parameters
from a chosen evolution of the physical system of interest. The notion  is broadly applicable for quantum, classical,
or stochastic dynamics, and embraces many  STA techniques. In this section 
we shall deal first with two simple examples, for motional and internal degrees of freedom, where more sophisticated or auxiliary 
concepts and formalisms, e.g. related to ``invariants'', ``couterdiabatic driving'', or ``fast forward'',  are not
explicitly used.   
Instead, the connection between dynamics and control is performed by solving for the control function(s)
in the effective equation of motion for a classical variable or a quantum mean value.  

It is often necessary or useful to go beyond this simple inversion approach for several reasons.   
For example:  the dynamical path of the system is not easy  
to design;  the inversion is non trivial; we are interested in finding state-independent protocols;
the direct inversion leads to unrealizable control functions; or we look for stable controls, robust with respect
to specific perturbations.
To address  these problems, in the following subsections several auxiliary concepts and
formal superstructures will be added to the simple inverse-engineering idea in the different STA approaches. 
A first hint on a family of inversion methods dealing with detailed dynamics beyond mean values is given 
in Sec. \ref{sssec:bmv}, after the two examples. 

\subsubsection{Quantum transport}
%
Finding the motion of a harmonic trap to transport a quantum particle so that it starts in the ground state and
ends up  in the ground state of the displaced potential amounts to solving 
inversely a classical Newton equation \cite{Schmiedl2009,Torrontegui2011}. 
The equation of motion of the particle position $x$ inside the potential reads
\begin{equation}
\ddot x + \omega_0^2x = \omega_0^2x_0(t),
\label{eqdoh}
\end{equation}
where $\omega_0$ is the angular frequency of the harmonic trap, $x_0(t)$ the instantaneous position of its center, and the dots 
represent, here and in the following, time derivatives. 
Equation (\ref{eqdoh}) describes a forced oscillator driven by the time-dependent force $F(t)=m\omega_0^2x_0(t)$. We can interpolate the trajectory $x(t)$ between the initial position of the particle and the desired final position $d$.  In addition, 
to ensure that the transport ends up at the lowest energy state, one needs to cancel out the first and second derivatives at the initial, $t=0$, and final time $t_f$ \cite{Torrontegui2011}. A simple polynomial interpolation of fifth degree can account for such boundary conditions \cite{Torrontegui2011},
\begin{equation}
x(t)= d \left[ 10\left(\frac{t}{t_f}\right)^3-15\left(\frac{t}{t_f}\right)^4+6\left(\frac{t}{t_f}\right)^5  \right].
\label{polyinterp}
\end{equation} 
Once $x(t)$ is defined, Eq. (\ref{eqdoh}) can be easily inverted to give the corresponding expression for the driving term $x_0(t)$. Note that there are infinitely many interpolating functions consistent with the boundary conditions at initial and final times. This freedom is quite typical of different STA methods and can be exploited to satisfy    
other conditions, e.g. minimizing average energy of a particle during displacement. More parameters can be added in the  interpolation functions  to minimize the quantity of interest \cite{Torrontegui2011}. 
For instance, the robustness against errors in the value of the angular frequency of the trap $\omega_0$ can be enhanced 
using a Fourier reformulation of the transport problem  \cite{Guery-Odelin2014_063425}, see an experimental application in  \textcite{An2016} and Sec. \ref{Fourier}. The simple example provided here was generalized to take into account anharmonicities  \cite{Zhang2015_043410} and in 3D to manipulate  Bose-Einstein condensates
with an atom chip \cite{Corgier2018}. Section \ref{ssec:robustness} provides a deeper view on the techniques developed to enhance the robustness of a given protocol.
%
%
%

\subsubsection{Spin manipulation}
A similar approach can be used for designing the magnetic field components to induce  a given trajectory of 
the mean value of  a spin 1/2 ${\bf S}(t)$ on the Bloch sphere \cite{Berry2009},
\begin{equation}
{\bf B}(t) = B_0(t) {\bf S}(t) + \frac{1}{\gamma}{\bf S}(t)  \times \partial_t {\bf S}(t), 
\end{equation}
where $\gamma$ is the gyromagnetic ratio and $B_0(t)$ any time-dependent function. This solution is 
found  by inverse engineering of the precession equation for the mean value of the spin,
\begin{equation}
\partial_t {\bf S}(t) = \gamma {\bf B}(t) \times {\bf S}(t). 
\end{equation}
In practice, we usually define the initial and target state and build up the solution by interpolation as before. Such an approach has been used for instance to manipulate a spin by defining the time evolution of the spherical angle of the spin on the Bloch sphere \cite{Vitanov2015,Zhang2017_15814}. The very same problem can be reformulated using other formalisms such as the Madelung representation \cite{Zhang2017_15814} that yields an equation of motion that can be readily reversed. 
Other formulations of the inverse engineering approach can be made by a proper shaping of the evolution operator \cite{Jing2013,Kang2016_30151} or by  time rescaling \cite{deLima2019}. 
This method has been applied to two \cite{Zhang2017_15814}, three \cite{Kang2016_30151,Kang2017_025201}, and four level systems \cite{Li2018_013830}. Inverse engineering has also been used for open quantum systems \cite{Jing2013,Impens2017}.

%
%
%

\subsubsection{Beyond mean values\label{sssec:bmv}}
Suppose now that a more detailed specification of the dynamics is needed 
and let us focus on 
closed, linear quantum systems.  
We shall design the unitary evolution operator $U(t)$ by specifying a complete basis
of  dynamical states $|\psi_j(t)\ra$ assumed to satisfy a  time-dependent Schr\"odinger 
equation driven by a -to be determined- Hamiltonian,  
\beq
U(t)=\sum_j |\psi_j(t)\ra\la\psi_j(0)|
\label{ut}
\eeq
(see other proposals for the form of $U$ in \textcite{Kang2016_30151}
and \textcite{Santos2018_015501}.). 
The corresponding Hamiltonian is given from the assumed dynamics by 
\beq\label{Hbasic}
H(t)=i\hbar \dot{U}U^\dagger.
\eeq
As in the previous examples, a typical scenario is that the initial and final Hamiltonians are fixed by the 
experiment or the intended operation. With these boundary conditions the functions  $|\psi_j(t)\ra$ at initial and final time
are usually chosen as the eigenstates of the corresponding Hamiltonians, but there is freedom to interpolate them, and therefore $H(t)$, in between.    

Several approaches depend on different ways to choose the orthogonal basis functions $|\psi_j(t)\ra$:  
(a) in the counterdiabatic driving approach they are instantaneous eigenstates 
of a reference Hamiltonian $H_0(t)$;  (b) in invariant-based engineering, they are  
eigenstates of the invariant
of an assumed Hamiltonian form;  (c) more generally they can be just  
convenient functions.    
In particular they can be parameterized  so that $H(t)$ obeys certain constraints,
such as making zero undesired terms or matrix elements.    
Note that (a), (b), and (c) are in a certain sense equivalent, as they can be reformulated 
in each other's language, and they all rely on Eq. (\ref{Hbasic}). 
For example, (b) or (c) do not explicitly need an $H_0(t)$, and (a) or (c) do not
explicitly need the invariants,   
but $H_0(t)$ or the invariants could be found if needed. In particular, any linear combination $\sum c_j |\psi_j(t)\ra\la\psi_j(t)|$ 
with constant coefficients $c_j$ is by construction an invariant of motion of $H(t)$. 
Inverse engineering may also  be based on partial information, such as e.g., using
a single function of the set, and imposing some additional condition on the Hamiltonian, 
for example that the potential is local (i.e., diagonal) and real in
coordinate space. These conditions are  the  essence of the (streamlined) version \cite{Torrontegui2012_013601} of the fast-forward approach
\cite{Masuda2008,Masuda2010}.  The following explores all these approaches in more detail.

\subsection{Counterdiabatic driving\label{ssec:cd}}
The basic idea of counterdiabatic driving is to add auxiliary interactions to some reference Hamiltonian $H_0(t)$ so that the dynamics follows exactly the (approximate) adiabatic evolution driven by  $H_0(t)$.  
An illustrative analogy is a flat, horizontal  road turn (the reference) that is modified by inclining the roadway surface about its longitudinal axis with a bank angle so that the vehicles can go faster without sliding off the road.    
After some precedents \cite{Unanyan1997,Emmanouilidou2000},  the counterdiabatic (CD) driving  paradigm was worked out and developed 
systematically by Demirplak and Rice \cite{Demirplak2003, Demirplak2005, Demirplak2008}
for internal state transfer using control fields, then rediscovered in a different but equivalent way as ``Transitionless Tracking'' 
by Berry \cite{Berry2009}, and used to design many control schemes after \textcite{Chen2010_123003}, unaware of the work of Demirplak and Rice, employed Berry's  method to 
control two- and three-level systems.\\    

{\it Berry's formulation.}
We start with Berry's formulation because it is somewhat simpler. 
In \textcite{Berry2009}, the starting point  is a reference Hamiltonian
\beq
\label{H0}
H_0(t)=\sum_n | n(t)\rangle  E_n(t) \langle n (t)|.
\eeq
We adopt for simplicity a notation appropriate for a discrete (real) spectrum and no degeneracies.\footnote{Generalizations for degenerate levels,    
relevant for example to speed up  holonomic quantum gates, may be found in \textcite{Takahashi2013_062117}, \textcite{Zhang2015_18414},  or \textcite{Karzig2015}.
Generalizations for non-Hermitian Hamiltonians $H_0(t)$ are discussed in Sec. \ref{nhh}.}   
A state $|n(0)\ra$ that is initially an eigenstate of $H_0(0)$ will continue to be so under slow enough driving,  
with the form
\beq
\label{aa}
|\psi_n (t) \rangle =  e^{i \xi_n (t)} |n(t)\rangle,
\eeq
where the adiabatic phases $\xi_n(t)$ are found by substituting  Eq. (\ref{aa}) as an ansatz into the time-dependent Schr\"odinger equation driven by $H_0(t)$,
\beq
\label{ad_phase}
\xi_n (t)=-\frac{1}{\hbar} \int^t_0 dt' E_n(t') +  i\int^t_0 dt' \langle n(t')| \partial_{t'} n(t') \rangle.
\eeq
We now seek a Hamiltonian $H(t)$ for which the above approximate states $|\psi_n(t)\rangle$ become the exact evolving states,
\beq
i\hbar\partial_t|\psi_n(t)\rangle=H(t)|\psi_n(t)\rangle.
\eeq
$H(t)$ is constructed using Eq. (\ref{Hbasic}) from the  unitary evolution operator
\beq
\label{eq:U}
U(t) = \sum_n e^{i \xi_n (t)} |n(t)\rangle \langle n(0)|,
\eeq
which obeys  $i\hbar\partial_t U(t)= H(t) U(t)$, so that 
an arbitrary state evolves as
\beq
\label{evol}
|\psi(t)\ra=\sum_n |n(t)\ra e^{i\xi_n(t)}\la n(0)|\psi(0)\ra.
\eeq
%
%
%
%
After substituting Eq. (\ref{eq:U}) into Eq. (\ref{Hbasic}), or alternatively 
differentiating Eq. (\ref{evol}), the Hamiltonian  becomes 
\beqa
\label{Berry Hamiltonian}
H(t)&=&  H_{0}(t) + H_{CD}(t),
\\
H_{CD}(t)&=& i \hbar \sum_n  \bigg[|\partial_t n(t)  \rangle \langle n(t) |
\nonumber
\\
&-&\langle n(t) | \partial_t n(t)  \rangle | n(t) \rangle \langle n(t) |\bigg],
\label{h1}
\eeqa
where $H_{CD}$ is hermitian and nondiagonal in the $|n(t)\rangle$ basis.

As a simple example of $H_0$ and $H_{CD}$, consider a two-level system with reference Hamiltonian 
\beq
\label{H0_Lie}
H_0(t)=\frac{\hbar}{2}\left ( \begin{array}{cc}
-\Delta(t) & \Omega_R(t) \\
\Omega_R(t) & \Delta(t) \end{array} \right ), 
\eeq
where $\Delta(t)$ is the detuning and $\Omega_R(t)$ is the real Rabi frequency.  
The counterdiabatic Hamiltonian has the form
\beq
\label{counter}
H_{CD}(t)=\frac{\hbar}{2}\left ( \begin{array}{cc}
0 & -i \Omega_a(t) \\
i \Omega_a(t) & 0 \end{array} \right ), 
\eeq
with $\Omega_a(t)=[\Omega_R(t) \dot \Delta(t)-\dot \Omega_R(t) \Delta(t)]/\Omega^2(t)$ and $\Omega(t)=\sqrt{\Delta^2(t)+\Omega_R^2(t)}$.   
See further simple examples in Appendix \ref{2gen}.

Coming back to the general expression  (\ref{Berry Hamiltonian}), we note that $H_{CD}$ is orthogonal to $H_0$ considering the scalar product 
$tr(A^\dagger B)$ of two operators. Using $d\la n(t)|m(t)\ra/dt=0$ it can be seen that $H_{CD}$ is also orthogonal to $\dot{H}_0$ \cite{Petiziol2018}.  
An alternative form for $H_{CD}$ is found  by differentiating $H_0(t)|n(t)\ra=E_n(t)|n(t)\ra$, 
\beq
H_{CD}=i\hbar\sum_{m\neq n}\sum_n \frac{|m(t)\ra\la m(t)|\dot{H}_0|n(t)\ra\la n(t)|}{E_n-E_m}, 
\eeq 
which, using the scaled time $s=t/t_f$, gives the scaling $H_{CD}\sim 1/t_f$.  
In general $H_{CD}(t)$ vanishes for $t<0$ and $t>t_f$, either suddenly or continuously at the extreme times, so that 
the $|n (t) \ra$ become
eigenstates of the full Hamiltonian at the boundary times $t=0^{-}$ and $t=t_f^{+}$. 

In terms of the phases the total Hamiltonian (\ref{Berry Hamiltonian}) can be written as 
\cite{Chen2011_062116}
\beq
\label{alternative1}
 H(t)= J(t)+i\hbar\sum_n|\partial_tn(t)\rangle\langle n(t)|,
\eeq
with 
\beq
\label{alternative2}
J(t)=-\hbar\sum_n|n(t)\rangle\dot\xi_n(t)\langle n(t)|.
\eeq
Subtracting $H-H_{CD}$ we get an alternative form of $H_0$ consistent with 
Eqs. (\ref{H0}) and (\ref{ad_phase}),   
\beq\label{h0xi}
H_0=\hbar \sum_n |n(t)\ra [-\dot{\xi}_n+i\la n(t)|\dot{n}(t)\ra]\la n(t)|.
\eeq
If we  write the {\it same} $H_0(t)$ (\ref{H0}) as before in  an alternative basis with  different phases, 
$H_0(t)=\sum_n|n'(t)\ra E_n(t) \la n'(t)|$, where 
$|n'(t)\ra=e^{i\phi_n}|n(t)\ra$, the formalism  goes through using primed functions  $\xi'_n$ and corresponding 
primed operators. (One example is $|n'(t)\ra=e^{\xi_n(t)}|n(t)\ra$ so that $\xi'_n(t)=0$.) 
It is easy to check though that the terms compensate so that $H'_{CD}(t)=H_{CD}(t)$, i.e.,
 a change of representation for $H_0(t)$  does not change $H(t)$, 
nor the CD-driving term, nor the physics. 

Quite a different issue is to change the physics when imposing a set of basis functions $\{|n(t)\ra\}$ and
a set of phases $\{\xi_n(t)\}$  which are not a priori regarded as adiabatic \cite{Chen2011_062116}. This procedure is 
essentially invariant-based 
inverse engineering since one is imposing some specific dynamics (thus some invariants)  without presupposing a given $H_0(t)$. 
$U(t)$ in Eq. (\ref{eq:U}) becomes the primary object  and the driving Hamiltonian is given by Eq. (\ref{alternative1}) with diagonal part  (\ref{h0xi}), which defines $H_0(t)$,  
and a coupling, non-diagonal  part (\ref{h1}). Now, changing the phases $\xi_n(t)$ 
modifies $H_0(t)$, 
and has an impact on the physical evolution of a general wavefunction $|\psi(t)\ra$. Of course the populations 
${\cal P}_n(t)\equiv |\la n(t)|\psi(t)\ra|^2={\cal P}_n(0)$ driven by (\ref{alternative1}) 
are not affected by phase shifts.  The $\dot{\xi}_n(t)$ can be optimized to minimize
energy costs \cite{Hu2018},  robustness against decoherence 
\cite{Santos2018_025301}, or intensity of the extra CD controls \cite{Santos2019}.  

A related procedure  is to set  the $|n(t)\ra$ but consider $H_0(t)$ and the  eigenvalues $E_n(t)$  controllable elements rather than given \cite{Berry2009}. For example, setting all $E_n(t)=0$ cancels the dynamical phase. That means that for a given set 
of states $|n(t)\ra$, $H_{CD}(t)$ alone (without an $H_0$) 
drives the same populations than $H(t)=H_0(t)+H_{CD}(t)$ for any choice of $E_n(t)$ \cite{Chen2010_123003}.\\

{\it The formulation of Demirplak and Rice.} 
To follow some past and recent developments and generalizations it is worth  finding the auxiliary Hamiltonian $H_{CD}$ by the equivalent formulation
of Demirplak and Rice \cite{Demirplak2003,Demirplak2005,Demirplak2008}, which is  
the zeroth  iteration, $j=0$, in Fig. \ref{fscheme}. 

In the following discussion we concentrate on $j=0$. Among the possible phase choices for the eigenstates of $H_0(t)$, we use now 
the one that satisfies the ``parallel transport'' 
condition $\la n_0(t)|\dot{n}_0(0)\ra=0$. Regardless of the ``working''  basis of eigenvectors ${|n(t)\ra}$ one starts with, the parallel transported  basis  is found as 
\beq
\label{patr}
|n_0(t)\ra=e^{-\int_0^t \la n(t')|\dot{n}\ra dt'}|n(t)\ra.
\eeq
Later on we shall see the consequences of applying or not applying parallel transport.
A dynamical solution driven by $H_0(t)$ is written as $|\psi_0(t)\ra$ and the instantaneous eigenvalues as 
$E_n^{(0)}(t)$. The extra notational burden  of the index $j=0$ 
may be ignored in many applications, in particular for ordinary CD driving,  but it will be useful to 
cope with higher order  suparadiabatic iterations later on.
  
The way to find the auxiliary driving term  is to 
express first the dynamics in an adiabatic frame. For that we 
apply the transformation $|\psi_1(t)\ra =A_0^\dagger (t)|\psi_0(t)\ra$ to the dynamical states driven by $H_0(t)$,
with  $A_0(t)=\sum_n |n_0(t)\ra\la n_0(0)|$.     
In the resulting adiabatic frame the coupling, diabatic  terms 
are made obvious, 
$-A_0^\dagger { K}_0 A_0$, where ${ K}_0=i\hbar \dot{A}_0A_0^\dagger=\sum_n |\dot{n}_0(t)\ra\la n_0(t)|$.  In this adiabatic frame we may (a) neglect the coupling (adiabatic approximation), or  (b) cancel the 
coupling by adding $A_0^\dagger { K}_0 A$ to the Hamiltonian, see again Fig. \ref{fscheme}. In the original, Schr\"odinger picture,  also referred to as laboratory frame hereafter, this addition amounts to  using $H=H_0+{ K}_0$.

An alternative useful form for $K_0(t)$ is \cite{Messiah1960}
\beq\label{kk}
K_0(t)=i\hbar\sum_n \frac{dP_n(t)}{dt} P_n(t), 
\eeq
where $P_n(t)=|n_0(t)\ra\la n_0(t)|=|n(t)\ra\la n(t)|$. 
$K_0$ may as well be written 
in an arbitrary basis of eigenvectors $|n(t)\ra$, i.e., not necessarily parallel transported, 
$K_0(t)=i\hbar \sum_n  |\dot{n}(t)\ra\la n(t)|-|n(t)\ra \la n(t)|\dot{n}\ra\la n(t)|=H_{CD}(t)$. 
This result proves that $K_0$  is identical to $H_{CD}$ in Eq. (\ref{h1}).  
Since the projectors $P_n(t)=|n_0(t)\ra\la n_0(t)|=|n(t)\ra\la n(t)|$ are invariant with respect  to phase choices of the eigenvectors, the form in which $H_{CD}$ is usually written
in Eq. (\ref{h1})  is invariant under different choices of phases for the eigenstates. 
In particular this implies that $K_0$ is purely non-diagonal (Kato's condition \cite{Kato1950,Demirplak2008})
in the arbitrarily chosen basis of eigenvectors of $H_0(t)$, $|n(t)\ra$.   

If we now return to the  parallel transported basis in the interaction (adiabatic) picture, the addition of $A_0^\dagger K_0 A_0$ cancels the couplings so the dynamics is trivially solved with  dynamical phase factors. In the Schr\"odinger picture,  
\beq
|\psi_0(t)\ra= e^{-\frac{i}{\hbar}\int_0^t E_n(t')dt'} |n_0(t)\ra \la n_0(0)|\psi_0(0)\ra,  
\eeq
which is exactly Eq. (\ref{evol}) as can be seen by using Eq. (\ref{patr}).  

If, instead of $A_0$, a more general transformation $\widetilde{A}=\sum_n |n(t)\ra\la n(0)|$ is used, 
the coupling term in the interaction picture becomes $-\widetilde{A}_0^\dagger \widetilde{K}_0\widetilde{A}_0$ 
where $\widetilde{K}_0=i\hbar\dot{\widetilde{A}}_0 \widetilde{A}_0^\dagger= \sum_n |\dot{n}(t)\ra\la n(t)|=K_0+i\hbar |n(t)\ra
\la n(t)|\dot{n}(t)\ra\la n(t)|$
and, 
its cancellation would lead in the laboratory picture to  a different state, 
\beq
\sum_n e^{-\frac{i}{\hbar}\int_0^t E_n(t')dt'} |n(t)\ra \la n(0)|\psi_0(0)\ra,  
\eeq
although the probabilities are not affected.

\textcite{Demirplak2008} also pointed out that the operator   
\beq
H_{CD}^{[n]}=i\hbar [dP_n/dt,P_n]_-,  
\label{onestate}
\eeq
which fulfils $P_mH_{CD}P_n=P_mH_{CD}^{[n]}P_n$ and $P_nH_{CD}P_m=P_nH_{CD}^{[n]}P_m$ for all $m$,
as well as $P_mH_{CD}^{[n]}P_m'=0$ for $m,m'\neq n$, uncouples the dynamics of level $n$.   
For $H_0+H_{CD}^{[n]}$, $e^{i\xi_n(t)}|n(t)\ra$ is an exact solution of the dynamics.   
This is an interesting simplification as we are often only interested in one state, typically the ground state. 
The last condition   imposed in \textcite{Demirplak2008},  $P_mH_{CD}^{[n]}P_m'=0$ for $m,m'\neq n$, 
is not really necessary for the uncoupling 
of the $n$-th level. Without it, 
a broad set of ``state-dependent'' CD operators  
$H_{CD}^{[n]}+Q_n B Q_n$, where $Q_n=1-P_n$ and  $B$ is any hermitian operator, can be generated. This multiplicity may be useful, 
and explains why different auxiliary state-dependent $CD$ terms have been proposed  \cite{Patra2017_125009, Setiawan2017}.   
\subsubsection{Superadiabatic iterations\label{sai}}

\begin{figure}[b]
\begin{center}
\includegraphics[height=12.cm,angle=0]{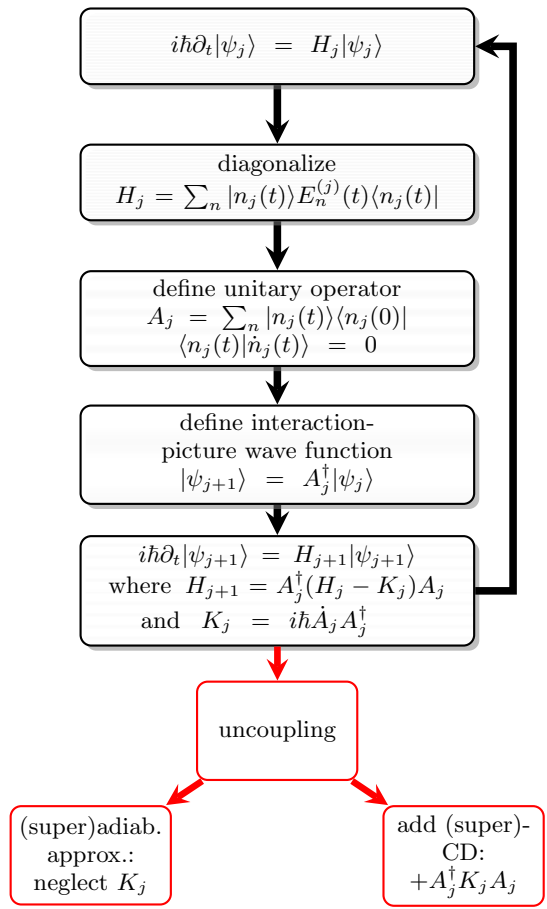}
\end{center}
\caption{Scheme for superadiabatic iterative interaction pictures. At the end of each iteration it is possible, apart from starting a new one,  
to either neglect the non-diagonal coupling in $H_{j+1}$ (adiabatic approximation if $j=0$ or superadiabatic approximation for $j\ge 1$), 
or add a term (CD for $j=0$ or super-CD otherwise)  to the Hamiltonian $H_{j+1}$ to exactly cancel the coupling.
\label{fscheme}}
\end{figure}

Let us recap before moving ahead. The adiabatic interaction picture  (IP) corresponds to expressing the quantum dynamics in the adiabatic basis of instantaneous 
eigenstates of $H_0(t)$. The dynamical equation in the adiabatic IP includes an effective Hamiltonian $H_1(t)$ with a diagonal (adiabatic) term and a coupling term, see once more Fig. \ref{fscheme}.  

We can repeat the sequence iteratively. In the first ``superadiabatic'' iteration, $H_1$ is diagonalized to find its instantaneous (parallel-transported) eigenstates $|n_1(t)\ra$. With the new ``superadiabatic basis'' a new IP is generated 
driven by an effective Hamiltonian $H_2$ with a diagonal part in the superadiabatic basis and a coupling term. This new coupling term, 
$-A_1^\dagger K_1 A_1$   may be (a) 
neglected  (superadiabatic approximation) or (b) cancelled by adding its negative,... and so on. 
The cancelling term to be added to $H_0$ in the Schr\"odinger picture is
$H_{cd}^{(j)} = B_jK_jB_j^\dagger$,
with $B_0 =1$ and $B_j =\prod_{k=0}^{j-1}A_k$ for $j\ge 1$.  Note that in general only the one generated in the zeroth iteration agrees with the 
standard CD term, $H_{cd}^{0}=H_{CD}$. 

The recursive iterations  were worked out by 
\textcite{Garrido1964}, without considering the  cancellations,  to find out generalizations of the adiabatic approximation.  \textcite{Berry1987} also used them to calculate a sequence of corrections to Berry's phase and introduced the name  
``superadiabatic transformations''.
\textcite{Demirplak2008} proposed to apply the  superadiabatic iterative frame to generate alternative (to the simple CD approach) higher order coupling-cancelling
terms.  
Later  \textcite{Ibanez2013} made explicit the conditions that the derivatives of $H_0(t)$ must satisfy at the time boundaries 
in order to really generate a shortcut to adiabaticity (rather than just a shortcut to superadiabaticity), i.e., a protocol that takes instantaneous 
eigenstates of $H_0(t=0)$ to corresponding eigenstates of $H_0(t_f)$. 

The naive expectation that each iteration will produce smaller and smaller couplings does not hold in general. They decrease 
up to an optimal iteration and then grow \cite{Berry1987}. Working with the optimal frame may or may not be worthwhile  depending on whether the boundary conditions 
for derivatives of $H_0(t)$ are fullfilled \cite{Ibanez2013}. 

An interesting feature of the superadiabatic sequence of coupling terms 
is that their operator structure changes with the iteration.  For example, for a  two-level  system with Hamiltonian $X\sigma_x+Z(t)\sigma_z$, $X$ constant,   where the $\sigma_{x,y,z}$ are Pauli matrices, the first 
(adiabatic) CD term $K_0$ is of the form $Y(t)\sigma_y$, see Eq. (\ref{counter}), whereas the second (first order superadiabatic) coupling term in the Schr\"odinger 
frame reproduces the structure of $H_0$ with  $x$ and $z$ components but not a $y$-component. (For three-level systems see \cite{Song2016_052324,Huang2016_105202,Kang2016_36737,Wu2017_34}.)
Unitarily transforming $K_0$, see Sec. \ref{beyond} below,  also provides a Hamiltonian without $y$-component, which is different from the term $H_{cd}^{(1)}=A_0K_1A_0^\dagger$ generated from the first superadiabatic iteration, see an explicit comparison in \cite{Ibanez2012_100403}, where $H_{cd}^{(1)}$ was shown to have smaller intensity than $H_{CD}$. 

To avoid confusion we discourage the use of the expression ``superadiabatic'' to refer to the regular CD approach 
(in fact a zeroth order in the superadiabatic iterative frame) or its unitarily transformed 
versions discussed below in Sec. \ref{beyond}.  
This use of the word ``superadiabatic'', as being equivalent to CD driving or even generically to all shortcuts  is, however, somewhat extended.\\                     

{\it{DRAG controls.}}
The ``Derivative Removal by Adiabatic Gate'' (DRAG) framework was developed to avoid diabatic transitions to undesired levels in  the context of superconducting 
quantum devices. It has been recently reviewed in \textcite{Theis2018} but we shall  sketch the main ideas here since it is related to the  
superadiabatic 
scheme as formulated e.g. in \textcite{Ibanez2013}.   A typical scenario is that 
the  (super)CD 
term $H_{cd}^{(j)} = B_jK_jB_j^\dagger$ is not physically feasible and does not match the controls in the lab. The DRAG approach addresses this problem by decomposing the controllable Hamiltonian that ``corrects'' the dynamics as $H^{\rm{ctrl}}=\sum_k u_k(t)h_k+\rm{h.c.}$  with control fields $u_k(t)$ and  coupling terms
$h_k$. In a given superadiabatic frame partial contributions to $u_k(t)$  are found by projecting  $H_{cd}^{(j)}$  into the assumed 
Hamiltonian structure. This generates an approximation to the exact uncoupling term so the diabatic coupling, 
even if not cancelled exactly, is reduced. 
An important point is that further iterations, in contrast with the bare superadiabatic iterations, 
typically converge, so that the coupling eventually vanishes. A  variant of this approach applies when the error terms are not independently controlled, because they all depend on some common control, e.g. a single laser field. Different perturbative approximations using power series in the inverse gap energies
were worked out systematically \cite{Theis2018}.

%

\subsubsection{Beyond the basic formalism\label{beyond}}
The  CD Hamiltonian $H_{CD}$ 
often implies different operators  from those in $H_0$, that typically may be 
%
hard or even impossible to generate in the laboratory. Moreover, a lot of spectral information is in principle used to build $H_{CD}$, specifically the 
eigenvectors of $H_0$,     
so a number of strategies, reviewed hereafter,  are put forward to avoid  
some terms in the auxiliary Hamiltonian, and/or the spectral information needed.    
Changing the phases $\xi_n(t)$ without modifying the $|n(t)\ra$ 
changes $H_0$ but not $H_{CD}$, so it is not enough for these purposes \cite{Ibanez2011_023415}.

\paragraph{``Physical'' unitary transformations.\label{utr}} 
A very useful  method to generate alternative, physically feasible shortcuts, from an existing shortcut generated by counterdiabatic driving or otherwise, 
is to perform physical, rather than formal, unitary transformations \cite{Ibanez2012_100403}, or corresponding canonical transformations in  classical
systems \cite{Deffner2014}. 
Given a Hamiltonian $H(t)$ that drives the wave function $|\psi(t)\ra$, the unitarily transformed state $|\psi'(t)\ra={\cal U}^\dagger(t)|\psi(t)\ra$
is driven by the Hamiltonian (the primes here distinguish the picture, they do not represent derivatives) 
\beqa\label{Hprime}
H'(t)&=&{\cal U}^\dagger (H-K) {\cal U},
\\
K&=&i\hbar  \dot{\cal U} {\cal U}^\dagger.
\label{Kdef}
\eeqa
If we set ${\cal U}(0)={\cal U}(t_f)=1$, then the wave functions coincide at the boundary times, $|\psi(0)\ra=|\psi'(0)\ra$
and $|\psi(t_f)\ra=|\psi'(t_f)\ra$. 
If, in addition, $\dot{\cal U}(0)=\dot{\cal U}(t_f)=0$, then also the Hamiltonians coincide at boundary times, $H(0)=H'(0)$ and 
$H(t_f)=H'(t_f)$.  

Note that here  
the alternative Hamiltonian form (\ref{Hprime}) and the state $|\psi'(t)\ra$ are not  just convenient 
mathematical transforms of $H(t)$ or $|\psi(t)\ra$ representing the same physics, as in conventional interaction or Heisenberg picture 
transformations. Instead, at intermediary times, $H'(t)$ and $H(t)$ represent indeed different (laboratory) drivings, 
and $|\psi(t)\ra$ and $|\psi'(t)\ra$ different dynamical states. This was emphasized in  \textcite{Ibanez2012_100403}
by calling the different alternatives ``multiple Schr\"odinger pictures". 

The art is to find a useful ${\cal U}(t)$ to make $H'(t)$ feasible. 
This approach has been applied in many works, see e.g.  \textcite{Hollenberg2012,Takahashi2015,Agundez2017,Sels2017};  
and 
several experiments \cite{Bason2012,Zhang2013_240501,An2016,Du2016}. 
\textcite{Deffner2014} applied it to generate feasible (i.e., involving local potentials in coordinate space, independent of momentum) Hamiltonians for scale-invariant dynamical processes. 
 
When $H(t)$ is a linear combination  of generators $G_a$ 
of some Lie algebra, 
\begin{eqnarray}\label{Lial}
\left[G_{b},G_{c}\right]=\sum_{a=1}^{N} \alpha_{abc} G_{a},
\end{eqnarray}
where the $\alpha_{abc}$ are the structure constants, 
${\cal U}$ may 
be constructed by exponentiating elements of the algebra and imposing the vanishing of the unwanted terms  \cite{Martinez-Garaot2014_053408}.     
To carry out the transformation an element $G$ of the Lie algebra of the Hamiltonian is chosen, 
\beq
\label{Lie_transformation}
{\cal U}(t)=e^{-i g(t) G},
\eeq
where $g(t)$ is a real function to be set.  
This type of unitary operator ${\cal U}(t)$ constitutes a ``Lie transform''.
Note that $K$ in Eq. (\ref{Kdef}) becomes $-\hbar \dot{g}(t)G$ and commutes with $G$ so    
$H'$ is given by 
\beqa
\label{HI_series}
{\cal U}^\dag (H-K) {\cal U}&=&e^{ig G}(H-K)e^{-i g G}
\nonumber
\\
&=&H-\hbar\dot{g}G+ig [G,H]-\frac{g^2}{2!}[G,[G,H]]
\nonumber
\\
&-&i\frac{g^3}{3!}[G,[G,[G,H]]] + \cdots
\label{tra}
\eeqa
which depends only on $G$, $H$, and its nested commutators with $G$, 
so it stays in the algebra. 
If we can choose $G$ and $g(t)$ so that the undesired generator components in $H(t)$  cancel out 
and the boundary conditions 
for ${\cal U}$ are satisfied, the method provides a feasible, alternative shortcut. 
A simple example for the two-level Hamiltonian is given in Appendix \ref{elt}. \textcite{Martinez-Garaot2014_053408,Kang2018} provide
more examples.\\

\begin{figure}[t]
\begin{center}
\includegraphics[height=5.0cm,angle=0]{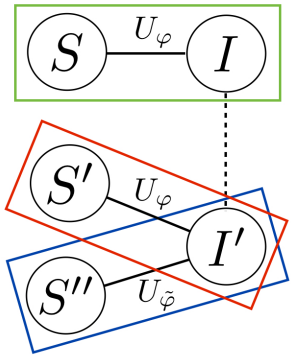}
\end{center}
\caption{\label{figurepictures}
(Color online)
Schematic relation between different Schr\"{o}dinger and interaction pictures. Each node corresponds also to different Hamiltonians. The rectangular boxes enclose nodes that represent the same underlying physics. The solid lines represent unitary relations for the linked states and the dashed line represents a non-unitary addition of an auxiliary term to the Hamiltonian.\label{msp}}
\end{figure}

{\it Interaction pictures.\label{pic}}
Frequently the CD terms are found formally in a transformed picture, $I$ in Fig. \ref{figurepictures},
which is only intended as a mathematical aid. The effective Hamiltonian $H_{I}$ in this picture, without the CD term, 
represents the same physics than some original Hamiltonian $H_S$ in the Schr\"odinger picture $S$ in Fig. \ref{figurepictures}. Once the CD term is added we get a new Hamiltonian 
$H_I'$ that, when transformed back gives $H_{S'}$. It happens often that      
simple-looking auxiliary terms in the transformed frames become difficult  to implement  in the laboratory frame. 
In Sec. \ref{mbsm} we shall comment on some $N$-body models  with easily solvable CD terms 
in a convenient transformed  frame but very hard to realize in the lab frame. 
Alternative unitary transformations $U_{\tilde{\varphi}}$, different from the one used to go between the $I$  and $S$ pictures, $U_\varphi$, may help to solve the problem,  
giving from $H_{I'}$ feasible shortcuts driven by a new lab frame Hamiltonian $H_{S''}$.   
\textcite{Ibanez2015} worked out this alternative route  for two-level systems when the rotating
wave approximation is not applicable, see also 
\textcite{Ibanez2011_013428,Li2017_30135,Chen2015_023405}. 
Physics beyond the rotating wave approximation is of much current interest due to the increasing use of strong fields and microwave 
frequencies, for example in Nitrogen vacancy (NV) centers. 

%
\begin{table*}[t]
\caption{Scheme for ``dressed-state'' driving \cite{Baksic2016}\label{tableBaksic}}. 
\begin{ruledtabular}
\begin{tabular}{llll}
{\rm{picture}}&wavefunction&unitary transformation&Hamiltonian 
\\\hline
Schr\"odinger   &        $\psi(t)$  &&$H=H_0+H_c$
\\
{\rm{first}} rotating picture (``adiabatic'') &       $\psi_I(t)=U^\dagger \psi(t)$&              $U=\sum_n |n(t)\ra\la n|$ & $H_I=i\hbar \dot{U}^\dagger U+U^\dagger HU$
\\
{\rm{second}} rotating picture (``dressed'')&   $\psi_{II}(t)=V^\dagger \psi_I(t)$& $V=\sum_n |\tilde{n}(t)\ra\la n|$ & $H_{II}=i\hbar\dot{V}^\dagger V +V^\dagger H_I V$
\end{tabular}
\end{ruledtabular}
\end{table*}

\paragraph{Schemes that focus on one state.}
A simplifying assumption  that helps to find 
simpler decoupling terms is to focus on only one state, typically the ground state $|0(t)\ra$, or a subset of states.
Explicit exact forms of the driving that uncouples that state were worked out quite early \cite{Demirplak2008}, see Eq. (\ref{onestate}), and more recently for 
systems described in coordinate space in \textcite{Patra2017_125009} or in a discrete basis \cite{Setiawan2017}. 
As for approximate schemes, \textcite{Opatrny2014} proposed to add to $H_0(t)$ the Hamiltonian $H_B(t)=\sum_k^K f_k(t) T_k$ using only feasible interactions (i.e., available experimentally) $T_k$, 
where the $f_k(t)$ are amplitudes found by  minimizing  the
norm of $(H_B- H_{CD})|0(t)\rangle$.
A related approach uses  Lyapunov control theory \cite{Ran2017}.  Similarly,
\textcite{Chen2016_052109, Chen2017_062319} achieved feasible auxiliary Hamiltonians 
by adding adjustable Hamiltonians that  
nullify unwanted nonadiabatic couplings for specific transitions. 

\paragraph{Effective counterdiabatic field.}
In \textcite{Petiziol2018,Petiziol2019} $H_0(t)=\sum_k u_k(t)H_k$  is written as before in terms of control functions $u_k(t)$
and available time-independent control Hamiltonians $H_k$.  Using control theory arguments it is found  that $H_{CD}$ necessarily belongs to the 
corresponding dynamical Lie algebra, i.e., the smallest algebra that contains the $-iH_k$ and the nested commutators. As well, the action
of $H_{CD}$ can be approximated by using only the initially available Hamiltonians in an ``effective counterdiabatic'' Hamiltonian $H_E=\sum e_k(t) H_k$. To 
implement $H_E$  
in \textcite{Petiziol2018,Petiziol2019}, the control functions $e_k(t)$ are chosen as periodic functions of period $T$ 
with the form of a truncated Fourier expansion, $\sum_k {\cal A}_k \sin (k\omega t)+{\cal B}_k \cos(k\omega t)$, where $\omega=2\pi/T$. 

The coefficients are determined by setting the first terms of the Magnus expansion generated by $H_E$ to match those of the desired evolution stroboscopically, at 
multiples of $T$, and interpolating smoothly in between.  Avoided crossing problems and entanglement creation are addressed with this technique. 
Note alternative uses of the Magnus expansion in the WAHWAH technique of \textcite{Schutjens2013}, aimed at producing fast pulses 
to operate on a qubit without interfering with other qubits in frequency-crowded systems, and in \textcite{Claeys2019}.

\paragraph{Dressed-states approach.} 
CD driving is generalized in \textcite{Baksic2016} by considering a dressed-states approach which uses 
three different dynamical pictures.  In this summary  the notation and even the terminology differs from 
\textcite{Baksic2016}, see Table \ref{tableBaksic}. 

First consider a 
Schr\"odinger picture description where the driving Hamiltonian is $H_0(t)+H_c(t)$ with reference Hamiltonian $H_0(t)$ as in Eq. (\ref{H0}),
and driven wavefunction $|\psi(t)\ra$.  $H_c(t)$ will be found later so that the total Hamiltonian
satisfies some conditions. Then a first rotating picture defined by 
$|\psi_I(t)\ra=U(t)^\dagger|\psi(t)\ra$ is introduced, where $U(t)=\sum_n |n(t)\ra\la n|$ and the ${|n\ra}$ are time-independent.  
A simple choice is $|n\ra=|n(0)\ra$ that insures $U(0)=U(t_f)=1$. Since $|n(t)\ra$ are instantaneous (adiabatic) eigenstates 
of $H_0(t)$ we may naturally call this picture the ``adiabatic frame'', with driving Hamiltonian $H_I(t)$, see Table \ref{tableBaksic}.
Then, a second rotating frame is introduced as $|\psi_{II}(t)\ra=V(t)^\dagger|\psi_I(t)\ra$, with $V(t)=\sum_n |\tilde{n}(t)\ra \la {n}|$, driven by 
$H_{II}(t)$. $\{|\tilde{n}(t)\ra\}$ are called dressed states and it is assumed that $V(0)=V(t_f)=1$. 
The method to generate shortcuts is to choose $V(t)$, i.e., the functions $|\tilde{n}(t)\ra$, and $H_c$ so that $H_{II}(t)$ is diagonal 
in the basis $\{|n\ra\}$, $H_{II}(t)=\sum_n |n\ra E_n^{II}(t) \la n|$. Back to the Schr\"odinger 
picture this means that no transitions occur among states 
$U(t)V(t)|n\ra=U(t)|\tilde{n}(t)\ra$.  

$V(t)$ is chosen  to ensure that $H_c(t)$ is feasible.
The two nested transformations make the method more involved and less intuitive than standard CD driving. To gain insight note that 
if $V(t)=1$ for all $t$ the method reduces to standard CD driving. One can think of $V$ as a way to add flexibility to the inverse engineering 
so as to drive states in the Schr\"odinger picture  along uncoupled  $U(t)V(t)|{n}\ra$ vectors rather than along vectors $U(t)|n\ra=|n(t)\ra$. The latter are given, up to phases, once 
$H_0(t)$ is specified, whereas $U(t)V(t)|n\ra$ may still be manipulated to find a convenient $H_c(t)$ and possibly minimize the occupancy of some  
state to be avoided, e.g. because of spontaneous decay \cite{Baksic2016}.      

For applications see \textcite{Baksic2017, Wu2017_46255, Zhou2017_330,Zhou2017_095202,Coto2017,Liu2017_062308}. 

\paragraph{Variational approach.\label{vara}} 
Motivated by difficulties to diagonalize $H_0(t)$ 
and the non-localities in the exact counterdiabatic Hamiltonian in many-body 
systems, \textcite{Sels2017, Kolodrubetz2017} 
developed a variational method to construct approximate counterdiabatic Hamiltonians without using spectral information. 
The starting point in \textcite{Sels2017} is a unitary transformation $U[\lambda(t)]=\sum_n |n[\lambda(t)]\ra \la n|$ 
to rotate the state $|\psi(t)\rangle$ that evolves under a time dependent Hamiltonian $H_0[\lambda(t)]$,
to the moving frame state  $|\tilde\psi(t)\rangle= U^\dagger[\lambda(t)]|\psi(t)\rangle$, which  satisfies the effective Schr\"odinger equation
\beq
\label{moving}
i\hbar\partial_t|\tilde\psi\rangle = \bigg({\tilde{H}}_0[\lambda(t)]-\dot\lambda{\tilde{A}}_{\lambda}\bigg)|\tilde\psi\rangle,
\eeq
where ${\tilde{H}}_0$ is diagonal in the $|n\ra$ basis and ${\tilde{A}}_{\lambda}$ is  the adiabatic ``gauge potential'' in the moving frame,
\beqa
\label{h0a}
{\tilde{H}}_0[\lambda(t)] &=& U^{\dag}  H_0[\lambda(t)] U =\sum_nE_n(\lambda)|n\rangle\langle n|, 
\nonumber\\
{\tilde{A}}_{\lambda}&=&i\hbar U^{\dag} \partial{\lambda}  U.
\eeqa
All nonadiabatic transitions are  produced by the gauge potential. In the counterdiabatic approach
the system is driven by  the Hamiltonian 
\beq\label{Hf}
 H(t)= H_0+\dot\lambda A_{\lambda}, 
\eeq
where $A_\lambda=U{\tilde{A}}_{\lambda}U^\dagger=i\hbar (\partial_{\lambda} U) U^{\dag}$ such that, in the moving frame $H_0^{eff}={\tilde{H}}_0$ is diagonal and no transitions are allowed. Up to now we have only introduced 
a new notation and terminology in Eqs. (\ref{moving}-\ref{Hf}).\footnote{We have assumed for simplicity a dependence on time via a single parameter $\lambda$. A more general dependence on a vector $\bm{\lambda}$ is worked out e.g. in \textcite{Deffner2014,Nishimura2018} which leads  to a decomposition of the CD term and a ``zero curvature condition'' among the CD components.} 
In the adiabatic limit $|\dot\lambda|\rightarrow 0$ the Hamiltonian (\ref{Hf}) reduces to the original one $ H(t)= H_0(t)$.
After differentiating the first equation in Eq. (\ref{h0a}), the gauge potential satisfies \cite{Jarzynski2013}
\beq
\label{gauge}
i\hbar(\partial_{\lambda} H_0+ F_{ad})=[ A_{\lambda}, H_0],
\eeq
where $ F_{ad}=-\sum_n\partial_{\lambda}E_n(\lambda)|n(\lambda)\rangle\langle n(\lambda)|$
is the adiabatic force operator. 
Since, by construction, 
$F_{ad}$ commutes with $H_0$, Eq. (\ref{gauge}) implies that
\beq
\label{gauge2}
[i\hbar\partial_{\lambda} H_{0}-[{A}_{\lambda}, H_0], H_0]=0,
\eeq
were the difficult-to-calculate force has been eliminated. 
This equation can be used to find the adiabatic gauge potentials directly without diagonalizing the Hamiltonian.
Moreover, solving this equation is analogous to minimizing the Hilbert-Schmidt norm of the operator
\beq
 G_{\lambda}=\partial_{\lambda} H_0+\frac{i}{\hbar}[ A^*_{\lambda}, H_0]
\eeq
with respect to $A^*_{\lambda}$, where $A^*_\lambda$ is a trial gauge potential.  
For an application to prepare ground states in a lattice gauge model see \textcite{Hartmann2018}. 

Potential problems with the above scheme are that it may be difficult to know what local operators should be included
in the variational basis, and moreover their practical realizability is not guaranteed.      
 To solve these two difficulties, \textcite{Claeys2019} expand the gauge potential  in terms of nested commutators of the Hamiltonian $H_0$ and the driving term $\partial_\lambda H_0$, 
even if these do not close a Lie algebra.  
Since the commutators also arise in the Magnus expansion in Floquet systems, they can be realized up to arbitrary order using Floquet-engineering, i.e.,  periodical driving \cite{Boyers2018}.
The expansion coefficients can either be calculated analytically or variationally. The method can be adapted to suppress excitations in a known frequency window.   
Application examples were provided for a  two- and a three-level system, where one, respectively two terms returned the exact gauge potential, and to a many-body spin chain, where a limited number of terms and an approximate gauge potential resulted in a drastic increase in fidelity. Further useful properties  of the expansion are that 
it relates the locality of the gauge potential to the order in the expansion, and that it remains well-defined in the thermodynamic and classical limits.

\paragraph{Counterdiabatic Born-Oppenheimer dynamics.}
\textcite{Duncan2018} propose to exploit the separation of fast and slow variables using the Born-Oppenheimer approximation. 
Simpler CD terms (compared to the exact CD term) for the fast and slow variables can be found in two steps avoiding the diagonalization of the full Hamiltonian.  The method is tested for two coupled harmonic oscillators and a system of two charged particles.    

\paragraph{Constant CD-term approximation.} 
\textcite{Oh2014} proposed, in the context of the adiabatic Grover's search algorithm  implemented by a two-level system, to substitute  the exact CD term  by a constant term. This method  improves the scaling of the nonadiabatic transitions with respect to the running time.
\textcite{Zhang2018_052323} performed an experiment with a single trapped ion choosing the constant term with the aid of a numerical simulation.  

In a different vein \textcite{Santos2018_025301} discuss the conditions for $H_0(t)$, and phase choice necessary in the unitary evolution operator (\ref{eq:U}) 
to implement an exact  constant driving.  

\subsection{Invariants and scaling laws\label{ssec:inv}}

Dynamical invariants and invariant-based engineering constitute a major route  to design STA protocols. 
The basic reason, already sketched in Eqs. (\ref{ut}) and (\ref{Hbasic}),  is that, in linear systems, 
determining the desired dynamics amounts to setting dynamical invariants of motion, and 
the Hamiltonian may in principle be found from them. 
This connection is quite general, and is also valid classically \cite{Lewis1982}, 
but it is mostly  applied for systems in which the Hamiltonian 
form and corresponding invariants are known explicitly as functions of auxiliary parameters that satisfy
auxiliary equations consistent with the dynamical equation  and the invariant.   
\subsubsection{Lewis-Riesenfeld Invariants\label{lri}}
Originally proposed in 1969, a Lewis--Riesenfeld invariant \cite{Lewis1969} for a Hamiltonian $H(t)$ is a Hermitian operator $I(t)$ which satisfies
\begin{equation}
\frac{d I}{d t}=\frac{\partial I}{\partial t}+\frac{i}{\hbar}\left[H,I\right]=0,
\label{eq:LR}
\end{equation}
so that the expectation values for states driven by $H(t)$ are constant in time. 
Since $I(t)$ is a constant of motion it has time--independent eigenvalues.
If $\left|\phi_{n}(t)\right\rangle$ is an instantaneous eigenstate of $I(t)$, a solution 
of the Schr\"{o}dinger equation, $i\hbar \partial_t \left|\psi_n(t)\right\rangle = H(t) \left|\psi_n(t)\right\rangle$, can be constructed  as $\left|\psi_{n}(t)\right\rangle=e^{i \alpha_{n}(t)} \left|\phi_{n}(t)\right\rangle$. Here  $\alpha_{n}(t)=(1/\hbar)\int_{0}^t \left\langle\phi_{n}(s)\right.\left|\left[i\hbar\partial_{s}-H(s)\right]\right|\left.\phi_{n}(s)\right\rangle ds$ is the Lewis--Riesenfeld phase. Hence a  general solution to the Schr\"{o}dinger equation can be written as
\begin{equation}
\left|\psi(t)\right\rangle=\sum_{n} c_{n} \left|\psi_{n}(t)\right\rangle,
\end{equation}
where the $c_{n}$ are independent of time. 

These invariants were originally used to solve for the state driven by a known  time-dependent Hamiltonian  \cite{Lewis1969}. In shortcuts to adiabaticity   this idea is reversed \cite{Chen2010_063002} and the Hamiltonian is found from a prescribed state evolution.
Formally,  a time evolution operator of the form
\begin{eqnarray}
U=\sum_{n} e^{i \alpha_{n}(t)}\ketbra{\phi_{n}(t)}{\phi_{n}(0)}
\end{eqnarray}
implies a 
Hamiltonian $H(t)=i \hbar \dot{U} U^{\dagger}$. This is the essence of invariant-based inverse engineering.

For a given Hamiltonian there are many possible invariants. For example, the density operator describing the evolution of a system is a dynamical invariant. The choice of which particular invariant to use is made on the basis of mathematical convenience. Invariants have also been generalized to non--Hermitian invariants and Hamiltonians \cite{Gao1992,Lohe2009,Ibanez2011_023415}, as well as open systems (see Sec. [\ref{sec:open}]). 
     
%
A connection to adiabaticity is asymptotic. In the limit of long operation times (i.e. adiabatic), Eq. \eqref{eq:LR} becomes $\left[H,I\right] \approx 0$ and the dynamics prescribed by invariant-based engineering will approach adiabatic dynamics driven by $H$. Hence for long times, $H(t)$ and $I(t)$ have approximately a common eigenbasis for all times. 

Other relations to adiabaticity do not need  long times nor common bases for $H(t)$ and $I(t)$ at all times.  In particular,  
demanding only that the invariant and the Hamiltonian commute at the start and the end of the process, i.e., $\left[I(0),H(0)\right]=\left[I(t_f),H(t_f)\right]=0$,  the eigenstates of the invariant and the Hamiltonian coincide at initial and final times, but may differ with each other at intermediate times
because the commutativity is not imposed.  If no level crossings take place, the final state 
will keep the initial populations for each $n$-th level, as in an adiabatic process, but in a finite  (short, faster-than-adiabatic)  time.     
This leaves freedom to choose how the state evolves in the intermediate time and then use Eq. \eqref{eq:LR} to find the Hamiltonian that drives  such a state evolution. The flexibility can be exploited to improve the stability of the schemes against noise and systematic errors, see Sec. \ref{ssec:robustness}.

The relation of  the invariant-based approach to CD driving \cite{Chen2011_062116} implies a different connection to adiabaticity.  So far we have not mentioned 
in this section  any ``reference'' $H_0(t)$, but,   
by reinterpreting $|\phi_n(t)\ra$ as eigenvectors 
of $H_0(t)$, and the Lewis-Riesenfeld phases as adiabatic phases, an  implicit  $H_0(t)$ operator may be written down using Eq. (\ref{h0xi}). Hence  
an implicit $H_{CD}(t)$ follows from subtraction, $H_{CD}(t)=H(t)-H_0(t)$. In invariant-based engineering, however, these implicit operators are not really used
and do not play any role in practice. As a consequence of the different emphases and construction recipes for $H(t)$,
given some initial and final Hamiltonians, STA protocols designed via CD or invariant approaches are often  very different.\\

{\it Lie Algebras.}
As first noted in \textcite{Sarandy2011}, invariant based inverse engineering can also be formulated in terms of dynamical Lie algebras. The assumption of a Lie structure is also used to classify
and construct dynamical invariants for four-level or smaller nontrivial systems for specific applications
\cite{Gungordu2012,Herrera2014,Kiely2016}.
%
The approaches in 
\textcite{Petiziol2018,Martinez-Garaot2014_053408}, already discussed, also make use of the Lie algebraic structure.  

\textcite{Torrontegui2014} provides a bottom-up construction procedure of the Hamiltonian using invariants and Lie-algebras. Let us assume that the Hamiltonian of a system $H(t)$ and the invariant $I(t)$ can be written as  linear combinations of Hermitian operators $G_{a}$ (generators),
\begin{eqnarray}
H(t)=\sum_{a=1}^{N} h_{a}(t) G_{a},
\;\;\;
I(t)=\sum_{a=1}^{N} f_{a}(t) G_{a},
\end{eqnarray}
that form a Lie algebra closed under commutation, see Eq. (\ref{Lial}). 
Inserting 
these forms into Eq. \eqref{eq:LR}, we get that
\begin{eqnarray}
\dot{f}_{a}(t)-\sum_{b=1}^{N}{\cal A}_{ab}(t)h_{b}(t)=0,
\end{eqnarray}
where the $N \times N$ matrix ${\cal A}$ is defined by
\begin{eqnarray}
{\cal A}_{ab} \equiv \frac{1}{i \hbar} \sum_{c=1}^{N} \alpha_{abc} f_{c}(t),
\end{eqnarray}
and the $\alpha_{abc}$ are the structure constants. 
In  vector form, 
\begin{eqnarray}
\partial_{t} \vec{f}(t)={\cal A} \vec{h}(t),
\label{invariant_vec_eq}
\end{eqnarray}
where the vectors $\vec{f}(t)$ and $\vec{h}(t)$ represent the invariant and Hamiltonian respectively. The inversion trick is to 
choose first 
the auxiliary functions $\vec{f}(t)$ (and therefore the state evolution) and infer $\vec{h}(t)$ from this. Technically the inversion requires 
introducing a projector ${\cal Q}$ for the null-subspace of ${\cal A}$ and the complementary projector ${\cal P}$. 
In ${\cal P}$-subspace a pseudoinverse matrix can be defined, and the ${\cal Q}$ component of $\vec{h}$ is chosen to make the resulting Hamiltonian realizable \cite{Torrontegui2014}. Examples for two ($\text{SU}(2)$ algebra) and three-level systems (with the four-dimensional algebra $\text{U3S3}$) were provided.  
\textcite{Levy2018} reformulated this formalism in terms of the density operator and used it to design robust control protocols against the influence of different types of noise. In particular, they developed a method to construct a control protocol which is robust against dissipation of the population and minimizes the effect of dephasing. 

\subsubsection{Examples of Invariant-based Inverse Engineering\label{examplesIBIE}}

Lewis-Riesenfeld invariants have  been used to design state transfer schemes in two-level \cite{Ruschhaupt2012,Martinez-Garaot2013,Kiely2014}, three-level \cite{Chen2012,Kiely2014,Benseny2017} and four-level systems \cite{Kiely2016,Gungordu2012,Herrera2014}, or Hamiltonians quadratic in creation and
annihilation operators \cite{Stefanatos2018_07129}.
They are also very useful to design  motional dynamics
in harmonic traps or otherwise.  
Many more applications specific for different system types may be found in  Sec. \ref{qt}. Here we provide two basic examples, the two-level model 
and the Lewis-Leach family of potentials for a particle of mass $m$ moving in 1D.

\paragraph{Two-level system}
We now present an example of a two-level system with a Hamiltonian given by
\begin{eqnarray}
H(t)=\frac{\hbar}{2}\left(\begin{array}{cc}
-\Delta(t) & \Omega_{R}(t)-i\Omega_{I}(t)
\\
\Omega_{R}(t)+i\Omega_{I}(t) & \Delta(t)
\end{array}\right)\!.\label{eq:1}
\end{eqnarray}
and an invariant of the form 
\begin{eqnarray}
I\left(t\right)=\frac{\hbar}{2}\left(\begin{array}{cc}
\cos\left[\theta\left(t\right)\right] & \sin\left[\theta\left(t\right)\right]e^{-i\alpha\left(t\right)}\\
\sin\left[\theta\left(t\right)\right]e^{i\alpha\left(t\right)} & -\cos\left[\theta\left(t\right)\right]
\end{array}\right).
\end{eqnarray}
From the equation that defines the invariant, Eq. \eqref{eq:LR}, $\theta(t)$ and $\alpha(t)$ must satisfy
\begin{eqnarray}
\dot{\theta}&=&\Omega_{I}\cos\alpha-\Omega_{R}\sin\alpha,\label{eq:4}\\
\dot{\alpha}&=&-\Delta-\cot\theta\left(\Omega_{R}\cos\alpha+\Omega_{I}\sin\alpha\right)\,.\label{eq:5}
\end{eqnarray}
The eigenvectors of $I\left(t\right)$ are 
\begin{eqnarray}
\left|\phi_{+}\left(t\right)\right\rangle &=&\left(\begin{array}{c}
\cos\left(\theta/2\right)e^{-i\alpha/2}\\
\sin\left(\theta/2\right)e^{i\alpha/2}
\end{array}\right)\,,\\
\left|\phi_{-}(t)\right\rangle &=&\left(\begin{array}{c}
\sin\left(\theta/2\right)e^{-i\alpha/2}\\
-\cos\left(\theta/2\right)e^{i\alpha/2}
\end{array}\right),
\end{eqnarray}
with eigenvalues $\pm \hbar/2$. The angles $\alpha$ and $\theta$ can be thought of as spherical coordinates on the Bloch sphere. The general
solution of the Schr\"{o}dinger equation is then a linear combination of the eigenvectors of $I\left(t\right)$
i.e. $\left|\Psi\left(t\right)\right\rangle =c_{+}e^{i\kappa_{+}(t)}\left|\phi_{+}\left(t\right)\right\rangle +c_{-}e^{i \kappa_{-}(t)}\left|\phi_{-}(t)\right\rangle $
where $c_{\pm}\in\mathbb{C}$ and $\dot{\kappa}_{\pm}\left(t\right)=\frac{1}{\hbar}\left\langle \phi_{\pm}\left(t\right)\right.\left|\left(i\hbar\partial_{t}-H\left(t\right)\right)\right|\left.\phi_{\pm}\left(t\right)\right\rangle\,.$
Therefore, it is possible to construct a particular solution 
\begin{eqnarray}
\left|\psi\left(t\right)\right\rangle
=\left|\phi_{+}\left(t\right)\right\rangle
e^{-i\gamma\left(t\right)/2}
\end{eqnarray}
where $\gamma = \pm 2 \kappa_{\pm}$ and 
\begin{eqnarray}
\dot{\gamma}=\frac{1}{\sin\theta}\left(\Omega_{R}\cos\alpha+\Omega_{I}\sin\alpha\right)\,.\label{eq:11}
\end{eqnarray}
Using Eqs. \eqref{eq:4},
\eqref{eq:5} and \eqref{eq:11} we can retrieve the physical quantities in terms of the auxiliary functions,
\begin{eqnarray}
\Omega_{R}&=&\cos\alpha\sin\theta\,\dot{\gamma}-\sin\alpha\,\dot{\theta}\label{eq:12}\,,\\
\Omega_{I}&=&\sin\alpha\sin\theta\,\dot{\gamma}+\cos\alpha\,\dot{\theta}\label{eq:13}\,,\\
\Delta&=&-\cos\theta\,\dot{\gamma}-\dot{\alpha}\label{eq:14}\,.
\end{eqnarray}
If the functions $\alpha,\gamma$, and $\theta$
are chosen with the appropriate boundary conditions,  different state manipulations are possible.  For example  $\theta\left(0\right)=0$
and $\theta\left(t_f\right)=\pi$ imply  perfect population inversion at a time $t_f$.
Note the freedom to interpolate along different paths. 

\paragraph{Lewis-Leach Family.\label{sLLf}}
Consider a one-dimensional Hamiltonian $H=p^{2}/2m+V(q,t)$ with potential \cite{Lewis1982}
\begin{eqnarray}
\label{LLf}
V(q,t)&=&-F(t)q+\frac{m}{2}\omega^{2}(t)q^{2} \nonumber\\
&+&\frac{1}{\rho(t)^{2}}U\!\left[\frac{q-q_{c}(t)}{\rho(t)}\right]+g(t).
\end{eqnarray}
These Hamiltonians have a quadratic in momentum invariant 
\beqa
I&=&\frac{1}{2m}\left[\rho\left(p-m\dot{q_{c}}\right)-m\dot{\rho}\left(q-q_{c}\right)\right]^{2}
\nonumber
\\
&+&\frac{1}{2}m\omega^{2}_{0}\left(\frac{q-q_{c}}{\rho}\right)^{2}+U\left(\frac{q-q_{c}}{\rho}\right), \label{LL_invariant}
\eeqa
provided the 
functions $\rho$, $q_{c}$, $\omega$, and $F$ satisfy the  auxiliary equations
\begin{eqnarray}
\ddot{\rho}+\omega^{2}(t)\rho &=& \frac{\omega^{2}_{0}}{\rho^{3}}, 
\nonumber \\
\ddot{q_{c}}+\omega^{2}(t)q_{c} &=& F(t)/m, \label{ermakov_friends}
\end{eqnarray}
with $\omega_{0}$ a constant. 
The first equation is known as the Ermakov equation 
\cite{Ermakov1880} while the second is the Newton equation of
motion for a forced harmonic oscillator. 
They can be found by inserting the quadratic-in-$p$ invariant, Eq. \eqref{LL_invariant}, into Eq. \eqref{eq:LR}. 
The properties of such invariants have also been formulated in terms of Feynman propagators \cite{Dhara1984}.

For  this family of Hamiltonians we can explicitly calculate the Lewis-Riesenfeld phase,
\begin{eqnarray}
&&\alpha_{n}(t)= \nonumber\\
&&-\frac{1}{\hbar}\! \int_{0}^{t}\!\! dt'\! \left[\frac{\lambda_{n}}{\rho^{2}}+\frac{m[\left(\dot{q_{c}}\rho-q_{c}\dot{\rho}\right)^{2}-\omega_0^2q_c^2/\rho^2]}{2 \rho^{2}}+g\right]\!\!,        
\end{eqnarray}
and the eigenvectors in coordinate representation,
\begin{eqnarray}
&&\phi_{n}(q,t)= 
\nonumber \\
&&\exp\left\{\frac{i m}{\hbar}\left[ \dot{\rho}q^{2}/2\rho+\left(\dot{q_{c}}\rho-q_{c}\dot{\rho}\right)q/\rho\right]\right\}\rho^{-1/2}\Phi_{n}\left(\frac{q-q_{c}}{\rho}\right), 
\nonumber \\
\end{eqnarray}
where $\Phi_{n}(\sigma)$ is a solution of the  stationary Schr\"{o}dinger equation
\begin{eqnarray}
\left[-\frac{\hbar^{2}}{2m}\frac{\partial^{2}}{\partial \sigma^{2}}+\frac{1}{2}m \omega^{2}_{0}\sigma^{2}+U\left(\sigma\right)\right]\Phi_{n}=\lambda_{n}\Phi_{n},
\end{eqnarray}
with $\sigma=({q-q_{c}})/{\rho}$.
This quadratic invariant has been instrumental in designing many of the schemes which manipulate trapping potentials for expansion/compressions, transport, launching/stopping or combined processes. 
It is key, for example, to manipulate the motion of trapped ions, see Sec. 
\ref{trapped}, and in proposals to implement 
STA-based interferometry \cite{Dupont-Nivet2016,Martinez-Garaot2018}.  

Designing first  the function $\rho(t)$ which determines the wavefunction width, and $q_{c}(t)$ (a  classical particle trajectory), the force $F(t)$  and $\omega(t)$  can be determined  using Eq. \eqref{ermakov_friends}.  
The boundary values of the auxiliary functions at the time limits are fixed to satisfy physical conditions and commutativity of 
$H$ and the invariant. The non-unique interpolation is usually done with polynomials or trigonometric functions.\\

{\it Expansions.}
Invariant based inverse engineering for Lewis-Leach Hamiltonians  was first implemented in \cite{Chen2010_063002} to cool down a trapped atom by expanding the trap. Expansions using invariant-based engineering were first implemented with ultracold atoms in a pioneering 
experiment on STA techniques by \textcite{Schaff2010,Schaff2011_113017}. Such cooling protocols have also been envisioned to optimize  sympathetic cooling \cite{Choi2011,Onofrio2017}. 
For very short processes, $t_f<1/(2 \omega_f)$ where $\omega_f$ is the trap frequency at the final time, $\omega^2(t)$ may become negative during some time interval. While this implies a transient repulsive potential, the atoms always remain confined \cite{Chen2010_063002}. 
A repulsive potential may or may not be difficult to implement depending on the physical setting. For example, the analysis in \textcite{Torrontegui2018} suggests that it is viable for trapped ions. 
   
Compared to the simplicity of  invariant-based engineering,  the CD approach for expansions/compressions provides, for  an $H_0$ characterized by some predetermined time-dependent frequency $\omega(t)$, a non-local, cumbersome counterdiabatic term $H_{CD}=-(pq+qp)\dot{\omega}/(4\omega)$ \cite{Muga2010}. 
However,  a unitary transformation produces a new shortcut with local potential and  modified frequency   \cite{Ibanez2012_100403}
\beq\label{omegap}
\omega'=\left[\omega^2-\frac{3\dot{\omega}^2}{4\omega^2}+\frac{\ddot{\omega}}{2\omega}\right]^{1/2},
\eeq  
see  further connections among the two Hamiltonians in \textcite{delCampo2013,Mishima2017}.

So far we have only considered  one-dimensional motion. 
Formally, the three coordinates in an ideal harmonic trap are uncoupled so expansion or transport processes can be treated independently. However, in many cold atom experiments changing the intensity of a laser beam affects simultaneously the longitudinal and transversal frequencies \cite{Torrontegui2012_033605}.\\

{\it Transport.}
Designing particle transport has been another major application of the invariant (\ref{LL_invariant}) \cite{Torrontegui2011,Tobalina2018,Ness2018}, see also Secs. \ref{trapped} and \ref{sOCT}. 
Note that $U$ in Eq. (\ref{LLf}) is arbitrary. Setting $\omega=\omega_0=0$,
and $\rho=1$,  
$F=m \ddot{q}_c$ plays the role of a ``compensating force'' that  cancels the inertial effects of a moving $U[q-q_c(t)]$ so that the wavefunction stays at rest in the frame moving with $q_c$.\footnote{For rigidly moving harmonic traps, a formal  alternative to the compensating force is to set $U=0$ and  $F=m\omega_0^2 q_0(t)$, keeping $\omega(t)=\omega_0$ as the trap frequency \cite{Torrontegui2011}. The two routes may be shown to be equivalent up to a gauge time-dependent term,  
see e.g.  \textcite{Tobalina2018}.}     
$q_c(t)$ can be chosen as an arbitrary function connecting the desired initial and final trap positions.  
The same solution is reached applying fast forward  \cite{Masuda2010}, or using unitary transformations combined with the  CD approach
\cite{Ibanez2012_100403}, which in principle provides the difficult-to-realize term $H_{CD}=p\dot{q}_c$.\footnote{See, however, \textcite{An2016} and Sec. \ref{SOC} for a possible realization.
Note that CD terms like $-(pq+qp)\dot{\omega}/(4\omega)$ or $p\dot{q}_c$ anticommute with the time reversal operator $\Theta$ 
\cite{Sels2017}. Microscopic irreversibility holds by changing the sign of the external forces in the backwards trajectories \cite{Campisi2011}. 
In the two examples above this means to change $\dot{\omega}\to-\dot{\omega}$ and $\dot{q}_c\to-\dot{q}_c$.}  

\textcite{Deffner2014}  discussed more generally that for Hamiltonians  of the form
\beq
\label{60}
H_0=\frac{p^2}{2m}+\frac{1}{\gamma^2}U\left(\frac{q-f}{\gamma}\right)
\eeq
the (nonlocal) CD term is 
\beq
\label{61}
H_{CD}=\frac{\dot{\gamma}}{2\gamma}[(q-f)p+p(q-f)]+\dot{f}p,
\eeq
and found the generic unitary transformation that provides  the local auxiliary terms that appear  in Eq. (\ref{LLf}). 
More general cases for multiparticle systems are discussed in the following subsection.    

\textcite{Tobalina2017} used the invariants to design shortcuts to adiabaticity for nonrigid driven transport and to launch particles in harmonic and general potentials.  Compared to rigid transport, nonrigid transport requieres a more demanding manipulation, but it also provides a wider range of control opportunities, for example to achieve narrow  velocity distributions in a launching process, suitable for accurate ion implantation or low-energy scattering experiments.
\subsubsection{Scaling laws\label{sssec:scaling}}
For many-body systems constructing the Lewis-Riesenfeld invariant is in general much more involved. In \textcite{Takahashi2017_012309} this difficulty is avoided for an infinite-range Ising model in a transverse field, by constructing an invariant using a mean-field ansatz.
However another approach is to exploit scaling laws. If the Hamiltonian fulfills certain scaling laws one can determine the invariant in a similar manner as in Sec. \ref{sLLf}. 
Specifically the Hamiltonian 
\begin{eqnarray}
H(t)&=& \sum_{i=1}^N \left\{\frac{{\bf p}_i^2}{2m}+U[{\bf{q}}_i,\lambda(t)] \right\}+\epsilon(t) \sum_{i<j} V({\bf{q}}_i-{\bf{q}}_j) \nonumber \\
&+&\sum_{i=1}^N\left[-\frac{m}{2}\frac{\ddot{\gamma}}{\gamma}({\bf{q}}_i-{\bf{f}}
)^2-m \ddot{\bf{f}} \cdot {\bf{q}}_i \right],
\end{eqnarray}
describing $N$ interacting particles with the following scaling laws \cite{Deffner2014}
\begin{eqnarray}
U[\bf{q},\lambda(t)] &=& U_0\{[{\bf{q}}-{\bf{f}}(t)]/\gamma(t)\}/\gamma(t)^2,\nonumber \\
V(\kappa {\bf{q}}) &=& \kappa^{-\alpha}V({\bf{q}}),\nonumber \\
\epsilon(t) &=& \gamma(t)^{\alpha-2},
\label{63}
\end{eqnarray}
where $U_0({\bf{q}})=U[{\bf{q}},\lambda(0)]$, has the invariant
\begin{eqnarray}
I&=&\sum_{i=1}^N \frac{1}{2m}\left[ \gamma ({\bf{p}}_i-m \dot{\bf{f}})-m \dot{\gamma}({\bf{q}}_i-{\bf{f}})\right]^2 \nonumber \\
&+&\sum_{i=1}^N U_0\left(\frac{ {\bf{q}}_i-{\bf{f}} }{\gamma}\right)+\sum_{i<j} V\left(\frac{ {\bf{q}}_i-{\bf{q}}_j }{\gamma}\right).
\end{eqnarray}

The method based on Lewis-Riesenfeld invariants is not applicable to non-linear equations such as the Gross-Pitaevskii equation, but 
the presence of scaling laws can still prove beneficial for inverse engineering. 
Scaling solutions for the Gross-Pitaevskii equation were first noticed in \textcite{Kagan1996,Castin96}
and were exploited to perform STA expansions in harmonic traps in \textcite{Muga2009}. 
For example the 1D equation 
\begin{eqnarray}
&& i \hbar \partial_t \psi(x,t)=\nonumber \\ 
&&\left[ - \frac{\hbar^2}{2m} \frac{\partial^2}{\partial x^2}+\! \frac{1}{2} m \omega^{2}(t) {x}^2+g(t) |\psi({x},t)|^2\right]\!\psi({x},t), \label{GPE}
\end{eqnarray}
with $g(t)=g_0/\rho(t)$  
has  scaling solutions of the form
\begin{eqnarray}
\psi({x},t)=\rho^{-1/2} e^{\frac{im\dot{\rho} x^2}{2\hbar \rho}}  e^{-i\mu \tau(t)/\hbar} \Psi({x}/\rho,0)
\end{eqnarray}
provided that the following consistency equations, including the Ermakov equation,  are fulfilled,
\begin{eqnarray}
&&\ddot{\rho}+\omega(t)^2\rho= \frac{\omega_0^2}{\rho^3},
\nonumber \\
&&\tau(t)=\int_0^t \frac{dt'}{\rho^2(t')},
\end{eqnarray}
and $\Psi(y,\tau)$ satisfies
\beq
i\hbar \frac{\partial \Psi}{\partial \tau}=-\frac{\hbar^2}{2m}\frac{\partial^2\Psi}{\partial y^2}+
+\frac{m\omega_0^2}{2}y^2\Psi+g_0|\Psi|^2\Psi.
\eeq
Different dimensions imply different scalings. In 1D and 3D traps, the 
STA scaling solutions for the frequency change are found for either a simultaneous change of the time dependence of the coupling, or a Thomas-Fermi type of regime, whereas 2D traps are privileged in this respect since none of these conditions are needed. 

These methods have been realized experimentally for 
trapped Bose-Einstein condensates \cite{Rohringer2015,Schaff2011_23001}. Scaling solutions also exist for a class of many-body systems including interacting quasi-1D Bose
gases \cite{Gritsev2010} and were applied to construct shortcuts for a square box \cite{delCampo2012_648}, 
and other trapping potentials \cite{delCampo2013,Deffner2014}.

Inverse engineering the transport of condensates  was worked out in \textcite{Torrontegui2012_013031} 
using as an ansatz  the scaling provided by invariant theory for linear dynamics. Scaling has also been used to 
design STA protocols for the     
fast expansion of a condensate in an optical lattice \cite{Yuce2012},  possibly pumped from a reservoir  \cite{Ozcakmakli2012}.\\
%

{\it Fermi gas.}
In \textcite{Papoular2015} a new class of exact scaling solutions is put to work  for 3D ``unitary'' Fermi gases and 2D weakly interacting Bose gases in  anisotropic time-dependent harmonic traps with initial and final traps having the same frequency ratios. These solutions may be useful to implement a 
microscope for quantum defects hosted by the cloud, avoiding the strong distortion due to free expansion. The proposal of \textcite{Papoular2015} is realized  experimentally in \textcite{Deng2018_013628,Deng2018_eaar5909} 
for  a 3D Fermi gas ``at unitarity''.

\subsubsection{Connection with Lax Pairs\label{sec:LaxPairs}}
Lax pairs were originally introduced by Peter Lax in 1968 \cite{Lax1968}. A completely integrable nonlinear partial differential equation (PDE) can be associated with a Lax pair and Lax pairs have been used  to find the solution $u\left(x,t\right)$ of the corresponding nonlinear PDE.
%
The key objective was to construct a pair of linear differential operators $L = L(u), M=M(u)$,
in such a way that the  equation
\begin{eqnarray}
\partial_{t}L(u) +\left[L(u),M(u)\right] =0\label{eq:LM}
\end{eqnarray}
is fulfilled if and only if $u$ is a solution of the  initial non-linear PDE.
Let $\psi(t,x)$ be the eigenvectors of $L(u)$,
\begin{eqnarray}
L(u)\psi = \lambda\psi, \label{eq:L}
\end{eqnarray}
where it follows from Eq. \eqref{eq:LM} that the eigenvalues $\lambda$ must be time-independent.
It also follows from Eq. \eqref{eq:LM} that
\begin{eqnarray}
\partial_{t}\psi = M(u)\psi. \label{eq:M}
\end{eqnarray}
This transforms the problem of solving the non-linear PDE for $u(x,t)$ to that of solving the linear equation  \eqref{eq:LM}
(resp. the linear equations \eqref{eq:L} and \eqref{eq:M}) which is often easier to solve than the initial non-linear PDE.
%

The relation between Lax pairs and shortcuts with counterdiabatic Hamiltonian was first noted in \textcite{Okuyama2016}.
The authors consider systems for which $H_0$ is the invariant for $H=H_0+H_{CD}$ 
 (see Sec. \ref{ssec:cd}). From the equation defining  the invariant 
 the counterdiabatic Hamiltonian can be determined by
\begin{eqnarray}
\partial_t H_0 =\frac{i}{\hbar} \left[H_0,H_{CD}\right]. \label{eq:LPCD}
\end{eqnarray}
By comparing Eqs. \eqref{eq:LPCD} and \eqref{eq:LM}, the connection with Lax pairs is  given by setting $L = H_0$ and $M = - \frac{i}{\hbar} H_{CD}$.
\textcite{Okuyama2016} considered first $H_0 = p^2 + u (x,t)$ and $H_{CD}$ containing third-order in $p$ terms. 
This results in the nonlinear Korteweg-de Vries equation for the physical potential $u (x,t)$.
The advantage is  that a complete set of solutions for this non-linear equation can now be found. Based on these solutions,
the exact counterdiabatic term for a particle in a hyperbolic Scarf potential was determined, and 
unitary transformations were applied to generate feasible auxiliary Hamiltonians avoiding the 
cubic-in-$p$ term. 
The authors discussed a spin lattice as a second example. When 
considering instead that the invariant takes the form $\gamma^2(t) H_0$ and $H_{CD}$  includes up to first orders in $p$, as in  Eq. (\ref{61}), the  scale-invariant potential (\ref{60}) follows. It is also possible to extend the approach to non-scale invariant systems.  

A  different but similar route is to regard 
Lewis-Riesenfeld invariants \cite{Lewis1969} and Lax pairs (the invariants were 
proposed one year after the Lax pairs)  
as two sides of the same coin \cite{Kiely2019},
relating Eqs. (\ref{eq:LR}) and   \eqref{eq:LPCD} by the alternative 
connection  $L = I$ and $M = - \frac{i}{\hbar} H$.
This approach would facilitate to extend the domain of invariant-based shortcuts, for example using  
cubic-in-$p$ invariants with feasible interactions. 
\subsection{Variational methods\label{ssec:variational}}
We have already discussed an STA variational approach in Sec. \ref{vara} 
\cite{Sels2017,Kolodrubetz2017} for  CD driving.  
Here we review other variational proposals. 

\textcite{Takahashi2013_315304,Takahashi2015} reformulated invariant-based engineering or counterdiabatic approaches in terms of 
a quantum brachistochrone variational problem, with the action
\begin{eqnarray}
S= \int_0^T dt \left( \cL_T+ \cL_S+ \cL_C  \right)
\end{eqnarray}
and  Lagrangians corresponding to the constraints for the process time  ($\cL_T$), the Schr\"{o}dinger equation ($\cL_S$), and additional experimental constraints ($\cL_C$). This formulation is used to examine the stability of the driving, noting that processes are stable against variations in operators which commute with $H_{CD}$ and unstable against variations in those that anticommute with it. This work has been extended to classical, stochastic finite-sized systems described by a continuous-time master equation to find an optimal transition-rate matrix \cite{Takahashi2016}.

A different variational approach is worked out in   
\textcite{Li2016_38258,Li2017_015005}. These works  consider a Bose-Einstein condensate (BEC) trapped in a harmonic trap.  Quantum  dynamical equations  follow by minimizing the action defined from a Lagrangian density, using some  ansatz for the wave function.
This allows one to cope with systems that would be otherwise difficult to treat (e.g. with no scaling laws available). However the quality of the approximate dynamics strongly depends on the ansatz chosen.

\textcite{Li2016_38258} use this approach to control matter waves in harmonic traps. 
Using as an ansatz a bright solitary wave solution, approximate auxiliary equations analogous to equations \eqref{ermakov_friends} are found. The Newton equation for the wave center remains the same, whereas 
the Ermakov equation is modified to include a term that depends on the non-linear coupling parameter.  
%
%
%
%
Using only a time dependent control of this parameter via a Feschbach resonance, the soliton wavefunction can be compressed/expanded in non-adiabatic time scales with high fidelity. 

Applying  the same concepts,
an efficient quantum heat engine running an Otto cycle with a 
condensate as its working medium is proposed in \textcite{Li2017_015005}. The engine strokes are done on a short timescale ensuring a large power output and high efficiency. Also, in \textcite{Fogarty2019} the variational approach is applied to design the 
time-dependent interaction strength between two  ultracold atoms so as to 
create entangled states.

\subsection{Fast Forward\label{ss:FF}}
The so-called fast-forward (FF) approach was  first derived to accelerate a given quantum dynamics 
by mimicking the effect of an FF button in  an audio or video player \cite{Masuda2008}. The original idea was to use  
a scaling transformation $\tau=\alpha t$, for some constant $\alpha$ (inhomogeneous scalings are also possible) so that the solutions $\psi_0(\tau)$ of 
$i\hbar \partial_\tau \psi_0(\tau)=H(\tau)\psi_0(\tau)$ are just scaled in time as $\psi_0(\tau)=\psi_\alpha(t)$ with respect of the solutions of 
$i \hbar \partial_t \psi_\alpha(t)=\alpha H(\alpha t)\psi_\alpha(t)$. 
This concept works formally and in some discrete Hamiltonians it is experimentally viable. However, if $H$ includes a kinetic energy it implies that 
the mass should be changed \cite{deLima2019}.   
\textcite{Masuda2008} solved this problem by modifying the potential but not the mass.
\subsubsection{The original formalism} 
Consider the time-dependent Schr\"odinger equation
\beq
i \hbar \frac{\partial \psi_0(\mathbf{r},t)}{\partial t} = -\frac{\hbar^2}{2m} \Delta\psi_0(\mathbf{r},t) + V_0(\mathbf{r},t)\psi_0(\mathbf{r},t)
\eeq
in a time interval $[0,T]$
with real potential $V_0$.  
To speed up the dynamics keeping the potential real, a non trivial phase factor that depends on the coordinates 
has to be added to the FF wavefunction,  
\beq
\psi_{FF}(\mathbf{r},t)=\psi_0(\mathbf{r},\Lambda(t))e^{i \varphi(\mathbf{r},t)},
\eeq
driven by the FF potential  
\begin{eqnarray}
V_{FF}(\mathbf{r},t)& = & -\hbar \ddot \Lambda \Sigma  (\mathbf{r},\Lambda(t)) - 2 \hbar[\dot \Lambda -1]\partial_t \Sigma  (\mathbf{r},\Lambda(t)) \nonumber \\ &-& \frac{\hbar^2}{2m}(\dot \Lambda^2-1) (\partial_{\bf r}\Sigma  (\mathbf{r},\Lambda(t)))^2 + V_0 (\mathbf{r},\Lambda(t)). \nonumber \\
\label{eq:vff}
\end{eqnarray}
The phase is given by  $\varphi(\mathbf{r},t)=[\dot \Lambda -1]\Sigma  (\mathbf{r},\Lambda(t))$, where $\Sigma$ is the phase of the reference  wavefunction,   
$\psi_0(\mathbf{r},t)=\widetilde{\psi}_0(\mathbf{r},t) \exp(i \Sigma (\mathbf{r},t))$.  
Imposing 
$\Lambda(0)=0$, $\Lambda(t_f)=T$, $\dot \Lambda(0)=\dot \Lambda(t_f)=1$, and $\ddot \Lambda(0)=\ddot \Lambda(t_f)=0$ implies that $V_{FF}(\mathbf{r},0)=V_0 (\mathbf{r},0)$ and 
$V_{FF}(\mathbf{r},t_f)=V_0 (\mathbf{r},T)$. Moreover  the additional phase $\varphi (\mathbf{r},t) $ vanishes at the boundary of the time interval $[0,t_f]$. For $\dot \Lambda>1$, the dynamics is accelerated while it is slowed down for $0<\dot \Lambda<1$. A negative $\dot \Lambda$ corresponds to a time reversed evolution.

The transposition of these results to speed up an adiabatic evolution is not direct because the adiabatic evolution 
only holds for an infinitely slow process that needs infinite magnification factors.   
A first but too naive approach would consist in considering the adiabatic wave function as a candidate for applying the previous fast-forward formalism. If the system is in the $n^{\rm{th}}$ eigenstate $\phi_n({\bf r}, \lambda )$ associated with  the eigenvalue $E_n( \lambda )$, where $ \lambda$ accounts for the parameters, the corresponding adiabatic wave function reads
\begin{equation} 
\psi_{\rm ad}({\bf r},t, \lambda(t) )=\phi_n({\bf r},\lambda(t) )e^{-i \varphi_{\rm dyn}(t)+i \varphi_{\rm ad}(t)},
\label{eq:ad}
\end{equation}
where $\varphi_{\rm dyn}(t)=\int_0^t E_n( \lambda (t'))dt'/\hbar$ is the dynamical phase.
The direct route  fails and a renormalization is needed 
modifying both the wave function and the Hamiltonian in a consistent manner to ensure that the wave function remains valid for a finite change of the parameters \cite{Masuda2010}. 

To highlight the smallness of the change of parameter, \textcite{Masuda2010} introduce the constant rate $\varepsilon \ll 1$ associated with $\lambda \to \lambda + \delta \lambda$ where $\delta \lambda = \varepsilon t$. A finite change of the parameter $\lambda$ during the time interval $[0,t_f]$ is given by
\begin{equation}
\lambda(\Lambda(t_f))-\lambda(0)= \varepsilon \Lambda(t_f).
\end{equation}
The expression for the FF potential 
to drive the regularized wave function
\begin{equation} 
\psi^{\rm (reg)}({\bf r},t, \lambda(t) )=\phi_n({\bf r},\lambda(t) )e^{-i \varphi_{\rm dyn}(t)+i \varepsilon\delta \theta({\bf r},t)}
\label{eq:ad2}
\end{equation}
finally reads \cite{Masuda2010}
\begin{eqnarray}
V_{FF} ({\bf r},t) & = & \dot \Lambda \varepsilon \delta V [{\bf r},\tilde \lambda_t]  + V_0[{\bf r},\tilde \lambda_t]  \hbar \ddot \Lambda \varepsilon \delta \theta [{\bf r},\tilde \lambda_t]
\nonumber \\
&-&   \hbar \dot \Lambda^2 \varepsilon^2 \frac{\partial\delta \theta}{\partial \lambda}\bigg|_{{\bf r},\tilde \lambda_t}  
- \frac{\hbar^2}{2m_0}   \dot \Lambda^2 \varepsilon^2 (\nabla \delta \theta)^2,
\end{eqnarray}
with $\tilde \lambda_t=\lambda(\Lambda(t))$, and 
\begin{eqnarray}
0 & = & |\phi_n|^2\Delta \delta \theta + 2{\rm Re}[\phi_n \nabla \phi_n^*]\cdot \nabla \delta \theta
 + \frac{2m}{\hbar}{\rm Re} \left[   \phi_n\partial_\lambda \phi_n^* \right],\nonumber \\
\delta V&=&- \hbar {\rm Im} \left[ \frac{\partial_\lambda \phi_n}{\phi_n} \right] - \frac{\hbar^2}{m} {\rm Im} \left[ \frac{\nabla \phi_n}{\phi_n}\right]\cdot \nabla \delta \theta.
\end{eqnarray}

As an application,  consider the 1D transport of a wave function with a moving confining potential: $V_0(x,t)=U(x-\varepsilon t)$. We readily find $\delta V=0$ and $\delta \theta=mx/\hbar$ for all $n$. The fast-forward potential is now  $n$-independent and reads 
\begin{equation}
V_{FF}(x,t)=U(x-x_0(t)) - mx\ddot x_0,
\end{equation}
with $x_0(t)=\varepsilon \Lambda(t)$ and terms that only depend on time have been dropped.  
This result is exactly the one found  by the ``compensating force approach'' within invariant-based inverse engineering \cite{Torrontegui2011}, see Sec. \ref{sLLf}: the inertial forces in the frame attached to the moving potential are compensated for by an appropriate  uniform time-dependent force. 

Let us underline some conceptual and methodological similarities and differences with the CD approach: even if the adiabatic 
states $\phi_n(\bf{r},\lambda(t))$ are used as a reference in the construction, now the auxiliary  potential in general depends on the specific $n$-th wavefunction  
used (except in the previous example and for the Lewis-Leach family as discussed below),  and the dynamical wavefunction differs along the dynamics with the adiabatic function 
more strongly than just by a (constant in ${\bf{r}}$) phase factor, because of the  position-dependent  phases 
in Eq. (\ref{eq:ad2}). 
Moreover, the FF method leads by construction to local and real potentials. By contrast, 
the CD recipe may lead to auxiliary non-local terms that depend on the momentum.    

\subsubsection{Streamlined fast-forward approach\label{sFF}}
The original fast-forward approach is somewhat involved. A simpler, more direct ``streamlined version'' of fast forward was proposed in \textcite{Torrontegui2012_013601} and developed further in \textcite{Torrontegui2013_033630,Martinez-Garaot2016}
without making explicit use of a slow reference adiabatic process. \textcite{Kiely2015} extended the streamlined version for a charged particle 
in an electromagnetic field.  

This streamlined version is essentially  inverse  engineering. For this discussion we use a 1D setting. 
The   key simplification is to inverse engineer 
directly the potential $V(x,t)$ from the given (desired) $\psi(x,t)=\rho(x,t)e^{i\phi(x,t)}$ wavefunction in the Schr\"odinger equation,
imposing $V(x,t)$ to be local, 
\begin{equation} 
V(x,t)= \frac{1}{\psi(x,t)} \left(\displaystyle i\hbar\frac{\partial \psi(x,t)}{\partial t} + \frac{\hbar^2}{2m}\frac{\partial ^2 \psi(x,t)}{\partial x^2} \right).
\label{eq1}
\end{equation}
Using the representation ($\rho$ and $\phi$ real)   
\beq
\label{m-p}
\psi(x,t)=\rho(x,t)e^{i\phi(x,t)},
\eeq
and 
imposing ${\rm Im}[V(x,t)]=0$, we get a continuity equation 
\begin{equation} 
\frac{1}{\rho}\frac{\partial \rho}{\partial t} + \frac{\hbar}{2m}\left( \frac{2}{\rho}\frac{\partial \phi}{\partial x}\frac{\partial \rho}{\partial x}  + \frac{\partial^2 \phi}{\partial x^2}  \right)=0,
\label{eq2}
\end{equation}
that links $\rho$ and $\phi$. In particular if $\rho(x,t)$ is given,  $\phi(x,t)$ cannot be arbitrary.  
The expression for the potential then reads
\begin{equation} 
V(x,t) = -\hbar \frac{\partial \phi}{\partial t} + \frac{\hbar^2}{2m}\left[  \frac{1}{\rho}\frac{\partial^2 \rho}{\partial x^2} - \left( \frac{\partial \phi}{\partial x}  \right)^2  \right].  
\label{eq3}
\end{equation}
Equation~(\ref{eq2}) can be integrated formally,
\begin{equation} 
\frac{\partial \phi}{\partial x} = - \frac{mu(x,t)}{\hbar},
\label{eq4}
\end{equation}
where $u$ plays the role of a  ``hydrodynamic velocity'',
\begin{equation} 
u(x,t) = \frac{1}{\rho^2(x,t)} \frac{\partial}{\partial t} \left( \int_0^x \rho^2(x',t)dx' \right).
\label{eq5}
\end{equation}
The potential $V(x,t)$ can therefore be inferred  from $\rho(x,t)$ as 
\begin{eqnarray} 
V(x,t)  & = &    m\frac{\partial }{\partial t} \int_0^x u(x',t)dx' +
\frac{\hbar^2}{2m}  \frac{1}{\rho(x,t)}\frac{\partial^2 \rho (x,t)}{\partial x^2} 
\nonumber \\  & - &  \frac{1}{2} mu^2(x,t) -\hbar \dot \phi_0(t),
\label{eqpot}
\end{eqnarray}
where $\phi_0\equiv\phi(x=0,t)$.
This is the central result in  \textcite{Martinez-Garaot2016}.

While the freedom to choose  $\rho(x,t)$ is very  welcome for applications that go beyond the 
speed up of adiabatic processes, for example, a transfer between the ground and the first excited state
of a harmonic oscillator, $\rho$ may of course also be chosen as an adiabatic function.
Then connections to the other STA approaches can be made.  An important fact is that 
for the Lewis Leach family of Hamiltonians, the fast-forward potential becomes $n$-independent \cite{Torrontegui2012_013601,Patra2017_125009} and       
the method provides the terms found in invariant-based inverse engineering. 
As well,    
the local FF potentials may be unitarily related to the non-local CD Hamiltonians. 
Moreover  
\textcite{Patra2017_125009} noticed the following  connection with CD driving: 
The (generally state-dependent) Hamiltonian 
\beq
H_{CD}(n)=-\frac{p u_n+u_n p}{2}
\eeq
where $u_n$ is the {{hydrodynamic velocity}} for the $n$-th adiabatic state,
acts  on the (parallel transported, real) $\la x|n(t)\ra$  exactly as $H_{CD}$ does. 
They also introduced an acceleration flow field $a(x,t)=-\partial u/\partial t+u \partial u/\partial x$
so that the FF potential for a real function $\rho(x,t)$ corresponding to some eigenstate 
of $H_0$, which includes kinetic energy and a reference potential $V_0(x,t)$, may be written compactly as 
\begin{equation}
V(x,t)=V_0(x,t)-m\int_0^x a(x',t) dx'.
\end{equation}
%
  
FF potentials may have divergences due to wavefunction nodes. Nevertheless, \textcite{Martinez-Garaot2016} demonstrated that in a  transition from ground to excited states, truncating the potential is a viable approximation.  
\subsubsection{Generalizations, terminology}
\textcite{Masuda2016_51} review  FF and CD approaches focusing on applications in molecular systems.    
The FF method has been extended to finite-dimensional Hilbert space with applications to drive two-level system \cite{Takahashi2014}
and three-level systems  \cite{Masuda2016_51}, or two-spin systems \cite{Setiawan2017} 
and spin clusters with different geometries \cite{Setiawan2019}. In a finite Hilbert space, the auxiliary potential may be constrained by some criterion 
(e.g. to be  diagonal). In particular  
\textcite{Setiawan2017}  find different state-dependent CD terms.

FF was applied  to manipulate Bose-Einstein condensates in optical lattices \cite{Masuda2014_033621,Masuda2014_063003,Masuda2018}, to accelerate the STIRAP protocol \cite{Masuda2015_3479,Masuda2016_51}, or to investigate the fast generation of entanglement in spin systems \cite{Setiawan2017}. 
The extension to  the classical realm is discussed in \textcite{Jarzynski2017}, see   Sec.~\ref{counterclas}.

The method has also been used to control a charged quantum particle that interacts with electromagnetic fields \cite{Masuda2011,Kiely2015,Masuda2015_11079}. 
In \textcite{Kiely2015} the fields found via invariant-based engineering and (streamlined) FF, to change the radial spread of the particle state in a Penning trap, are shown to be equivalent.  
\textcite{Khujakulov2016,Nakamura2017} accelerate the  tunneling of a charged particle 
treating  the wave function phase differently than \textcite{Masuda2010}.

We conclude the section with a comment on terminology. Some authors, see e.g. \textcite{Bukov2019,Villazon2019},  qualify as FF  any Hamiltonian 
$H[\lambda(t)]$ for which no terms are added and only the parameter $\lambda$ (possibly multidimensional) is shaped in time to get to target states with unit fidelity.  
While this notion is in agreement with some of the results of the FF methodology described above, it generally covers a different domain.   
For example,  in \textcite{Masuda2015_3479,Masuda2016_51} a discrete  FF approach is set where actually an additional control parameter  with respect to the CD driving term  is added. 
%
%

\begin{figure}[t]
\begin{center}
\includegraphics[height=2.cm,angle=0]{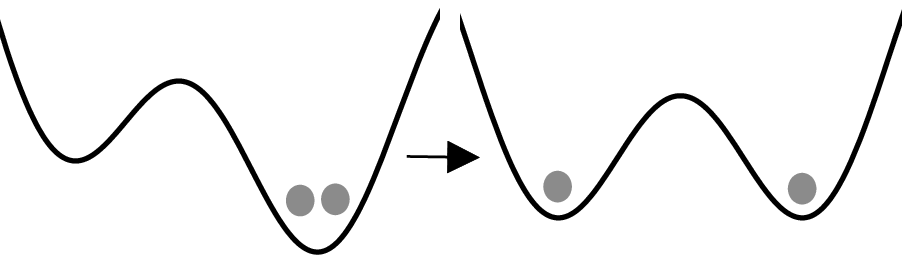}
\includegraphics[height=2.cm,angle=0]{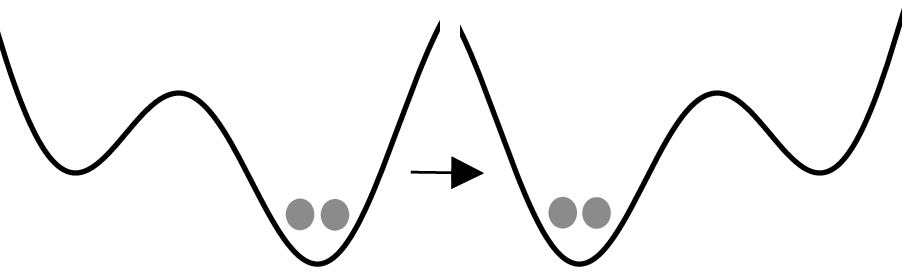}
\end{center}
\caption{\label{fFAQUAD}
(Color online) Splitting and cotunneling operations on two interacting bosons performed by FAQUAD in \textcite{Martinez-Garaot2015_043406}.}
\end{figure}

\subsection{FAQUAD and related approaches}\label{FAQUAD}
Fast quasiadiabatic (FAQUAD) and related approaches depart from the core of paradigms discussed so far in that 
they are intended from the start as approximate methods to balance two conflicting aims: shortening the process time 
and keeping the process as adiabatic as possible with respect to the actual Hamiltonian, not with respect to a reference Hamiltonian. For Hamiltonians that depend on one control parameter
$\lambda(t)$, 
the strategy  is to distribute its rate of change $\dot{\lambda}$ so that diabatic transitions are equally likely along the process. 
In this way 
$\dot{\lambda}$ slows down at and near avoided crossings but allows for fast changes away from them, to make diabatic transitions weak along the whole process. 

The specific methods differ on: the exact recipe used to 
distribute diabaticity ``homogeneously'' from $t=0$ to $t_f$, which leads to different $\lambda(t)$;
and on the spectral information needed to implement them.  This latter aspect sets a hierarchy of complementary approaches: the maximum 
information corresponds to FAQUAD, which needs the eigenfunctions and eigenvalues; the ``local adiabatic'', ``uniform adiabatic'', or ``parallel adiabatic transfer''  approaches imply an intermediate level which only requires eigenvalues; finally, phenomenological approaches may be based on knowing only the location of avoided
crossings  
\cite{Amin2008_130503,Xing-Xin2013}.
 Complex systems, and in particular many-body systems to implement quantum adiabatic computing 
and quantum annealing constitute a natural domain and motivation to develop effective methods
that need little or no spectral information \cite{Albash2018,Tian2018}. 
Of course there is also a domain of simpler systems for with 
more information-consuming approaches are useful and applicable.\\   

{\it FAQUAD.}
The idea of keeping the adiabaticity parameter constant along trap expansions, 
i.e. such that $\dot{\omega}/\omega^2=c\ll 1$  
had been applied in a number of works, \cite{Kastberg1995, Chen2010_063002,Bowler2012,Torrontegui2012_033605,Martinez-Garaot2013},
and was generalized for other systems in the 
FAQUAD approach developed by \textcite{Martinez-Garaot2015_043406}.  
In the simplest two-level scenario with instantaneous eigenvalues $E_{1,2}(t)$ and eigenvectors 
$\phi_{1,2}(t)$,  
\beq
\label{f_adiabatic}
\hbar \left |\frac{ \langle \phi_1(t)|\partial_t \phi_2(t)\rangle}{ E_1(t)-E_2(t)} \right |=
\hbar \left |\frac{ \langle \phi_1(t)|\frac{\partial H}{\partial t}| \phi_2(t)\rangle}{ [E_1(t)-E_2(t)]^2} \right | =c,
\eeq 
and, as $\lambda=\lambda(t)$, the chain rule gives    
\beq
\label{de_faquad}
\dot \lambda\!=\!\mp\frac{c}{\hbar} \!\left | \frac{ E_1(\lambda)-E_2(\lambda)}{ \langle \phi_1(\lambda)|\partial_\lambda \phi_2(\lambda)\rangle} \right |\!
\!=\!\mp\frac{c}{\hbar}\!\left | \frac{ [E_1(\lambda)-E_2(\lambda)]^2}{ \langle \phi_1(\lambda)|\frac{\partial H}{\partial \lambda}|\phi_2(\lambda)\rangle} \right |\!,
\eeq
where $\mp$ applies to a monotonous decrease or increase of $\lambda(t)$.  
Equation (\ref{de_faquad}) must be solved with the boundary conditions $\lambda(0)$ and  $\lambda(t_f)$, which fixes $c$ and the integration constant. 
This technique has been applied 
to accelerate processes described by two and three level systems such as 
cotunneling and splitting of two bosons in a double well,  see Fig. \ref{fFAQUAD}, to 
generate  macroscopically entangled states in a Tonks-Girardeau gas \cite{Martinez-Garaot2015_053406},  to design optical waveguide devices \cite{Martinez-Garaot2017,Chung2017,Liu2017_205501}, and quantum neural networks  
\cite{Torrontegui2019}.\\

\begin{figure}[t]
\begin{center}
\includegraphics[height=2.7cm,angle=0]{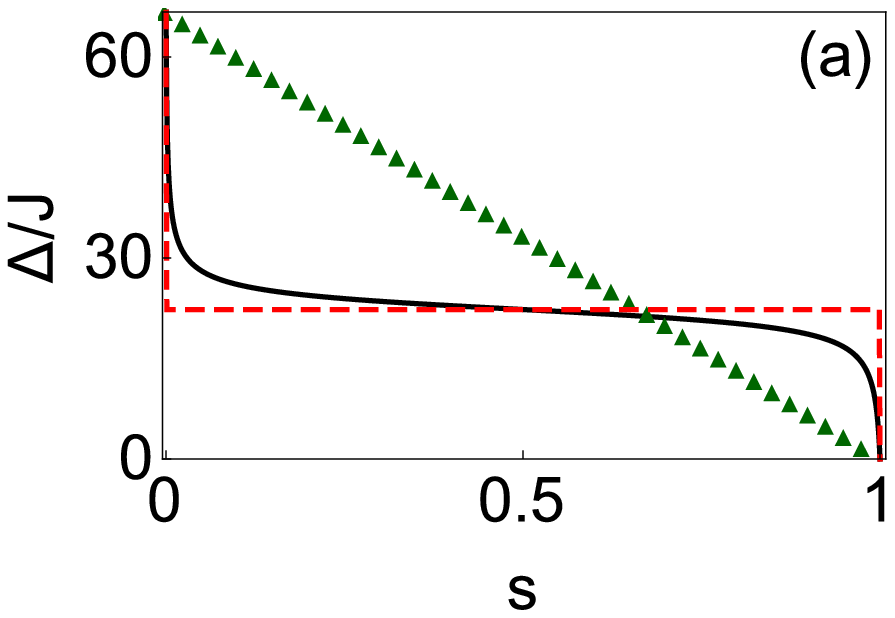}
\includegraphics[height=2.7cm,angle=0]{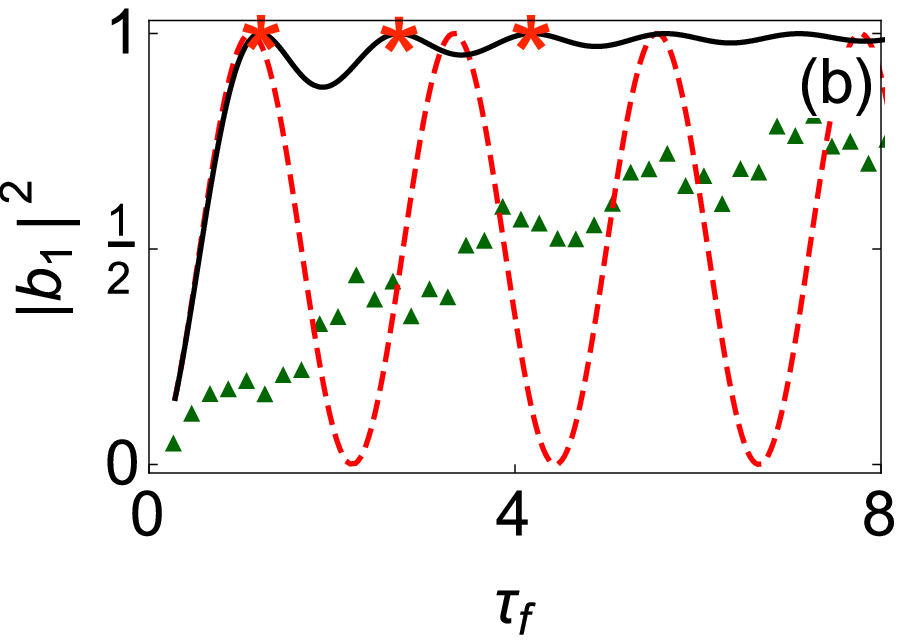}
\includegraphics[height=2.7cm,angle=0]{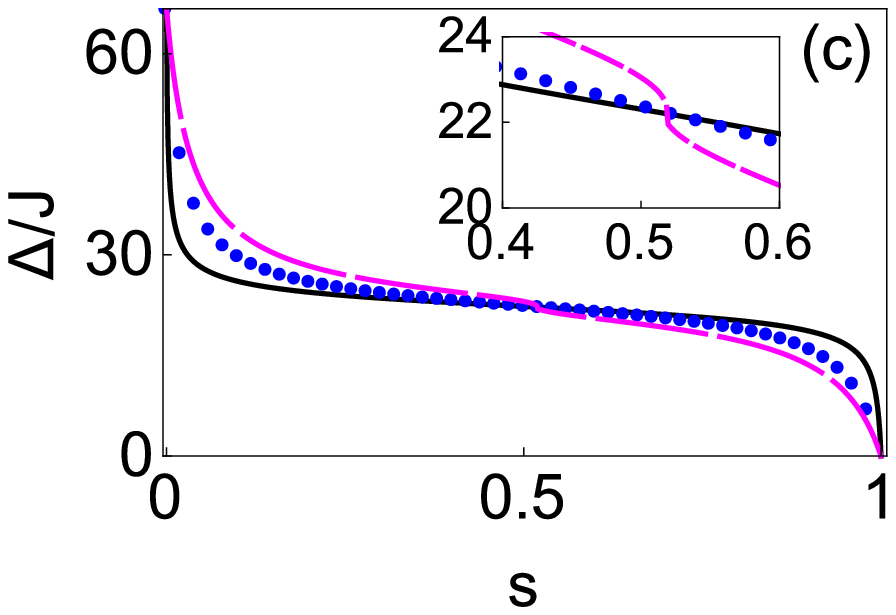}
\includegraphics[height=2.7cm,angle=0]{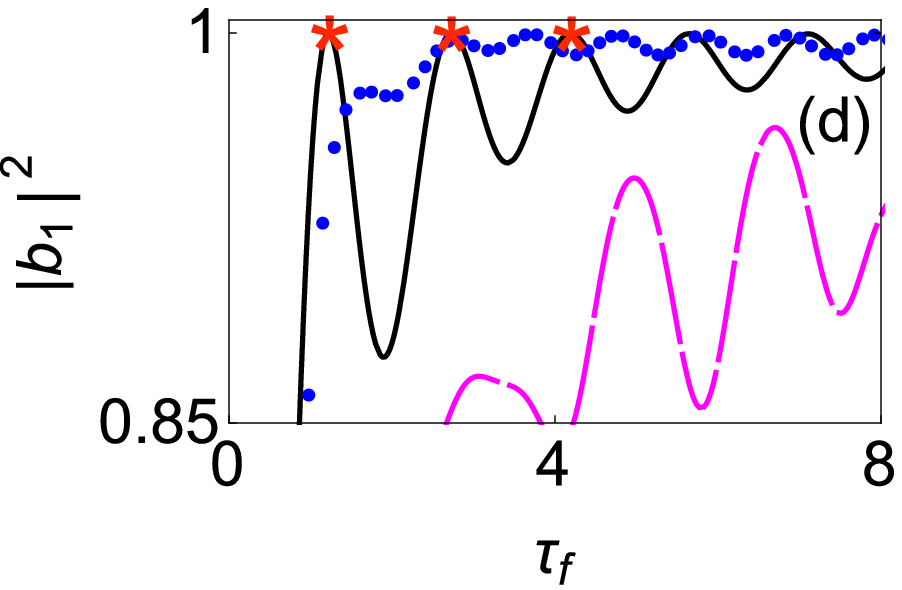}
\end{center}
\caption{\label{f2FAQUAD}
(Color online). (a) Bias versus $s=t/t_f$ for linear-in-time bias (green triangles), $\pi$-pulse (short-dashed red line), and 
FAQUAD (solid black line).  
(b) Final ground state population $|b_1(t_f)|^2$ vs.   
$\tau_f=Jt_f/\hbar$ for linear-in-time bias (green triangles), $\pi$-pulse
(short-dashed red line), and FAQUAD (solid black line).
(c) Bias vs. $s$ for FAQUAD (solid black line), local adiabatic  approach (blue dots), and uniform adiabatic  approach  (long-dashed magenta line).  
The inset amplifies a kink of the uniform adiabatic approach.
(d) $|b_1(t_f)|^2$ vs. $\tau_f=Jt_f/\hbar$ for FAQUAD (solid black line), 
local adiabatic approach (blue dots), and uniform adiabatic approach  (long-dashed magenta line). The
stars in (b) and (d) correspond to integer multiples of the characteristic
FAQUAD time scale $2\pi/\Phi$.
$\Delta(0)/J=66.7$, $U/J=22.3$.    From \textcite{Martinez-Garaot2015_043406}.}
\end{figure}

{\it Other approaches.}
In the ``parallel adiabatic transfer'' technique \cite{Guerin2002,Guerin2011} the level gap is required to be constant, which prevents it from being applicable when the initial and final gaps are different. 

The ``uniform adiabatic'' method developed by \textcite{Quan2010} relies on a comparison of {\it transition} and {\it relaxation} time scales
and proposes, instead of Eq. (\ref{de_faquad}), 
\beq
\dot \lambda
=\mp\frac{c_{UA}}{\hbar}\left | \frac{ [E_1(\lambda)-E_2(\lambda)]^2}{\partial [E_1(\lambda)-E_2(\lambda)]/\partial \lambda} \right |.
\eeq

The ``local adiabatic''  approach \cite{Roland2002,Richerme2013} gives an equation similar to Eq.~(\ref{de_faquad}),  without the factor 
$\langle \phi_1(\lambda)|\frac{\partial H}{\partial \lambda}| \phi_2(\lambda)\rangle$. 
This leads to a different constant, $c$, and time dependence of the parameter, $\lambda(t)$, and therefore different minimal times.
There are many works applying  the local adiabatic method  in the context of adiabatic quantum computation \cite{Albash2018}. For the Grover problem to find a marked item in an unsorted database of $N$ items, 
the schedule provided by the local adiabatic approach provides the best asymptotic scaling of the time needed with respect to $N$ \cite{Albash2018}.      
\textcite{Schaller2006} generalized the local-adiabaticity condition taking into account high-order powers of $[E_1(\lambda)-E_2(\lambda)]$. Moreover, 
\textcite{Wiebe2012,Kieferova2014} proposed a method combining linear  local adiabaticity and the boundary cancellation methods. 
In boundary cancellation methods \cite{Morita2007,Rezakhani2010} the diabatic transitions at the time boundaries 
are suppressed by imposing  vanishing derivatives at the boundaries $\dot H[\lambda(t)]=0$ at $t=0$ and $t=t_f$.  
Recently, \textcite{Stefanatos2019_012111,Stefanatos2019_00166} propose a modified FAQUAD protocol in which the adiabaticity parameter $c$ is not held constant but follows  a simple ``on-off'' modulation found by Optimal Control Theory. 

\textcite{Martinez-Garaot2015_043406} compared local-adiabatic,  uniform-adiabatic, and FAQUAD approaches, also with 
a $\pi$-pulse and a linear ramp,  for a
two-level population inversion.
The model uses a bare basis 
$|1\rangle=\left ( \scriptsize{\begin{array} {rccl} 1\\0 \end{array}} \right)$ and $|2\rangle=\left ( \scriptsize{\begin{array} {rccl} 0\\1 \end{array}} \right)$,  
so that a  time-dependent state is 
$
|\Psi(t)\rangle=b_1(t)|1\rangle+b_2(t)|2\rangle
$
and 
\beq
H_0=\left ( \begin{array}{cc}
0 & -\sqrt{2}J \\
-\sqrt{2}J & U-\Delta
\end{array} \right),
\eeq
where the bias $\Delta=\Delta(t)$ is the control parameter, and $U>0$, $J>0$, are constant.  
The goal is to drive the eigenstate from $|\phi_1(0)\rangle=|2\rangle$ to $|\phi_1(t_f)\rangle=|1\rangle$. To design the reference adiabatic protocol 
\textcite{Martinez-Garaot2015_043406} impose on $\Delta(t)$ the boundary conditions
$\Delta(0)\gg U,J$ and $\Delta(t_f) =0$.
%
The FAQUAD protocol is shown in Fig. \ref{f2FAQUAD} (a) compared to a linear-in-time $\Delta(t)$ and a 
constant $\Delta=U$. 
The final ground state populations $|b_1(t_f)|^2$ versus the dimensionless final time $\tau_f=Jt_f/\hbar$ 
are shown in Fig. \ref{f2FAQUAD} (b).  
For $\Delta=U$ between 0 and $t_f$, 
Rabi oscillations occur (short-dashed red line in Fig. \ref{f2FAQUAD} (b)) due to
interference between two dressed states.  
By contrast the 
FAQUAD process is quasi-adiabatic and is dominated by one dressed state. The existence of special process
 times with perfect fidelity can be also understood as an 
interference phenomenon with characteristic period $T=2\pi/\Phi$, the minimal time for fidelity one, where 
$\Phi=\frac{\hbar}{t_f}
\int_0^{t_f} dt E_{\rm{gap}}(t)$ and $E_{\rm{gap}}$ is the gap between instantaneous levels.
Fig. \ref{f2FAQUAD} (b) also shows the poorer results of the linear ramp for $\Delta(t)$. 
Figs. \ref{f2FAQUAD} (c) and (d) compare FAQUAD, local adiabatic, and uniform adiabatic approaches. 
FAQUAD gives the best behaviour at short times, and the local adiabatic method achieves better 
population stability 
after a few oscillations.     
\subsection{Optimal Control and Shortcuts to adiabaticity \label{sOCT}}
Optimal Control Theory (OCT) is a widely used method  \cite{Pinch1993,Kirk2004} to find control 
parameter trajectories that minimize a given cost function 
(\emph{global constraint}) 
and obey some specific boundary conditions. According to the Pontryagin maximum principle, such extremal solutions satisfy the equations of a generalized Hamiltonian system.  When more constraints are imposed, the quantity to be minimized should be adapted with some weights for the different constraints. Optimal Control Theory yields analytically solutions only for low dimensional systems. Very often  the OCT solutions
are found by discretizing the problem and implementing numerical approaches such as dynamic programming, the gradient ascent, or Krotov algorithms. 

In contrast, STA  techniques are not built, in general,  upon a minimization principle. However, they serve a similar objective, to drive the system 
towards the desired states in a short amount of time. 
The  solutions found for the parameter trajectories are by construction typically analytical, continuous, and well adapted to introduce many \emph{local constraints} (for instance the successive derivative at the initial and final time of the dynamical quantity of interest). Comparing directly  both approaches is therefore somewhat misleading. Actually, a class of STA solutions depending on a free parameter can be used to minimize a given cost function yielding results close to those of Optimal Control Theory. 

In fact  there are many examples  in which STA and OCT methods are usefully combined to get nearly optimal protocols by minimizing a cost function of interest or 
to accomodate for extra constraints in a reduced space made of analytical solutions originating from an STA approach. Such a hybrid strategy has been explicitly worked out to engineer spin like systems  \cite{Hegerfeldt2013,Sun2017,Zhou2017_330}, to minimize final excitation after a fast transport in the presence of anharmonicities \cite{Torrontegui2012_013031,Zhang2016}, to ensure fast transport with extra relevant constraints (e.g. minimum transient energy, bounded trap velocity, or bounded distance from the trap center) \cite{Torrontegui2011,Chen2011_043415,Stefanatos2014_733,Alonso2016,Amri2018}, to ensure a fast and robust shuttling of an ion with noise \cite{Lu2014_063414},  to perform fast expansions \cite{Salamon2009,Stefanatos2010,Stefanatos2013,Lu2014_023627,Boldt2016,Stefanatos2017_4290,Plata2019},
or  to drive a many-body Lipkin-Meshkov-Glick system \cite{Campbell2015}. 

In  \textcite{Mortensen2018} the perturbative approach to enhance robustness described in Sec. \ref{ssec:robustness} 
was combined with optimal control. A cost function is defined to achieve low peak laser power and stability against systematic error (or scaling) of the control functions. Shortcut  schemes for the $\Lambda$-system were  found which minimize this function. This strategy is less computationally expensive than other  optimal control methods, since the Schr\"{o}dinger equation does not need to be solved for every run. Normally the fidelity is part of the cost function, but this is unnecessary for STA methods.

To transport a particle in a harmonic trap of angular frequency $\omega_0$, OCT 
provides  a minimum transport time $t_f$ for a fixed transport distance $d$ using a ``bang-bang'' like solution for which the acceleration is changed abruptly from a constant value $\omega_0^2\delta$ to the opposite value $-\omega_0^2\delta$ at $t_f/2$ and where the relation between the final time and the parameters is $t_f=(2/\omega_0)(d/\delta)^{1/2}$ \cite{Chen2011_043415}. This kind of solution was implemented experimentally  to transport  a cold cloud of atoms trapped in a moving optical tweezer in twice the oscillation period  \cite{Couvert2008}.  

The challenges and prospects of Optimal Control Theory for quantum systems which overlap with those of shortcuts to adiabaticity  are discussed in the review articles \textcite{Glaser2015,Koch2016}.
%

%
%
%

\subsection{Robustness\label{ssec:robustness}}
Adiabatic processes posses a natural robustness to parameter variations. As long as the parameters vary slowly
enough,  there are
many smooth adiabatic paths to the same final result,  ignoring phases. However, this robustness does not apply to all imperfections, for example adiabatic wavepacket splitting is very sensitive to asymmetries in the potential \cite{Torrontegui2013_033630}. Indeed  adiabatic drivings are prone to decoherence, excitations and particle loss, due to the accumulation of noisy perturbations during  long process times. STA methods prove useful as they reduce the detrimental cumulative effect of noise,  but   they require specific control of the parameters for intermediate times,  
so in general will not have such a natural stability against smooth parameter variations.  

Nevertheless, the flexibility of STA methods can be exploited to improve robustness 
against external influences and imperfections.
Within the class of control schemes that work perfectly in the ideal, noiseless setting, the objective is to find the 
most robust one versus the relevant imperfections/noises,  singled out or combined,  of a given experiment.

\subsubsection{Error sensitivity and its optimization using perturbation theory}
In several works, see e.g.  \textcite{Torrontegui2012_013031,Choi2012}, 
the effect of perturbations and imperfections on STA protocols is analyzed.
We can go  further and actively improve or even maximize the
robustness of the control schemes.

The starting point  to do so \cite{Ruschhaupt2012} is to first design a class of shortcut schemes which
fulfill the wanted control task with fidelity one without perturbations. The next step is to define an error sensitivity $q \ge 0$ with respect to the relevant error source;
this is done by using perturbation theory to define a series expansion of the fidelity $F(\lambda)$
in terms of the error parameter, $q$ being (minus) the coefficient of the quadratic term. Depending on the nature of the error, systematic or stochastic, 
$q$ may be found using a  Schr\"odinger equation or 
a master equation.

This principle has been applied to many physical systems and different sources
of errors and imperfections. In the following, we review some of these works
beginning with the  simple example of  a two-level system.

\paragraph{Illustrative example: Control of a Two-level System}


\textcite{Ruschhaupt2012} examined 
population inversion in a two-level quantum system subjected to different systematic and noisy errors.
The starting point is a two-level Hamiltonian 
\begin{eqnarray}
H_0 (t)= \frac{\hbar}{2} \left(\begin{array}{cc} -\Delta(t) & \Omega_R (t) - i
  \Omega_I (t)\\
\Omega_R (t) + i \Omega_I (t) &  \Delta(t)
\end{array}\right).
\end{eqnarray}
Following the invariant-based inverse engineering, see Sec. \ref{examplesIBIE}, the control parameters
$\Omega_R(t), \Omega_I(t), \Delta(t)$ 
can be calculated from the auxiliary functions $\theta, \alpha, \gamma$ via Eqs. (\ref{eq:12},\ref{eq:13},\ref{eq:14}), 
with the boundary conditions
$\theta(0)=0$ and $\theta(T)=\pi$. 
By realizing these control functions exactly the population would be 
inverted in the unperturbed, error-free case along a family of solutions
for the parameter paths.  

For systematic errors, for example if atoms at different positions are subjected
to slightly different fields due to the Gaussian profile of the
laser, the actual, experimentally implemented 
Hamiltonian is $H_0 + \beta H_1$, where
$H_1 (t)= H_0 (t)\big|_{\Delta\equiv 0}$
and $\beta$ is the (dimensionless) amplitude of the relative systematic error in $\Omega_R$ and $\Omega_I$.
To give a specific example, consider now only systematic errors in the Rabi frequency. 
Using time-dependent perturbation theory, the population to be in the excited state $P_2 (\beta)$ can be expressed as
$P_2 (\beta) = P_2 (0) -  q_S \beta^2 + ...$
where the noise sensitivity is found to be 
\begin{eqnarray}
q_S=-\frac{1}{2} \left. \frac{\partial^2 P_2}{\partial \beta^2}\right|_{\beta=0}=\left|\int_0^{t_f} dt e^{-i \gamma} \dot{\theta} \sin^2\theta \right|.
\end{eqnarray}

One simple choice that gives  $q_S=0$ is 
\begin{eqnarray}
\Omega_R &=& \frac{\pi}{t_f} \sqrt{1+16\sin^6\!\left(\frac{\pi t}{t_f}\right)}, 
\Omega_I = 0, \nonumber \\
\Delta &=& -\frac{8\pi}{t_f}\sin\!\left(\!\frac{\pi t}{t_f}\!\right)\!\sin\!\left(\!\frac{2\pi t}{t_f}\!\right)\! \frac{1+4 \sin^6\!\left(\!\frac{\pi t}{t_f}\!\right)}{1+16 \sin^6\!\left(\!\frac{\pi t}{t_f}\!\right)}.
\end{eqnarray}

\begin{figure}[t]
\begin{center}
\includegraphics[height=4.6cm,angle=0]{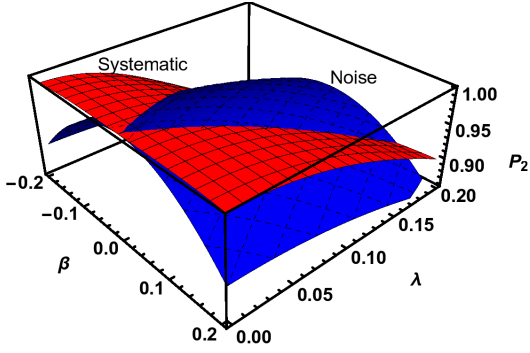}

\end{center}
\caption{\label{fig:robustness}
(Color online) Excitation probability $P_2$ versus noise error (strength $\lambda$) and 
systematic error
parameter (strength $\beta$); Noise-error-optimized STA (blue) and 
systematic-error-optimized STA (red). Adapted from \textcite{Ruschhaupt2012}}
\end{figure}

We can now consider amplitude noise. We assume that $\Omega_R$
and $\Omega_I$ are affected independently with the same strength parameter $\lambda^2$. This is motivated
by the assumption that in principle two lasers may be used to implement the real and the
imaginary part of the Rabi frequency with the same intensity. The corresponding master equation
with only noise error (no systematic error) is

\begin{eqnarray}
 \frac{d}{dt} \rho =&-&\frac{i}{\hbar} [H_0,\rho] \nonumber\\
                                    &-& \frac{\lambda^2}{2 \hbar^2} \left([H_{2R},[H_{2R},\rho]]\!+\! [H_{2I},[H_{2I},\rho]]\right)\!,
\label{masterfinal}
\end{eqnarray}
where $H_{2R} (t)= H_0 (t)\big|_{\Delta\equiv\Omega_I\equiv 0}$, and $H_{2I} (t)= H_0 (t)\big|_{\Delta\equiv\Omega_R\equiv 0}$. Note that STA methods for systems explicitly coupled to an external bath are discussed in further detail in Sec. \ref{thermostat}.
The noise sensitivity $q_N$ is defined as
\begin{eqnarray*}
q_N := 
%
- \left.\frac{\partial P_2}{\partial (\lambda^2)}\right|_{\lambda=0},
\end{eqnarray*}
where $P_2 = \braket{2}{\rho(t_f)|2}$ is the probability of the excited state at final time $t_f$,
i.e. $P_2 \approx 1 - q_N \lambda^2$. 
Using time-dependent perturbation theory for the master equation,
this sensitivity can again be calculated in terms of the auxiliary functions  which define the invariant and the state evolution. The 
transient values of these functions can then be optimally chosen to minimize $q_{N}$, 
while keeping the boundary conditions fixed to ensure perfect state transfer
without noise.

In Fig. \ref{fig:robustness}, two STA schemes optimized for 
noise-error resp. systematic error are shown, 
see \textcite{Ruschhaupt2012} for details. Clearly,  
different sources of imperfection
need different optimized STA schemes.
For a recent combination  of this perturbative approach
with optimal control theory in the context of topologically protected gates see \textcite{Ritland2018}.

\paragraph{Optimization using perturbation theory in other settings.}


The optimization of  robustness of shortcut schemes using the perturbative approach has
been applied to many different systems and systematic error and noise types.
Dephasing noise and systematic frequency shift for the two-level system have been examined in \textcite{Lu2013}, where
$\Gamma_{d}:=\gamma_{d}\sigma_{z}$ is the
noise operator 
\cite{Sarandy2007}, whereas  \textcite{Ruschhaupt2014} address  bit-flip noise, with 
$\Gamma_{b}:=\gamma_{b}\sigma_{x}$ being  the
noise operator.
The perturbative  approach has also been used to work out stable single and two-qubit gates
\cite{Santos2018_015501}, 
and to  designed schemes to suppress unwanted transitions \cite{Kiely2014,Yu2018,Yan2019_8267}.
It may also be combined with Optimal Control Theory \cite{Mortensen2018}, see Sec. \ref{sOCT}.  
There are as well many works studying and improving robustness in transport problems with respect to anharmonicities 
\cite{Chen2011_043415,Torrontegui2011,Zhang2015_043410,Zhang2016}, and noise \cite{Lu2014_063414,Lu2018},  
see Sec. \ref{trapped}.

The results of \textcite{Ruschhaupt2012} were extended in \textcite{Daems2013} in two ways. 
First, higher derivatives in the error were also considered and second,
the optimization of the auxiliary function was done by starting with an ansatz
with free parameters which were optimized numerically (while it was still
possible to derive the optimal scheme in \textcite{Ruschhaupt2012} analytically). In addition, 
the absolute systematic error in the detuning was examined. The results of \textcite{Daems2013} 
for population inversion have been used experimentally to rephase atomic coherences in a $\text{Pr}^{3+}\text{:Y}_{2}\text{SiO}_5$ crystal \cite{Van-Damme2017} and also applied   to create  a superposition states
with a controlled relative phase in a two-level system  \cite{Ndong2015}.\\

\begin{figure*}[t]
\begin{center}
a)\, \includegraphics[height=3.3cm,angle=0]{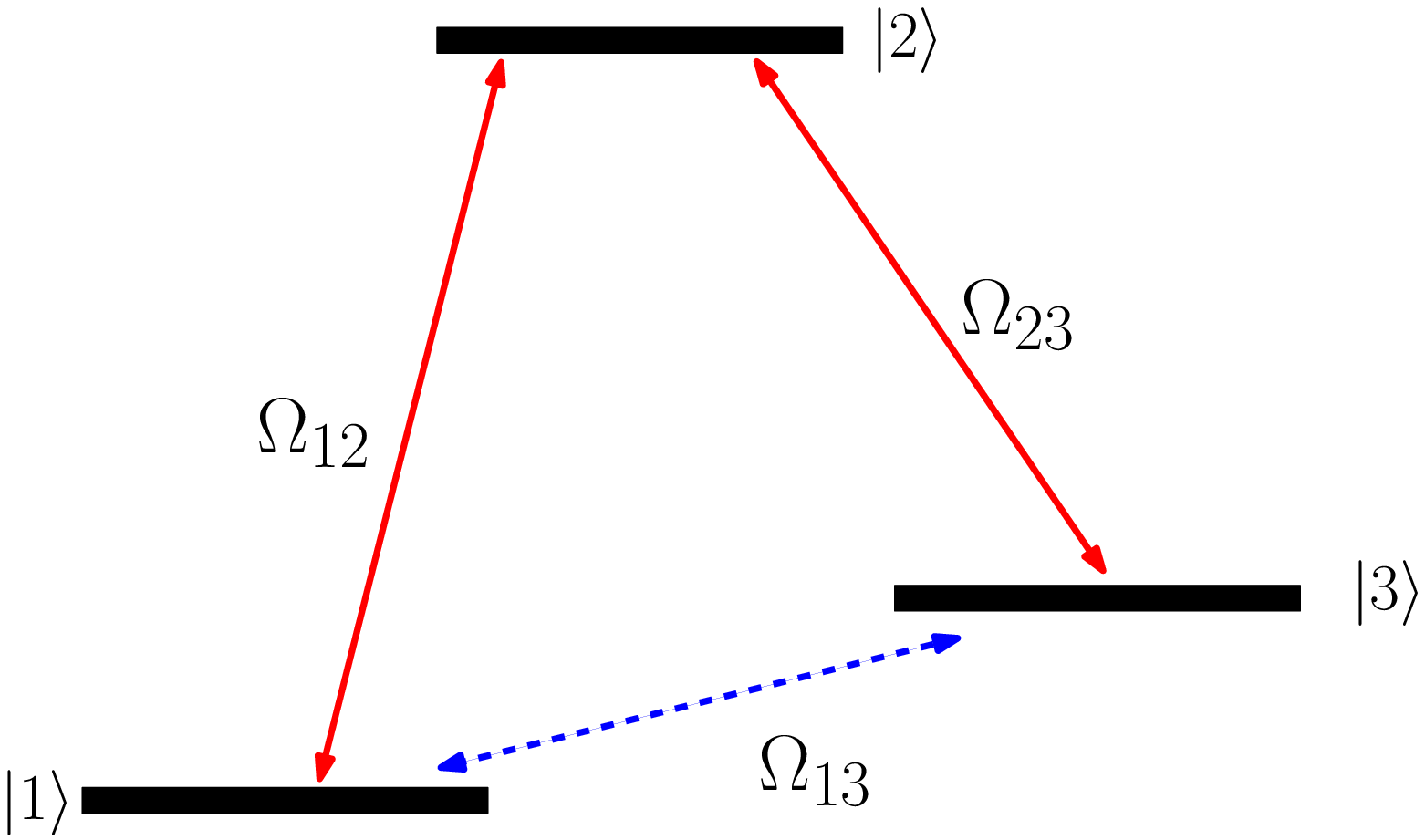}
b)\, \includegraphics[height=3.3cm,angle=0]{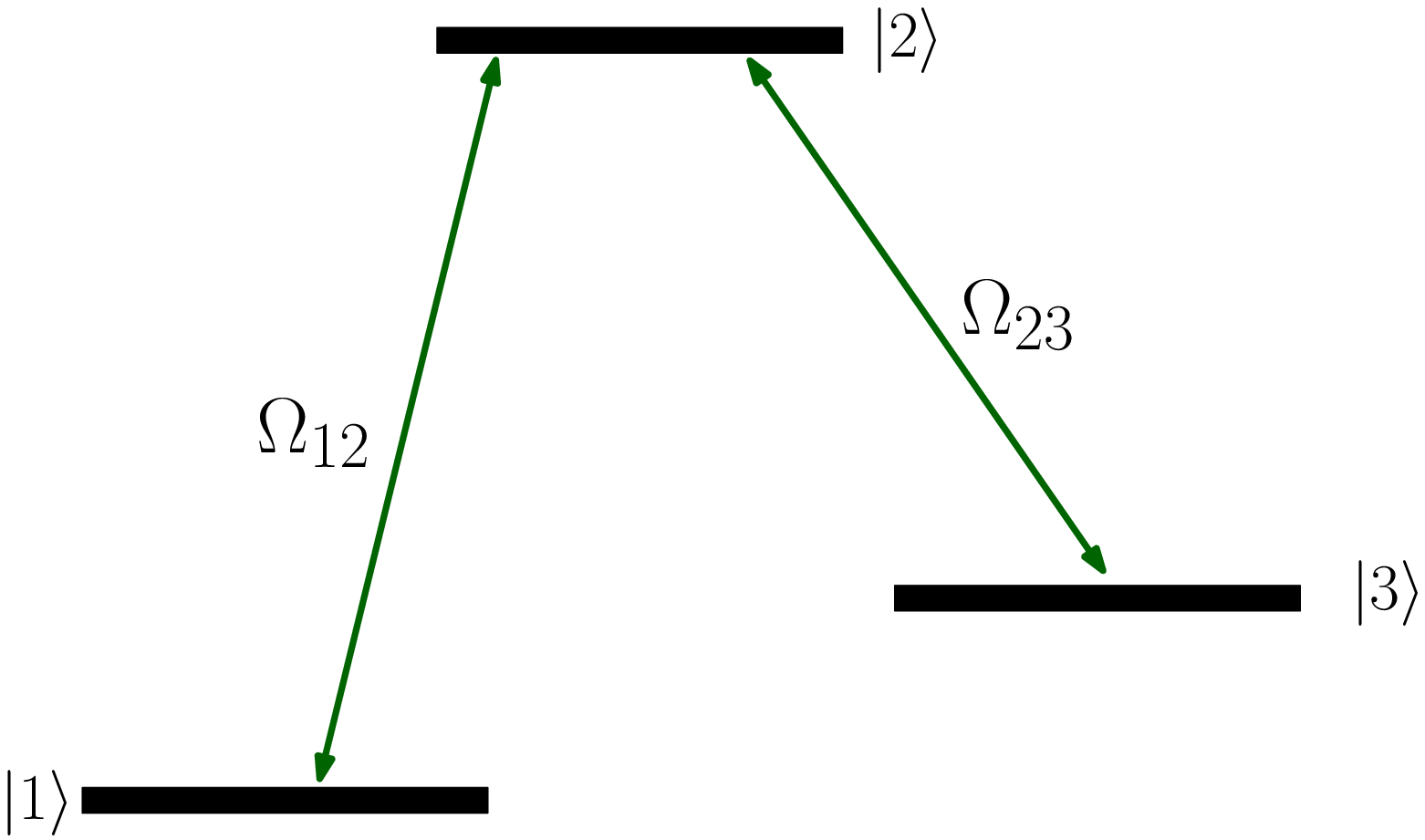}
c)\, \includegraphics[height=3.3cm,angle=0]{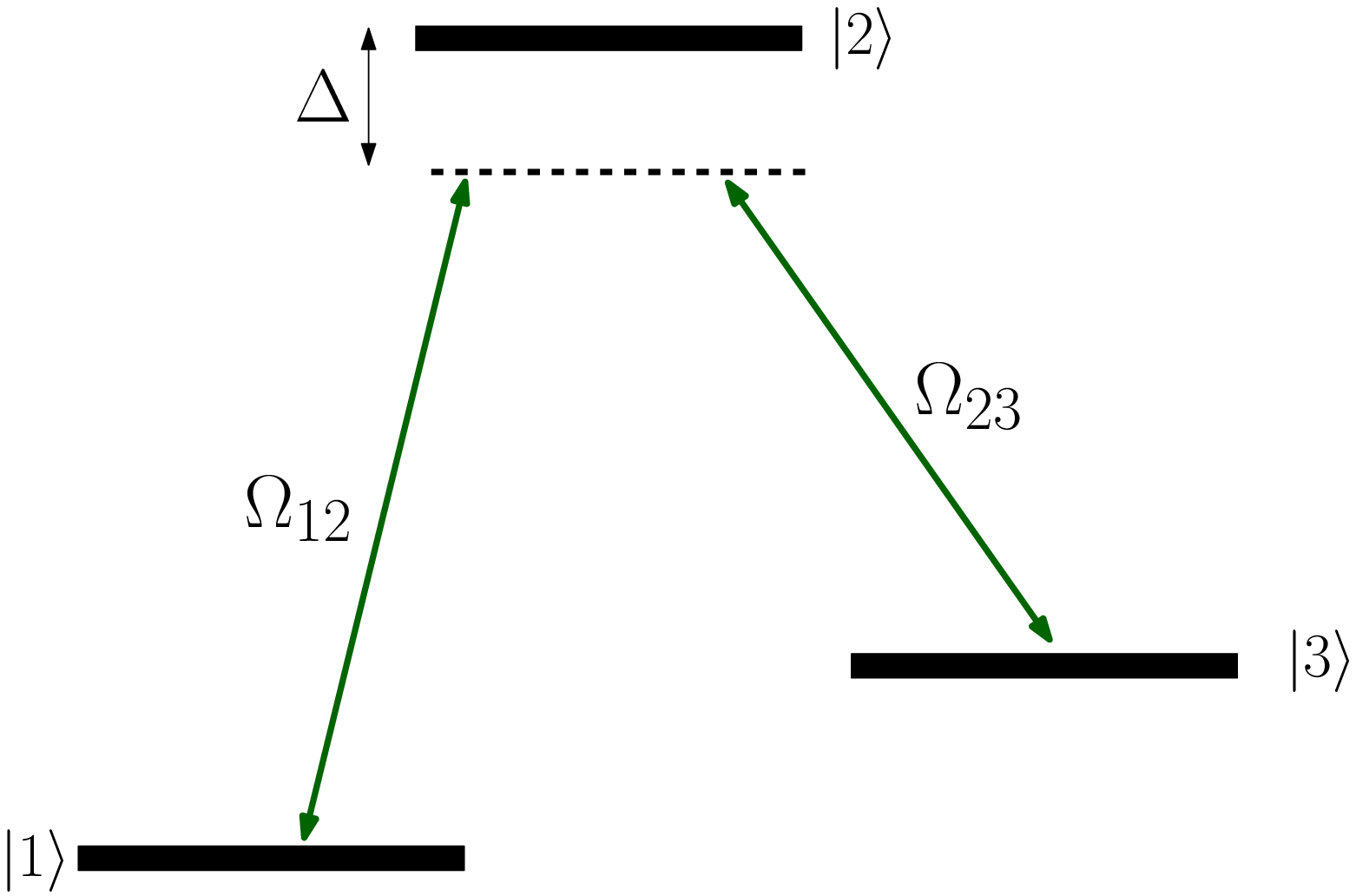}
\end{center}
\caption{\label{figure3level}
(Color online)
Different strategies for STA in three-level systems. (a) Apply counterdiabatic STA: initial couplings (red, solid lines), additional required STA coupling (blue, dotted line). (b) Apply invariant-based inverse engineering STA: modified STA couplings (green, solid lines). (c) Detuned couplings: use STA techniques after mapped to a 2-level system: modified STA couplings (green, solid lines).}
\end{figure*}

{\it Dirac systems.}
An application of the perturbative technique has been demonstrated for Dirac systems \cite{Song2017}. A  
plethora of natural or artificial systems obey the Dirac equation in certain conditions, with a proper reinterpretation of symbols. 
The new physical platforms for Dirac dynamics (trapped ions, optics, superconducting circuits) are easier to manipulate than relativistic particles. In trapped ions, for example, the effective (simulated) mass, speed of light, or electric field may be changed in time. This rich simulation scenario opens prospects for finding and implementing new or exotic effects and carrying out fundamental studies. Shortcuts to adiabaticity offer a suitable framework for the task \cite{Muga2016}.
For example  \textcite{Deffner2016} used the fast-forward  technique  to suppress production of pairs (transitions among positive and negative energy solutions) in fast processes. 

The goal of \textcite{Song2017}  is instead to induce a fast and robust population inversion among the bare levels on a 1+1-dimensional Dirac equation 
for a charged particle simulated by ultracold trapped ions,  designing a  simulated electric field $\alpha_t$. The problem is that the coupling between momentum and internal levels in the Dirac equation changes with the momentum.
For each plane wave, there is a 
momentum-dependent Hamiltonian, 
\beq
H_{p_0}=\left(\begin{array}{cc}
mc^2&cp_0+\alpha_t
\\
cp_0+\alpha_t&-mc^2
\end{array}
\right),
\label{diracH}
\eeq
but a  robust population inversion should be independent 
of the momentum within the momentum spread of the wave packet. This is achieved by considering the $p_0$-dependent  part in Eq. (\ref{diracH}) as a perturbation and designing the time dependences of the other elements in the Hamiltonian using the approach  in \textcite{Ruschhaupt2012}.


\subsubsection{Other  approaches}
%
%
%
In \textcite{Guery-Odelin2014_063425} a Fourier method is used to find transport protocols (for either a single particle or BEC) which are robust with respect to spring-constant errors, see more details in Sec. \ref{Fourier}. It exploits the fact that the final excitation energy can be expressed as the Fourier transform of the trap acceleration \cite{Reichle2006}. It avoids  perturbation approximations and can also be applied to   transport  non-interacting particles of different species. Its connection to flatness based control in mathematics is discussed in Sec. \ref{Fourier}.  

The use of perturbation theory or iterative methods is also avoided in \textcite{Zhang2017_15814,Levy2018}. Using the fast-forward approach (see section \ref{ss:FF})  a magnetic field is determined which prescribes specific stable dynamics of a single or two interacting spins \cite{Zhang2017_15814}. In \textcite{Levy2018}, protocols are designed for both a two-level system and a harmonic oscillator, which are stable against Markovian noise sources. They are designed using dynamical invariants and are made robust by enforcing that the invariant approximately commutes with the noise operator during the process. 

The effect of the environment can be also reduced by choosing the phases in the --noise-free-- evolution operator $U(t)=\sum_n e^{i\xi_n(t)}|n(t)\rangle \langle n(0)|$ to design the driving Hamiltonian \cite{Santos2018_025301}.

\textcite{Boyers2018} propose protocols based on ``Floquet engineering'', with periodic drivings where the coefficients 
are adjusted by matching an effective Floquet Hamiltonian found by a Magnus expansion with CD-driving Hamiltonians. 
For a qubit inversion the method is resilent to noise because the spectral bandwidth of the protocol (centered around the Floquet frequency) 
is separated from the spectral bandwidth of the noise around zero, as long as the noise is perturbative with respect to the driving.      

\subsection{Three-level systems\label{3ls}}
Few-level models are essential  to understand and manipulate actual or artificial atoms. 
We have presented several examples of shortcuts  applied to two-level models and in this section we review 
three-level models. STIRAP is a basic 
adiabatic method of reference, see \textcite{Vitanov2017} for a recent review, to transfer  population  among the two ground states of a $\Lambda$-configuration (or the extreme states in ladder systems) without populating the excited state. Being an adiabatic process, it can be  sped up with STA techniques. 
Here are three possible STA approaches to STIRAP speedup, see Fig. \ref{figure3level}:

\paragraph{Apply counterdiabatic shortcuts to the full three-level $\Lambda$ system.}
We assume a three-level $\Lambda$ system consisting of two ``ground levels'' $\ket{1}$ and $\ket{3}$ and a central excited level $\ket{2}$ coupled with time-dependent terms $\Omega_{12}$  and $\Omega_{23}$. We refrain from specifying by now the exact  nature of these couplings  which  will depend very much on the system. A first strategy is to apply the counterdiabatic STA technique directly to this $\Lambda$-system to speed up the STIRAP transfer. This strategy leads to an additional coupling between
the two levels $\ket{1}$ and $\ket{3}$  \cite{Unanyan1997,Demirplak2003,Chen2010_123003}. 
The improved robustness of STA schemes  compared with different STIRAP protocols was shown in \textcite{Giannelli2014}, and the effect of decay and dephasing was studied in \textcite{Issoufa2014}, who observed that the latter has more 
effect on the final fidelity than the former.  The robustness of this scheme with respect to energies fluctuations, e.g. due to collisions of a solute with a solvent, was examined in \textcite{Masuda2015_244303}.

In practice the additional coupling $\Omega_{13}$ can be implemented in some but not in all systems, e.g. because of selection rules due to  symmetry of the states or the necessary phase of the term.  In fact \textcite{Vitanov2019} proposes a method for efficient optical detection and separation of chiral molecules based on the phase sensitivity of the approach.
 
An example of a physical system to which this first strategy has been applied is ``Spatial Adiabatic Passage'' \cite{Menchon-Enrich2016}
in which  three wells play the role of  the three internal states. The additional imaginary coupling may be  implemented using a magnetic field that induces a  complex tunneling term \cite{Benseny2017}. 
In Nitrogen-Vacancy electronic spins, this additional coupling has been experimentally implemented mechanically via a strain field  \cite{Amezcua2017,Kolbl2019}. In a superconducting transmon with a three-level ladder configuration, the auxiliary field to induce a fast transition from the ground to the second excited state is 
achieved with a two-photon microwave pulse to circumvent the forbidden transition \cite{Vepsalainen2018}.  

As a generalization, a discrete  FF approach can be set to accelerate the STIRAP protocol with an additional control parameter  with respect to the CD solution \cite{Masuda2015_3479,Masuda2016_51}.
The following  paragraphs discuss alternative STA routes
when the new required coupling is not easy to implement or too weak,
e.g. a magnetic dipole transition.

\paragraph{Applying invariant-based inverse engineering shortcuts to the full three-level $\Lambda$ system.}
A second strategy is to apply the invariant-based inverse engineering to the $\Lambda$ system.  \textcite{Chen2016_033405}
apply this strategy  to a Hamiltonian with  resonant couplings that imply an $SU(2)$ dynamical symmetry, 
and build different protocols that may or may not populate level $\ket{2}$,   without the need for an additional coupling between $\ket{1}$ and $\ket{3}$. 
Interestingly, to achieve the same fidelity, less intensity is required when the intermediate level $\ket{2}$ is populated. 
This means  that protocols that populate level $\ket{2}$ may be considered as useful alternatives for certain systems and sufficiently short process times.
Moreover, \textcite{Chen2016_033405} put forward the concept  of
invariant-based ``multimode driving'', where the dynamical  state is a combination of invariant eigenstates rather than 
just one of them, as it had been customary in previous works.  
  
Related to this approach are  also the speeded-up STIRAP protocols  based on the dressed-state approach in \textcite{Baksic2016},  
which were used in experiments with  nuclear spins \cite{Coto2017,Zhou2017_330}.

\paragraph{Use STA techniques after mapping to a two-level system.}
The third strategy is based on mapping or approximate the three-level system to a two-level system and then applying two-level STA techniques.
In particular when the  middle level $\ket{2}$ is  detuned it can be adiabatically eliminated. 
The counterdiabatic techniques of STA can be then applied to this effective two-level system and the resulting pulses
can be  mapped back to the three-level system. No additional coupling is required between the metastable states
$\ket{1}$ and $\ket{3}$, and the existing couplings are only modified.
This approach is exploited theoretically in \textcite{Li2016_063411} and, using cold $^{87}\rm{Rb}$ atoms, it was experimentally implemented in \textcite{Du2016}.

\subsection{Motional states mapped into a discrete system} 
In trapped systems a  simplifying route to apply  shortcuts to control motional degrees of freedom  
is to discretize first the quantum system into a finite number of localized states,  that could be time dependent. Then the previous methods can be applied.   
Ideally the resulting STA protocol should be translated to the original setting to check its performance, 
or resistance to noise and perturbations,  although this  step is not always realized. 

We provide here some examples of approximations in terms of two, three and four states:  

-{\it Two states}: Wavepacket splitting operations were modeled by  systems of two time-dependent states in \textcite{Torrontegui2013_033630}, and multiplexing/demultiplexing of harmonic oscillator vibrational states in 
\textcite{Martinez-Garaot2013}. Two-level models are also used to study spin dynamics in a quantum dot with spin-orbit coupling \cite{Ban2012_206602}.

-{\it Three states}: In  ``Spatial Adiabatic Passage'', analogous to STIRAP   \cite{Vitanov2017}, a particle may tunnel between three wells. 
The system is approximated by a three-state system and shortcuts may be applied \cite{Benseny2017}. \textcite{Martinez-Garaot2014_053408} provides other examples, such as wavepacket splitting in three wells or operations on two-interacting  bosons in two wells.  

-{\it Four states}: \textcite{Kiely2016, Kiely2018} model the motion of an ultracold atom in a lattice  
by a four-state system, and apply   
invariant-based STA techniques to create exotic angular momentum states
of ultra-cold atoms in an optical lattice. 
In \textcite{Li2018_113029} the four-level model takes into account both motional  and internal aspects, 
representing up/down spin states in two different wells.

\section{Applications in quantum science and technology \label{qt}}
This section is organized by system type. 
A number of tables group together articles, otherwise disperse in different subsections, 
according to transversal criteria: ``gates'' in Table \ref{tablegates}, ``transport'' in  Table \ref{tabletransport}, and ``experiments'' in  Table \ref{tableexp}. 

\subsection{Trapped ions\label{trapped}}
Trapped ions constitute one of the most developed physical platforms to implement fundamental quantum
phenomena and quantum information processing. 
Since many ions in a single trap are difficult to control, a way towards large-scale computations with many qubits is a divide-and-conquer scheme \cite{Wineland1998,Kielpinski2002}, where ions are shuttled around in multisegmented Paul traps, while  keeping just a few ions in each processing site. Apart from shuttling, complementary operations such as separating and merging ion chains, trap rotations, and expansions or compressions of ion chains are  needed. Coulomb interactions, and controllable external effective potentials created by radio frequency or DC electrodes determine the motion of the ions and the corresponding Hamiltonians, which can be approximated by quadratic forms near equilibrium. 
\subsubsection{Dynamical normal modes\label{dnm}}
Dynamical normal modes are a useful generalization of ordinary normal modes  for  time-dependent,  
quadratic Hamiltonians \cite{Palmero2014},  or in the  small-oscillations regime for non-harmonic ones.
They are independent harmonic motions  which describe the dynamics of an effective time-dependent harmonic oscillator. 
Dynamical normal modes help to describe the motion in a simple way but also to inverse engineer the potentials to achieve fast motions without final excitation.  
The invariant-based engineering of the time-dependence of the potentials is a natural route for that end. An important difference with respect to 
inverse engineering a single harmonic oscillator is that several time-dependent harmonic oscillators for the different dynamical modes have to be engineered simultaneously with common 
control functions.

 For example, to transport a chain of two ions, the position of the external harmonic trap is a common control function, and  we cannot engineer a different trap position for each mode. 
The way to solve this type of  inversion problem is to increase the number of adjustable parameters in the ansatzes for the auxiliary functions, see Sec. \ref{sLLf},  so that all the boundary conditions of the auxiliary functions of all modes are satisfied simultaneously, either exactly or via minimization subroutines.   

Dynamical normal modes have been used to speed up, via invariants, the  transport of two or more (possibly different) ions   \cite{Palmero2013,Palmero2014,Lu2015},  and to design trap expansions or compressions of ion chains in a common trap \cite{Palmero2015_053411},  ion separation  \cite{Palmero2015_093031}, 
and two-ion phase gates driven by spin-dependent forces \cite{Palmero2017}. However, for some operations a point transformation\footnote{In a point transformation new coordinates depend only on old ones} to define the dynamical modes does not exist.

\begin{figure}[t]
\begin{center}
\includegraphics[height=5.0cm,angle=0]{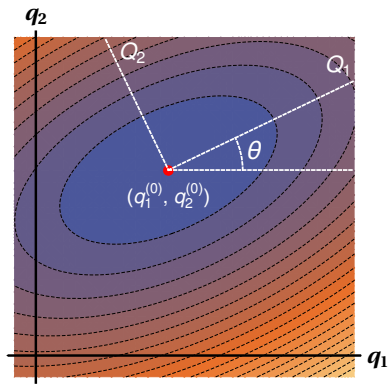}
\end{center}
\caption{\label{ellipse}
(Color online) 
Schematic representation of isopotential curves of a
mass-weighted potential in the two-dimensional configuration space
of laboratory-frame coordinates ${q_1 ,q_2 }$. These curves are ellipses
centered at the moving equilibrium position $(q_1^{(0)} ,q_2^{(0)})$ with the 
orientation of the principal axes given by the angle $\theta$. The dynamical normal mode coordinates 
are ${Q_1,Q_2}$. From \textcite{Lizuain2017}.}
\end{figure}

For 2D systems the condition that an uncoupling point transformation exists is simply that the principal axes of the equipotential ellipses, see Fig. \ref{ellipse},  do not rotate in the laboratory frame  \cite{Lizuain2017}. If  the ellipses are translated, expanded or compressed, dynamical normal modes may be defined via point transformations. An operation for which the point transformation does not exist in general is the separation of two different ions 
in an external potential $\alpha(t) x^2+\beta(t) x^4$, except when the ratio $\beta^3(t)/\alpha^5(t)$ is a constant.    
Similarly, for the rotation of a two-dimensional anisotropic trap holding one ion there is no point transformation leading to normal modes. 
A way out is to compensate for the inertial potential
proportional to the angular momentum with an (effective) magnetic field. This is similar to the inertial force compensation with
a homogeneous force in transport, see Sec. \ref{sLLf}. 
Finally, it is also possible to consider generalized transformations mixing coordinates and momenta
\cite{Lizuain2019}.   
\subsubsection{Ion transport\label{iont}} 
Among the different operations on ions addressed by STA methods, 
ion transport is the most studied both theoretically \cite{Palmero2013,Palmero2014,Furst2014,Lu2014_063414,Lu2015,Lu2018,Pedregosa-Gutierrez2015,Li2017_3272,Tobalina2018} and experimentally in Paul traps \cite{Bowler2012,Walther2012,Kamsap2015,Alonso2016,An2016,Kaufmann2018}. 
The first two experiments were done simultaneously in Boulder  \cite{Bowler2012}, using a Fourier transform technique,
and Mainz \cite{Walther2012}, optimizing some driving protocols, to transport one or two ions diabatically 
for 300 to 400 $\mu$m on a 5 to 10 $\mu$s time scale (a few oscillation periods) achieving final excitations below one motional quantum. 
Later, \textcite{Kamsap2015} transported  large ions cloud using numerical simulations to control the dynamics. In 2016, other shortcut techniques were used to improve transport experiments, for example the bang-bang method using nanosecond switching of the trapping potentials in \textcite{Alonso2016}. 
The experiment in \textcite{An2016} 
simulated CD-driven transport  in an interaction picture with respect to the harmonic oscillation, and also performed the compensating force approach as unitarily equivalent transport in the interaction picture. The driving forces were induced optically rather than by varying voltages of control electrodes.
Finally \textcite{Kaufmann2018} used recently invariant-based inverse engineering to design ion transport, and measured internal infidelities less than $10^{-5}$, an important prerequisite for the success of quantum-information-processing schemes that rely on ion transport. 

On the theoretical side,  
\textcite{Furst2014} used Optimal Control Theory  and the compensating force approach to analyze the transport of an ion in realistic conditions for state-of-the-art 
miniaturized ion traps. Given the simplicity of the approach and results, they considered the compensating-force to be the method of choice for current experimental settings. Anharmonicity was found to play no significant role.  This paper sets the relation between the desired trap trajectory and the 
voltages applied in the control electrodes. The model was later applied by \textcite{Tobalina2018} to analyze the energy cost of ion transport    

\textcite{Lu2014_063414,Lu2018} analyze the effect of different colored noises on single-atom transport and how to mitigate their effect. Dynamical and static sensitivities are distinguished, according to their dependence or independence with trap motion. They behave in opposite ways with respect to transport duration, which implies   
a transition  between the dominance of the dynamical sensitivity at short times and of the static one at large times. The crossover is important, as it demonstrates that 
the widespread  expectation that shorter STA times are always more robust versus noise (a behavior  that holds for the static but not for the dynamic sensitivity)   is not necessarily true, and that optimal times exist with respect to robustness.   
\textcite{Li2017_3272} proposed trigonometric protocols that 
minimize the (classical) excitation due to anharmonicities within a perturbative approach.  

Two-ion transport was addressed in \textcite{Palmero2013,Palmero2014}.  Also, \textcite{Lu2015} designed optimal transport of two ions under slow spring-constant drifts.
Designing fast transport of two different ions is challenging  because  the simple compensating force approach is not 
possible if only  forces induced by the electrodes are applied. They only depend on the charges, whereas the compensating forces should depend on the mass \cite{Palmero2014}. Dynamical normal modes can however be defined so the problem is solved using invariants \cite{Palmero2014}.    
Two-ion transport may also be performed with ``spin-dependent'' optically induced forces that may be different for different internal states. 
An interesting application  is the implementation of fast phase gates in which different  phases are imprinted depending 
on the internal states because of the different motions induced. Invariant-based design of the ion trajectories guarantees a 
robust phase because of its geometric nature and its  independence on the motional state
\cite{Palmero2017}. Similar ideas may be applied to design a single-ion driven interferometer to measure unknown small forces  \cite{Martinez-Garaot2018}. 
Interferometry driven by STA trajectories (using ions or neutral atoms \cite{Navez2016,Martinez-Garaot2018}) offers, compared to the usual schemes where the systems evolves freely along separated branches,
the possibility to control the timing and the sensitivity, 
absence of wavepacket dispersion, and robustness versus initial motional states.      
\begin{figure*}[t]
\begin{center}
\includegraphics[height=2.0cm,angle=0]{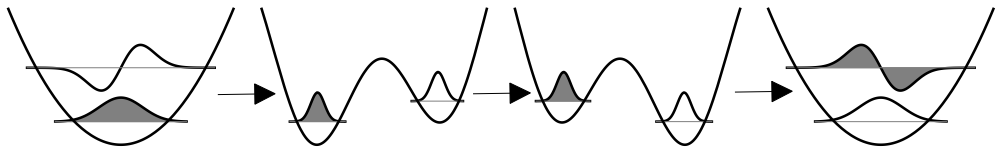}
\end{center}
\caption{\label{figuremulti}
Vibrational level inversion by STA processes based on trap deformations: demultiplexing, bias inversion, and multiplexing. From \textcite{Martinez-Garaot2013}}
\end{figure*}

\subsubsection{Other operations\label{oo}\\}  
{$^{}$}\vspace*{.01cm}\\

{\it Fast ion separation.}
Separating two ions, or more generally a chain,  is a delicate operation and STA-enhanced experiments \cite{Bowler2012,Ruster2014} provide excitations above one motional quantum per ion. 
\textcite{Bowler2012} used a FAQUAD approach, and \textcite{Ruster2014} optimized control parameters. 
In theoretical works \cite{Home2005,Nizamani2012,Kaufmann2014,Palmero2015_093031} the control process is modeled as a time-dependent evolution of the parameters in the external confinement potential $\alpha(t)x^2+\beta(t)x^4$. The difficulties come from the change of sign of $\alpha(t)$ from positive (in the initial trap) to negative values (to form the central barrier).  
When the harmonic confinement vanishes, the values of $\beta$ are bounded by experimental limitations. Thus  the confinement becomes  weak, levels get close to each other, 
and the ions suffer heating.  
If the ions are different, further difficulties arise as a consequence of the general absence of normal modes based on point transformations \cite{Lizuain2017}, as commented above, see  Sec. \ref{dnm}.\\   

{\it Ion expansions/compressions.}
\textcite{Palmero2015_053411} analyzed by invariant-based inverse engineering how to   expand/compress ion chains of equal or unequal ions, 
and \textcite{Torrontegui2018} studied how to speed up  a single-ion 
heat pump. This work analyzes the possibility to implement repulsive potentials. They are found to be feasible in the radial direction by turning 
of the radiofrequency drive   and relying on the Direct-Current  (cap) electrodes. Earnshaw's theorem does not allow for absolute minima so that a minimum 
in the axial  direction of the trap corresponds to a maximum in the radial direction.\\   

{\it Penning traps.}
Shortcuts have been applied as well to Penning traps: 
\textcite{Kiely2015} designed via invariants a non-adiabatic change of the magnetic field strength to change the radial spread without final excitations, and  \textcite{Cohn2018}   implemented experimentally a protocol  
to produce entangled states in a Dicke model realized in a two dimensional array 
of trapped ions.\\

{\it Rotations.}
\textcite{Palmero2016} provided invariant-based shortcuts to perform fast rotations of an ion in a 1D  trap.\footnote{Video in {https:$//$www.youtube.com/watch?v$=$ToQXnd$\_$FdUw}} \textcite{Lizuain2019} design fast rotations of a 2D anisotropic trap  to produce a rotated version of an arbitrary initial state when the two normal frequencies are commensurate. 
\subsection{Double wells\label{dw}} 
A double well is a useful potential to test fundamental quantum physics and applications in interferometry or quantum and classical information processing.  Shortcuts using invariants have been proposed for several operations involving double wells and motional states, for example to  split  one-particle wavefunctions in linear or nonlinear settings
\cite{Torrontegui2013_033630,Martinez-Garaot2016},  to speed up the cotunnelling of two interacting bosons
\cite{Martinez-Garaot2015_043406},
to split ion chains as discussed in Sec. \ref{oo}, or to model the erasure of a bit with a Brownian particle \cite{Boyd2019}.
Applications involving many-body systems are 
reviewed in Sec. \ref{mbsm}.  
Other STA operations in double wells are vibrational-state inversion and multiplexing/demultiplexing. 

Vibrational mode 
multiplexing is the spatial separation of the vibrational modes of a harmonic trap. For the first two modes 
the separation is done by transforming smoothly the harmonic trap into a biased  double well. \textcite{Martinez-Garaot2013} maps the STA process from a two-level model into a realizable potential  in coordinate space designing the time-dependence of two control parameters. A fast inversion of the double-well bias, so that the lower well becomes the upper one and viceversa, 
can be performed by noticing that the operation,  in an independent-well regime,  amounts to a transport process, so that the compensating force approach can be applied 
\cite{Martinez-Garaot2015_053406} to design the time dependence of the bias. Combining sequentially multiplexing, bias inversion, and demultiplexing, 
leads to a fast inversion of vibrational levels using only a transient trap deformation and no excited internal states, 
see Fig. \ref{figuremulti}.  Interestingly,  there is no adiabatic 
path connecting the ground and the excited states, but  a combination of STA processes leads to the desired state.   
The explanation is that inverting the bias  is in fact adiabatic within the approximation of independent wells, in other words, during a 
time scale which is short  compared to the duration of the process. 

\textcite{Bucker2011,Bucker2013} have independently performed motional-state
inversions by shaking  anharmonic potentials guided by an OCT algorithm to  produce twin atom beams and interferometry applications, but  
a smooth STA-based deformation similar to  Fig. \ref{figuremulti} has not yet been implemented experimentally.  
\subsection{Cavity quantum electrodynamics\label{cqe}}
{\it Entangled state preparation in cavity quantum electrodynamics.} 
In 
cavity quantum electrodynamics  atoms and light are confined such that the quantum states of the atom are protected by
allowing only controllable transitions compatible with the modes of the cavity and thus isolating the atom 
from any electromagnetic environment other than the cavity itself.
These systems have recently opened up new prospects to  implement  large-scale quantum computation and to generate  nonclassical states.

An early proposal to produce entangled states faster than adiabatically by moving atoms in and out an optical cavity is in \textcite{Marr2003}: the proposal  was to use  
cavity leakage  to stabilize the desired (adiabatic-like) time evolution by damping away population in unwanted states. However a consequence is that 
the success rate decreases.  
  
In the post-2010 era the use of  shortcuts to inverse  engineering the dynamics of two atoms inside a cavity was discussed first  by \textcite{Lu2014_012326}. They used the counterdiabatic driving formalism to create maximal entanglement between the two atoms but, due to the complexity of the resulting counterdiabatic Hamiltonian, they had to introduce an 
alternative, physically feasible Hamiltonian that needs  auxiliary internal levels, an extra laser field, and an extra cavity mode. The same group tried a different strategy using invariants   to accelerate the state transfer between two three-level atoms in a cavity quantum electrodynamics (QED) system \cite{Lu2014_105201}.

\textcite{Chen2014_033856} 
used invariants combined quantum Zeno dynamics (QZD)  to simplify the dynamics and speed up the population transfer between two atoms trapped in a cavity. The quantum Zeno effect inhibits transitions by frequent measurements so that  
the system evolves in the so-called Zeno subspace   \cite{Facchi2002}. 
This quantum Zeno dynamics  can be also achieved via a strong continuous coupling  \cite{Facchi2002}. In particular    
\textcite{Muga2008} discuss the mapping between discrete and continuous interactions.
Following \textcite{Chen2014_033856}, many other papers have used the QZD approximation to decouple different Hilbert subspaces by assuming that the atom-cavity coupling is much larger than the driving field Rabi frequencies.
The simplified effective Hamiltonians were controlled with the use of different STA techniques to design fast and robust protocols against decoherence produced by atomic spontaneous emission and cavity leakage, see for example \textcite{Wu2017_46255}. 
\textcite{Chen2014_115201} extended the idea of combining QZD and STA methods to control the dynamics of atoms trapped in distant cavities connected through a fiber. 

A requirement for the success of quantum mechanics in information processing tasks is scalability towards multiparticle ($N>2$) setups.
The macroscopic character of cavity QED setups favors the interaction among qubits and thus scalability. Numerous setups and control designs have been proposed to generate  large entangled states
using STA methods and the QZD assumption, 
in particular large-$N$ entangled W states  
\cite{Huang2015, Huang2016_171, Wang2016_162, Song2016_3169, Chen2016_140, Kang2016_052311, Kang2016_36737, Yu2017, Yang2018},
or GHZ states \cite{Chen2015_15616, Huang2016_25707, Ye2016, Zhang2017_015202, Huang2016_105202, Xu2017,Shan2016}. 
Moreover, using the QZD scheme several authors presented various STA setups to create 3D-entanglement between atoms individually trapped in distant optical cavities connected by a fiber \cite{Liang2015_5064,Lin2016, Wu2016_33669,Wu2016_2026} or between two atoms trapped in a single cavity \cite{He2016,Yang2017}.

Shortcuts have also been proposed to generate  other less common entangled states, such 
as  large NOON states of two sets of $\Lambda$-atoms in distant cavities using invariants \cite{Song2016_4159}, 
and, using CD driving, 
a three-atom singlet in a common cavity \cite{Chen2016_22202}, tree-type 3D entangled states \cite{Wu2016_33669, Wu2017_34}, two-atom qutrit entanglement \cite{Peng2017}, or maximally entangled states of two  Rydberg atoms \cite{Zhao2017}.\\

{\it Other applications.}
Different STA-enhanced quantum gates were designed within the QZD condition.   For example, invariant-based inverse engineering was used to develop Toffoli gates \cite{Song2014}, phase gates \cite{Chen2015_012325,Liang2015_032304}, CNOT gates \cite{Liang2015_23798}, or swap gates \cite{Liang2015_115201}. 
\textcite{Wu2017_294}  use the dressed-state method  \cite{Baksic2016} to design a fast CNOT gate in a cavity QED system which consists of two identical five-level atoms  in two single-mode optical cavities connected by a fiber.   
  
As well, shortcuts  have been proposed to produce  single photons on demand in an atom-cavity system approximated by three levels \cite{Shi2014}. 
A system of distant nodes  in two-dimensional networks
(cavities with a $\Lambda$-type atom) is approximated by a three-level $\Lambda$ system in \textcite{Zhong2016}
and then STA techniques are applied  to achieve  fast information transfer.\\ 

{\it Optomechanical systems.}
Shortcuts have been applied as well to optomechanical systems.  One of the early applications of the invariant-based approach in \textcite{Chen2010_063002}
was to cool down a mechanical resonator in a cavity optomechanical system with external optical fields \cite{Li2011}.\footnote{\textcite{Zhang2013_142201} proposed instead electromechanical cooling.} For more recent applications see \textcite{Zhou2017_095202,Chen2018_023841,Zhang2018_00102}.

\begin{widetext}

\begin{table*}

\caption{Quantum logic gates designed using STA methods\label{tablegates}\\CD: Counterdiabatic driving; OCT: Optimal control theory; QED: Quantum electrodynamics; QZD: Quantum Zeno dynamics; NV: Nitrogen vacancy.}. 
\begin{ruledtabular}

\begin{tabular}{llll}
Reference&Gate type&System&Method
\\\hline
\textcite{Martinis2014}&Phase&Superconducting Xmon transmon&Optimization
\\
\textcite{Song2014}&Toffoli&Cavity QED&QZD+Invariants
\\
\textcite{Santos2015}&Universal gates&$N$ qubits&CD driving
\\
\textcite{Chen2015_012325}&Phase&Cavity QED&QZD+Invariants
\\
\textcite{Liang2015_032304}&Phase&Cavity QED&QZD+Invariants
\\
\textcite{Liang2015_23798}&CNOT&Cavity QED&QZD+Invariants 
\\
\textcite{Liang2015_115201}&Swap&Cavity QED&QZD+Invariants 
\\
\textcite{Zhang2015_18414}&Non-Abelian geometric&Superconducting transmon&CD driving
\\
\textcite{Santos2016}&N-qubit&Four-level system&CD driving
\\
\textcite{Song2016_023001}&1- and 2-qubit holonomic&NV centers&CD driving
\\
\textcite{Liang2016}&Non-Abelian geometric&NV centers&CD driving
\\
\textcite{Palmero2017}&Two-qubit phase&Two trapped ions&Invariants
\\
\textcite{Du2017}&Non-Abelian geometric&NV centers&CD driving
\\
\textcite{Wu2017_294}&CNOT&Cavity QED&QZD+Dressed-state scheme
\\
\textcite{Santos2018_015501}&Single- and Two-qubit&Two- and Four-level system&Inverse  engineering
\\
\textcite{Liu2018}&Non-Abelian geometric&NV centers&Invariants
\\
\textcite{Wang2018_065003}&Single-qubit&Superconducting Xmon qubit &CD driving
\\
\textcite{Shen2018}&Two-qubit controlled phase&Two Rydberg atoms &Invariants
\\
\textcite{Ritland2018}&Majorana&Top transmon&OCT for noise cancelling 
\\
\textcite{Li2018_113029}&1-Qubit gate\&transport&Double quantum dot&Inverse engineering
\\
\textcite{Yan2019_080501}&Non-Abelian geometric&Superconducting Xmon qubit &CD driving
\\
\textcite{Lv2019}&Non-cyclic geometric&Two-level atom&CD driving 
\\
\textcite{Santos2019}&Single qubit& Nuclear Magnetic Resonance& CD driving
\\
\textcite{Qi2019}&Single and double-qubit holonomic&Rydberg atoms&CD driving
\end{tabular}
\end{ruledtabular}
\end{table*}
\end{widetext}

\subsection{Superconducting circuits\label{sc}}

Superconducting circuits have recently made rapid progresses and become a leading architecture for quantum technologies \cite{Wendin2017}.
In this context STA techniques were independently developed under the names ``Derivative Removal by Adiabatic Gate'' (DRAG) for single qubits \cite{Motzoi2009}, and 
Weak anharmonicity with average Hamiltonian (WAHWAH) \cite{Schutjens2013} for multi-qubit setups and multilevel systems (qutrits), see also \textcite{Theis2016,Theis2018,Lu2017} and the discussion in Sec. \ref{sai}, focusing on avoiding unwanted transitions to spectrally neighboring energy levels. 

Other techniques for sped-up manipulations take into account the peculiarities of 
the experimental settings, such as small nonlinearities in the qubits (implying that 
unwanted transitions are not necessarily off-resonant), or a need for smooth  pulses (versus square pulses 
common in nuclear magnetic resonance).  In particular \textcite{Martinis2014} considered a two-level model 
where only the $\sigma_z$ term changes to achieve fast gates with the $\sigma_x$ term constant in 
$H=H_x\sigma_x+H_z\sigma_z$. 
Relating the  error in the gate operation to the Fourier transform  of the (properly scaled) rate of change of the polar angle $\theta=\arctan(H_x/H_z)$,
optimal protocols were  found, minimizing the integrated error for any time larger than some chosen time.

Other works apply the standard STA methods, in particular CD driving,  
to transfer information between distant nodes of flux qubits  in annular and radial superconducting networks
\cite{Kang2017_1700154};  to complete Bell-state analysis for two superconducting-quantum-interference-device qubits
\cite{Kang2017_022304};  or to measure the Berry phase \cite{Zhang2017_042345} in a phase qubit. 

Further applications of CD shortcuts include the proposal or realization of gates.  
\textcite{Zhang2015_18414} proposed  holonomic one and two-qubit gates based on four-level systems in  superconducting transmons. Experiments were performed to implement single-qubit quantum gates in a superconducting Xmon  system \cite{Wang2018_065003} with a hybrid CD+DRAG approach, and non-Abelian geometric gates  with a ladder three-level system \cite{Yan2019_080501}. Superconductors also play a role as a possible platform to realize  
topological quantum information processing based on ``braiding'' non-Abelian quasiparticles. \textcite{Karzig2015} explore CD protocols to realize these braiding operations in finite time. 

Interfacing different architectures to make better use of their optimal features may be important to jump from proof-of-principle to practical technologies. 
In particular, superconducting circuits may be combined with opto-mechanical systems producing hybrid quantum systems. \textcite{Zhou2017_095202} 
propose a protocol to efficiently convert microwave to  optical photons, enabling the transmission of information through optical fibers with minimal loss.
One more  application of shortcuts in superconducting circuits is to create photonic cat states.  
Stored in high-Q resonators these states could lead to  efficient universal quantum computing \cite{Puri2017}.  
\subsection{Spin-orbit coupling\label{SOC}}
{\it Discrete models.}
Coherent spin manipulation in quantum dots 
via electric, magnetic, and spin-orbit coupling (SOC) control is one of the  avenues to solid-state based quantum information. 
STA methods have been proposed to speed up operations and fight decoherence also in this context. 
In particular, STA are very  welcome to speed up double quantum dots with SOC control, since strong fields nontrivially slow down the operations  involving 
spin-orbit coupling and tunneling \cite{Khomitsky2012}.  

The interplay between motional and spin degrees of freedom is typically modeled 
by an effective discrete Hamiltonian where STA techniques are applied. The physics behind the model, however, makes the exercise nontrivial, 
as the different matrix elements are not always controllable independently by the available external manipulations. 
An example of these physical constraints  and a way out is provided in    
\textcite{Ban2012_206602,Ban2012_249901}, in which the spin dynamics in a quantum dot with spin-orbit coupling and a weak magnetic field 
is controlled by time-dependent external electric fields only. The dependence of the effective $2\times 2$ Hamiltonian on the electric field makes a CD approach non-viable with electric control. However invariant-based inverse engineering is applicable and circumvents the difficulty.  

Two electrons in a double dot offer more freedom since different electric fields can be applied to each dot. Thus CD driving combined with 
a unitary transformation could be applied to induce fast singlet-triplet transitions \cite{Ban2014}, again within a $2\times 2$ Hamiltonian modeling.    

Synthetic spin-orbit coupling in ultracold atoms and condensates is also of interest to control internal and motional states. 
Invariant-based protocols to simultaneously control the internal (related to its pseudospin-$1/2$) and motional  states of a spin-orbit-coupled BEC in Morse potential are also studied in \textcite{Ban2015} by a $2\times 2$ effective Hamiltonian.

A $4\times 4$ Hamiltonian is used in \textcite{Li2018_113029} to model and design the   transport of a qubit 
encoded in the electron spin among two quantum dots, performing simultaneously an arbitrary  qubit rotation (gate).
The transfer may be extended sequentially to a chain of dots. 
These processes need 
time-dependent control of the spin-orbit and inter-dot tunneling coupling. The dynamical  
engineering of the four-level system applies the technique developed in  \textcite{Li2018_013830} based on the geometry of 4D rotations.\\

{\it Models with a continuum.}
Models that retain motional (1D, in $x$ direction) and internal degrees of freedom 
of the electron without discretization began with   
\textcite{Cadez2013}, who considered an electron with spin-orbit coupling in a moving harmonic quantum dot,   
\beq
H(t)=\frac{p_x^2}{2m^*}+\frac{m^* \omega^2}{2}[x-\xi(t)]^2+p_x(\alpha \sigma_y-\beta\sigma_x),
\eeq
where $m^*$ is the effective electron mass, $\xi(t)$ is the time dependent position of the harmonic trap, and 
$\alpha$ and $\beta$ are Rashba and Dresselhaus spin-orbit couplings. The dynamics of this model is exactly solvable 
via invariants, and fast bang-bang trap trajectories leading to spin control (e.g. spin flip) without final excitation were found   
for $\beta=0$.  
In \textcite{Cadez2014}, $\beta$ is fixed to zero while $\alpha$ remains time dependent. The dynamics can still be found exactly and expressions are given for dynamical and geometrical phases in closed loop trajectories of the control parameters.  
\textcite{Chen2018_013631} translate these ideas to synthetic spin-orbit coupling in BECs and apply inverse engineering of 
$\xi(t)$ and $\alpha(t)$ $(\beta=0)$ from auxiliary Newton-like equations for the center-of-mass position of the condensate and its spin precession.      
Again spin is flipped fast by proper design of the controls. 

{Controllable linear-in-momentum interactions to implement  direct (not unitarily transformed) CD-driven transport in the lab frame, i.e., with 
the form $H_{CD}=p_x \alpha(t)$, with $\alpha=\dot{q}_c$, see \textcite{Torrontegui2011} and Sec. \ref{sLLf}, 
may in principle be implemented  by synthetic spin-orbit coupling for one of the spin components \cite{Tobalina2018}.
Note that the  change of  sign for the other 
component is crucial to determine the possible applications.  
This dependence precludes, for example, transporting a qubit, but it is useful to set different paths in interferometry \cite{Martinez-Garaot2018}.
\subsection{Nitrogen-vacancy centers\label{NVc}}
Quantum information processing with nitrogen-vacancy (NV) centers is appealing because of the  possibility to operate with qubits at room temperature. 
Yet,  decoherence is still a problem  and several STA protocols have been implemented in experiments or proposed theoretically.

The counterdiabatic driving approach was applied to design (``universal'') sets of fast and robust non-Abelian geometric gates: \textcite{Song2016_023001} propose to use four NV centers around a whispering-gallery mode microsphere cavity; 
while \textcite{Liang2016} make use of a single NV center coupled to a $^{13}\rm{C}$ nuclear spin both modeled as two-level systems;  
and \textcite{Du2017} a three-level scheme for the NV center  where two of the couplings are due to microwave fields, and the third coupling 
is mechanically induced.\\   

Experiments have shown the flexibility of NV centers to implement STA protocols for discrete systems:  
\textcite{Kleissler2018} implement 1-qubit holonomic gates proposed by \textcite{Liang2016} in an effective two-level system driven by a microwave field with controllable, time-dependent  detuning, Rabi frequency, and phase;  \textcite{Liu2018} used the eigenbasis of the dynamical invariant $I(t)$ associated with $H(t)$ as the auxiliary basis for constructing geometrical gates; and 
in \textcite{Zhou2017_330}, stimulated Raman adiabatic passage in a three-level system 
was sped up using dressed state driving  in terms of the original controls of the reference Hamiltonian. 

The possibility of physically implementing two microwave couplings and one mechanical coupling among three  levels is one of the interesting features of STA applied to NV centers \cite{Du2017,Amezcua2017}. 
Thanks to this structure \textcite{Kolbl2019} use CD protocols experimentally to implement the 
initialization, readout, and coherent control of 3-level dressed states. These states offer efficient coherence protection,
better than the one achieved by two-level systems.   
\subsection{Many-body and spin-chain models\label{mbsm}}
Many-body systems display controllable emergent properties and phenomena  
potentially useful in metrology and quantum simulation or computation, as well as 
in applications such as quantum light generation, memory devices, or precise sensing and communications. 
Varying control parameters slowly is one of the key tools to prepare and manipulate quantum many-body systems. In particular,  adiabaticity plays  
a central role to understand or implement the quantum Hall effect, topological insulators, adiabatic computing, certain quantum phases, 
see \textcite{Bachmann2017} and references therein, or to maximize entanglement \cite{Dorner2003}.

\begin{table*}
\caption{\label{table1}Works where STA methods are applied to LMG-like models}
\begin{ruledtabular}
\begin{tabular}{lll}
Reference&Reference Hamiltonian&Comment
\\\hline
\textcite{Julia-Diaz2012}&$U(t)J_z^2-2JJ_x$&Invariants, large $N$.
\vspace*{.02cm}\\
\textcite{Yuste2013}&$UJ_z^2-2J(t)J_x$&Invariants, time-dep. mass 
\vspace*{.02cm}\\
\textcite{Takahashi2013_062117}&$-(2/N)[J_x(t)S_x^2-
J_y(t)S_y^2+\gamma S_y^2]-2h(t)S_z$&Large $N$, CD  
\vspace*{.02cm}\\
\textcite{Campbell2015}&$-(2/N)(S_x^2+\gamma S_y^2)-2h(t) S_z$&Approx. CD
\vspace*{.02cm}\\
\textcite{Opatrny2016}&$A_c(t)H_c+A_n(t)H_n$; $H_{c}=\vec{c}\cdot \vec{J}$, $H_n=(J_z-nI)^2$&Approx. CD
\vspace*{.02cm}\\
\textcite{Hatomura2017}&$-[(2J)/N]S_z^2-2\Gamma(t) S_x-2hS_z$&Mean field, CD
\vspace*{.02cm}\\
\textcite{Takahashi2017_012309}&$f(t)\{-[(2J)/N] S_z^2 - 2hS_z\}-2\Gamma(t) S_x$&Mean field, invariants
\vspace*{.02cm}\\
&$-f(t)[(2J)/N]S_z^2-2\Gamma_x(t)S_x-2\Gamma_y(t)S_y$&
\vspace*{.02cm}\\
&$f(t)(-\sum J_{ij}\sigma_i\sigma_j^2-2hS_z)-2\Gamma(t) S_x$&
\vspace*{.02cm}\\
\textcite{Hatomura2018_015010}&$-[(2J)/N]S_z^2-2\Gamma_x(t)S_x$&Mean field, CD
\end{tabular}
\end{ruledtabular}
\end{table*}

All the above sets a strong motivation for developing STA approaches in many-body systems. 
In some cases ``exact'' shortcuts may be found, for example   
after having applied mean-field theories, or semiclassical approximations for large number of particles $N$, or due to exact solvability,   
as for self-similar dynamics for specific interactions \cite{Muga2009,delCampo2011_031606,delCampo2012_648,delCampo2013,Deffner2014}. 
\textcite{Rohringer2015} demonstrated scaling behaviour and shortcuts for expansions and compressions 
of phase-fluctuating quasi-1D 
Bose gases, and more recently \textcite{Deng2018_013628,Deng2018_eaar5909} in a three-dimensional anisotropic ``unitary Fermi gas''. 

Indeed, early STA experiments  showed  that many-body systems may benefit from STA  techniques: 
expansions of clouds of cold thermal atoms were handled via invariants in the independent-atom approximation
\cite{Schaff2010}, and interacting Bose Einstein condensates could be expanded fast using scaling in the mean-field approximation 
\cite{Schaff2011_23001,Schaff2011_113017}.  

However, adiabaticity is often problematic for many-body systems making STA challenging. 
The difficulties are illustrated by the ``orthogonality catastrophe'':  two ground states for two slightly different values of a control parameter $\lambda$ may become orthogonal in the thermodynamic limit so that 
``strict'' adiabaticity breaks down in essentially zero time. For certain drivings this occurs even with a finite gap. 
Milder definitions of adiabaticity are possible \cite{Bachmann2017} refering to local variables rather than to the global $N$-body wavefunction.

The difficulties to implement adiabatic drivings are also evident in phase transitions across a quantum critical point,
which lead to excitations for any finite crossing rate of the parameter.     
CD drivings can be found for the family of models which are solvable by a transformation into independent fermions, as for the 1D Ising model
in a traverse field \cite{delCampo2012_115703},\footnote{Other solvable model where  the CD driving involves as well many-body interactions, 
 is the one-dimensional Kitaev honeycomb model \cite{Kyaw2018}.
It has been proposed to generate highly entangled ``cluster states'' needed to implement ``measured based quantum computation''.}  
\begin{equation}
H=-\sum_{n =1}^N (\sigma_n^x \sigma_{n+1}^x+g\sigma_n^z).
\end{equation}
After a Jordan-Wigner transformation, the Hamiltonian is decomposed into a series of independent Landau-Zener  Hamiltonians for which the counterdiabatic driving is known.  
However in spin space $H_{CD}$  is  
highly non-local,  it involves long-range, multi-body interactions, increasingly important and divergent near the critical point \cite{delCampo2012_115703,delCampo2013,Damski2014,Damski2015,delCampo2015,Duncan2018}.  For specific models the non locality may be circumvented. Thus \textcite{Okuyama2016} achieve a local CD for the Toda lattice using the machinery of Lax pairs, and \textcite{Takahashi2013_062117}
for the XY model uses a ``fixed-point condition'' where $H_{CD}=0$.  
  
Systematic efforts have been done to find approximate shortcuts based on truncation of the CD terms \cite{delCampo2012_115703,Damski2014} by restricting the range of the interaction, 
imposing adiabaticity only locally via local interactions \cite{Mukherjee2016},  or optimizing an approximate ansatz for  $H_{CD}$ based on simple
(in particular two-body) auxiliary fields. This optimization requires spectral and wavefunction knowledge that is available for finite systems of experimental interest \cite{Saberi2014}.  
Such detailed information is however typically not available for larger systems so       
another major research thread is to avoid using explicit spectral information to construct the shortcut.
A variational approach  proposed in \textcite{Sels2017}, see Sec \ref{beyond}, moves  in that direction. In particular, the aim of adiabatic computing is to find the ground state, which encodes the solution of a computational problem,  
by adiabatic following, precisely because it cannot be calculated. \textcite{Yoshimura2015} proposed to apply a method to estimate the probability to be in the ground state from time-dependent measurements without knowing 
the Hamiltonian or its eigenfunctions, in order to optimize approximate shortcuts.   
Another  phenomenological way out for some applications may be to optimize control parameters experimentally. For example  
\textcite{Cohn2018}  propose an optimization of a bang-bang protocol for the external parameter to 
produce entangled states in a Dicke model realized in a two-dimensional array 
of trapped ions in a Penning trap. \\

{\it Josephson junctions and Lipkin-Meshkov-Glick-like models.\vspace*{.2cm}} 
Much work has been done to apply shortcuts to many-body models which are either similar to or particular cases of the  Lipkin-Meshkov-Glick (LMG)
model 
\beq
H_{LMG}=\hbar[\Omega J_z+W(J_x^2+J_y^2)+V(J_x^2-J_y^2)], 
\eeq
where $J_x=\frac{1}{2}(a^\dagger b+ab^\dagger),\, J_y=\frac{1}{2i}(a^\dagger b-a b^\dagger)\, J_z=\frac{1}{2}(a^\dagger a-b^\dagger b)$
are ``pseudospin" operators, with    
$a$, $b$ being annihilation operators for two sites or two internal states. The same type of Hamiltonian appears in spin models where the 
operators $J_\alpha$ are to be interpreted as global angular momentum operators, usually denoted  
instead by $S_{\alpha}=\frac{1}{2}\sum_i \sigma^i_\alpha$, $\alpha=x,y,z$, where $\sigma^{i}$ is the Pauli spin operator for site $i$.  
Table \ref{table1} depicts some of the original Hamiltonians for which shortcuts  were developed.    
     
Bosonic Josephson junctions were  treated  in \textcite{Julia-Diaz2012} to generate spin-squeezed states. 
The junction was modeled with a two-site Bose-Hubbard Hamiltonian, see Table \ref{table1}. The ground state for $U>0$ (antiferromagnetic LMG model) is unique. 
In the large-$N$ limit and in the Fock basis of boson imbalance between the two wells the system may be treated semiclassically i.e., considering a continuous rather than discrete population imbalance, for time dependent $U(t)$ and fixed $J$.  
The model simplifies to a single particle in an oscillator with time-dependence frequency. Invariants can then be used to design shortcuts and produce spin squeezed states. A similar approach was later applied when the time dependence is in $J$, which is better suited for ``internal junctions'' where the connected states are not at different locations but in different internal states \cite{Yuste2013}.  

Other works have treated the two-site Bose-Hubbard Hamiltonian for low particle numbers, specifically $N=2$:   \textcite{Opatrny2014}
propose it as a toy model for a transition between a superfluid (maximally delocalized)  state and a
Mott-insulator ground state,  
and explain the difficulties to physically implement the CD Hamiltonian\footnote{Suppose that $A$ and $B$ are feasible operators 
in same $H_0$ and that $H_{CD}$ needs $[A,B]$, which is difficult to implement directly.  
According to the Baker-Hausdorff-Campbell
formula, 
applying the sequence of operators $bB\to aA\to -bB\to -aA$, where $a$ and $b$ are coefficients,
during time intervals $\Delta t$ amounts to applying $iab[A,B]\Delta t$ for a time $\Delta t$, 
$e^{iaA\Delta t} 
e^{ibB\Delta t} e^{-iaA\Delta t} e^{-ibB\Delta t}= e^{-ab[A,B]\Delta t^2}+O(\Delta t^3)$. Thus, in principle implementing $ab\Delta t [A,B]$ is possible with $\pm aA$ and $\pm bB$,
but the problem lies in the scaling of the coefficients with $\Delta t$.
For a given target value of $c=ab\Delta t$, then $ab=c/\Delta t$, which leads to intense pulses. 
If high intensities are  available the 
dynamics with $H_0$ becomes more adiabatic so that the CD term is not really needed \cite{Opatrny2014}.};  
\textcite{Martinez-Garaot2014_053408} apply Lie-algebraic methods to this problem, see  Sec. \ref{utr}, to get  STA 
without experimentally unfeasible generators; and  
\textcite{Stefanatos2018_055009} optimizes shortcuts via optimal control 
to maximize entanglement. 

\textcite{Opatrny2016} looked for maximally spin squeezed ``Dicke states'' for finite $N$ 
approximating the CD terms as in \textcite{Opatrny2014}, and discussed how to implement compensating terms that go beyond 
quadratic order 
in the collective spin operators by means of suitable time sequences switching between quadratic operators.

The negative non-linear coupling coefficient (ferromagnetic model) was studied in \textcite{Takahashi2013_062117} by providing the CD term for the ground state in the large-$N$ thermodynamic limit. It generally diverges at the critical point. However, protocols satisfying a ``fixed-point'' condition such that $H_{CD}=0$ were shown to be feasible in some instances.  \textcite{Campbell2015} used   approximate CD terms identified by retaining dominant few-body terms, 
that may be optimized, rather than calculated from spectral information,  with the aid of the instantaneous ground state. By applying a small longitudinal field that avoids the critical point, 
a mean-field prescription was applied using invariant-based engineering in \textcite{Takahashi2017_012309}, and counter-diabatic driving  in \textcite{Hatomura2017}.   
\textcite{Hatomura2018_015010} applied the semiclassical large-$N$  approximation to generate a  CD term, which, corrected by finite-size terms, 
avoids divergence at the critical point to generate cat states. Formally the process amounts to performing a transition from a single to a double well
assisted by CD driving. The effect of particle losses is studied in \textcite{Hatomura2019}.\\

{\it State transfer.}
Fast quantum state transfer in linear chains has been also addressed theoretically with STA concepts and techniques.  
\textcite{Agundez2017} used a CD approach plus unitary transformation in a spin chain 
and  \textcite{Wang2016_062338} speed up a slow protocol by scaling up the Hamiltonian. A
similar method was applied in \textcite{Ren2017}
to cut a chain into two pieces. \textcite{Ban2019} proposed to inverse engineer 
the tunneling barriers to transfer two-electron entangled states 
from one edge of an array of quantum dots described by the Hubbard model to the other. 
Finally,     
\textcite{Longhi2017} proposed a
nonadiabatic fast protocol of robust excitation transfer in a non-Hermitian Hatano-Nelson tight-binding linear chain assisted by gain and loss gradients to cancel nonadiabatic transitions providing a fast state transfer in coupled-resonator optical waveguide structures.\\ 
 
\begin{figure}[t]
\begin{center}
\includegraphics[height=5.4cm,angle=0]{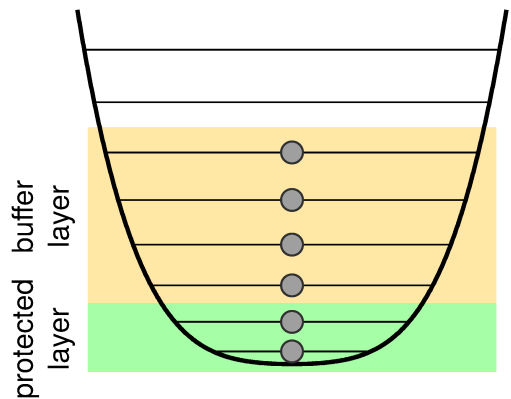}
\end{center}
\caption{\label{buffer}
(Color online) To prevent transitions from the protected layer  to higher-lying states during changes of the external parameters,
a buffer layer of fermions is added.  
The Pauli principle prevents the protected fermions  from accessing any level in the buffer zone. Adapted from 
\textcite{Dowdall2017} copyright American Physical Society.}
\end{figure}

{\it{1D Tonks-Girardeau  gases and  fermionic systems}}.
Under 1D effective confinement, the control and dynamics of non-interacting fermions or Tonks-Girardeau  gases can be studied with solvable models that ultimately rely on the Slater determinant built with orbitals for single particle dynamics.  
Shortcuts have been applied to them in several works with different aims:   
\textcite{Martinez-Garaot2015_043406} with the aid of the FAQUAD approach and  \textcite{Schloss2016} using invariants devised protocols to 
prepare macroscopic entangled states (NOON states) stirring a  Tonks-Girardeau gas on a ring. The observation that leaking between two occupied states does not influence the fidelity of the gas, and that only leaks into modes above the Fermi level do,   
was systematically exploited and developed in   
\textcite{Dowdall2017} for more general objectives to control fermionic systems. 

\textcite{Dowdall2017} propose to use  Pauli blocking for speeding up adiabatic
evolution of the ground state of a number of ``protected fermions'' by using an additional layer of buffer particles,
see Fig. \ref{buffer}. The protected fermions cannot make a transition to higher-lying excited states because these are occupied,
and  the fidelity for the final state of the protected particles 
increases with the number of buffer particles.  
\subsection{Metrology\label{metrology}}
Quantum systems offer, via interference, entanglement, and squeezing many interesting possibilities
for metrological applications and for improving measurement sensitivities.  Allowing for time dependence in the Hamiltonian 
can only enhance the possibilities to find useful protocols \cite{Pang2017} so  shortcuts to adiabaticity  will also play an 
important role in metrology.   
We have already seen along the review several examples: In Sec. \ref{trapped} we mentioned how 
STA-enhanced interferometry, in which the states are STA-driven along the interferometer branches, can increase sensitivities 
and decrease decoherence effects. Also, Secs. \ref{cqe},  \ref{sc}, or \ref{mbsm} provide examples where shortcuts  are used to create 
in different platforms states of metrological interest  such as spin-squeezed states, NOON states, and other entangled states. 

Here we underline the  strong  link between CD driving and optimal protocols for quantum metrology.    
\textcite{Pang2017} pointed  out that fundamental precision limits valid for time-independent Hamiltonians can be 
dramatically violated with appropriate time-dependent control.  In particular the Fisher information,  
whose inverse square root limits the precision to measure  some parameter $g$ in the Hamiltonian $H_g$ for arbitrary estimation strategies, 
can beat the limits valid for time-independent Hamiltonians.  The information on $g$ 
is transferred to a time evolution of a state measured at some time $T$. The upper bound for the Fisher information can be realized for superpositions states whose components evolve at all times along the maximal and minimal eigenvalues of $\partial_g H_g(t)$. Here a slight modification of CD driving comes in handy. Similarly to Sec. \ref{ssec:cd},
a Hamiltonian term $H_c(t)$ is added to $H_g(t)$ so that the dynamics follows exactly the eigenstates $|\psi_k(t)\ra$ of $\partial_g H_g(t)$ (rather than the eigenstates of a zeroth Hamiltonian $H_0(t)$ as in regular CD driving). Allowing for a freedom to choose phases $\theta(t)$ for each mode    
$H_{tot}=H_g(t)+H_c(t)$ may be written as 
\beq
H_{tot}=-\sum |\psi_k(t)\ra \dot{\theta}(t)\la \psi_k(t)| +i\hbar\sum|\partial_t\psi_k(t)\ra\la \psi_k(t)|,
\eeq
with  exactly the same form as Eqs. (\ref{alternative1}) and (\ref{alternative2}), except for the meaning of the wavefunctions.   
The way  to find $g$ is in fact iterative, assuming first an estimate for $g$ 
used to improve the estimate and so on. 
All this opens a very exciting avenue since   
the possible difficulties to implement counterdiabatic driving terms as well as the arsenal of remedies described in Sec. \ref{ssec:cd} are applicable.

\section{The energy cost of STA, engines, and the third law\label{ec}}
\subsection{Energy costs\label{ecproper}}
In human affairs, shortcuts to reach a place or a goal may cost money or consume energy, 
although there is no universal recipe or law on the costs applicable to all circumstances. 
Some shortcuts are really for free, or almost,  so they save time and resources, but others need a toll.  
Finding the ``energy cost'' of shortcuts to adiabaticity  is a very relevant and indeed, lately,  popular goal. 
It is tempting to consider shortcuts to be energetically a free lunch since, 
by definition, the final energy is the same as for a slow adiabatic process. Many works, however 
put forward different associated costs and imply that STA driving  can be hardly for free. 
In general the term ``cost'' has been used in a loose 
and heuristic way, without a fundamental analysis that justifies its suitability as an energy consumption. 
Actually most studies ``define'' rather than ``find'' the cost so     
that  different definitions of ``cost'' given so far are not necessarily in conflict. 
As long as we leave aside
the propriety of the term,  we may regard them as different aspects of the system energies or interactions, 
such as their evolution (transient values or time averages), excitations, fluctuations and flows, ``intensities'', or inequalities for  several times and energies involved.

In an early study 
on the expansion/compression of a particle in a time-dependent harmonic oscillator \cite{Chen2010_053403}, 
the cost was related to the time average of the particle energy. This average and its scaling with process time $t_f$ depend strongly on the STA applied. If 
the shortcut  makes use of a properly chosen $\omega(t)$ (e.g. designed via invariants) and no extra 
Hamiltonian terms,
it was found by Euler-Lagrange optimization that for the $n$-th eigenstate 
the time averaged energy obeys $\overline{E_n}>(2n+1)\hbar/(2 \omega_f t_f^2)$ for $(\omega_0/\omega_f)^{1/2}\gg1$ and $t_f\ll(\omega_0\omega_f)^{-1/2}$, where 
$\omega_0$ and $\omega_f$ are initial and final (angular) trap frequencies.   
Realizing this bound is indeed possible but at the price of infinite instantaneous power at boundary times \cite{Cui2015}.   
The bound is relevant even in anharmonic traps since the trap depth $D$ should be larger than $\overline{E_n}$, which sets a scaling 
$t_f\gtrsim [\hbar/(\omega_f D)]^{1/2}$ for the minimal process time \footnote{The energy in CD driving scales differently. For example, choosing a linear ramp for the reference $\omega(t)$,
$\overline{E_n}$ is independent of $t_f$, whereas the time-averaged variance (for all monotonous $\omega(t)$) goes as
$t_f^{-1}$.}.  The time average of the standard deviation of the energy was found to scale 
also as ${t_f}^{-2}$, tighter than the bound  $\gtrsim t_f^{-1}$  that follows from the Anandan-Aharonov (AA) relation \cite{Anandan1990},
\beq
\tau\geq \frac{\hbar}{\overline{\Delta E}}{\cal L}, 
\eeq
setting $\tau=t_f$, 
where
\beqa
{\cal L}&=&\arccos|\la \psi(0)|\psi(\tau)\ra|,
\\
\overline{\Delta E}&=&\frac{1}{\tau}\int_0^{\tau} (\la H^2\ra-\la H\ra^2)^{1/2} dt.
\eeqa
This is one of the first applications of  
``quantum speed limits'' to shortcuts to adiabaticity,\footnote{The AA relation has been later renamed a ``Mandelstam-Tamm-type'' relation, even if Mandelstam
and Tamm did not consider time-dependent Hamiltonians \cite{Mandelstam1945}.} 
see for review \textcite{Deffner2017}. Many other applications have followed \cite{Santos2015,Campbell2017,Abah2017} using AA and/or 
a Margoulis-Levitin type of relation \cite{Deffner2013_010402}. 
To be noted is that a naive extension of the AA relation substituting $\overline{\Delta E}$ by the time-averaged  energy $\la H(t)\ra$ is not valid in general
for time-dependent Hamiltonians. 
Instead, a valid Margoulis-Levitin-type relation is \cite{Deffner2013_335302,Santos2015}
\beq
\tau\geq \hbar \frac{|\cos{\cal L}-1|}{\frac{1}{\tau}\int_0^\tau |\la \psi(0)|H|\psi(t)\ra|}.
\eeq

As a more recent example of other definitions, the average power computed as the 
total work divided by the process 
time was defined as the cost in \textcite{Herrera2014}.
The cost has also been linked to the accumulated \cite{Zheng2016_042132} or time-averaged  \cite{Santos2015,Santos2016,Coulamy2016,Santos2018_025301} 
Frobenius norm of the Hamiltonian 
$||H(t)||=\sqrt{tr[H(t)^2}]$  or of some $n$-th power depending on the setting. 
This norm does not exist for many commonly found Hamiltonians such as the one for the harmonic oscillator. 
If it exists and $H=H_0+H_{CD}$, the time average for $H$, $\Sigma_{H}(t_f)=\frac{1}{t_f}\int_0^{t_f} \sqrt{tr[H_0^2+H_{CD}^2]}dt$, is larger than the 
one for the reference protocol $\Sigma_0(t_f)=\frac{1}{t_f}\int_0^{t_f} \sqrt{tr[H_0^2]}dt$, which
suggests that CD driving always implies an additional  cost. 
Choosing the phases $\xi_n(t)$ of the evolved states in the unitary evolution operator (\ref{eq:U}) has an impact on the energy cost. Specifically $\Sigma_{H}(t)$
is minimized by setting $\dot{\xi}_n(t)=-i\la \dot{n}(t)|n(t)\ra$ \cite{Santos2018_025301}. In the framework of Eqs. (\ref{Berry Hamiltonian}-\ref{h0xi}) 
this amounts to applying $H_{CD}$ alone, $H=H_{CD}$.

Some authors \cite{Zheng2016_042132,Campbell2017} define {\it differential costs},  instantaneous or accumulated, in terms of the CD term only, 
ignoring $H_0$. 
In particular, existence problems are circumvented by considering the state-specific counterdiabatic term $H_{CD}^{[n]}$
in Eq. (\ref{onestate}). Note that $||H_{CD}^{[n]}||=\hbar\sqrt{2 \la \dot{n}|\dot{n}\ra}$ exists in systems where $||H||$ does not.  
\textcite{Campbell2017} applied  ``quantum speed-limit inequalities'' to driving by $H=H_0+H_{CD}^{[n]}$
combining different  norm types (trace norm and Frobenius norm).
Interestingly, Demirplak and Rice already considered $||H_{CD}||$ and $||H_{CD}^{[n]}||$ as a measure of the ``intensity'' 
of the couterdiabatic  terms and used them to find ``minimal'' CD terms. The state-specific $H_{CD}^{[n]}$ is thus less intense (costly) than  
the general purpose $H_{CD}$ \cite{Demirplak2008}.  

Other ``differential costs'' have been defined using different references. 
In an STA  protocol driven by $H(t)$ with instantaneous eigenvalues ${\sf{E}}_n(t)$
and an initial state which is diagonal in the ${|{\sf{n}}(t)\ra}$ eigenbasis of $H(t)$, with initial probabilities $p_n(0)$, 
\textcite{delCampo2014} define a work distribution and the corresponding average work as 
$\la W\ra=\sum_{k,n}[{\sf{E}}_k(t)-{\sf{E}}_n(0)] p_{nk}^t p_n^0$, which is $\la H(t)\ra-\la H(0)\ra$ for such states. 
$p_{nk}^t=|\la {\sf{k}}(t)|U(t)|{\sf n}(0)\ra|^2$ is the probability for the system to start at $|{\sf n}(0)\ra$ and be found at $|{\sf k}(t)\ra$ at time $t$. Similarly,  they define the ``adiabatic work'' $\la W_{ad}(t)\ra=\sum ({\sf{E}}_n(t)-{\sf{E}}_n(0)) p_n^0$ 
and suggest as a ``pragmatic'' definition of cost the time-average of the differential $\delta W(t)=
\la W(t)\ra-\la W_{ad}(t)\ra$. A lower bound $\sim t_f^2$ for STA processes with inverse engineered 
time-dependent frequencies was found in \textcite{Cui2015}. 

In CD-driven processes,  the eigenenergies of $H(t)$, ${\sf E}_n(t)$,  differ in general from the eigenenergies 
of $H_0(t)$, $E_n(t)$. Similarly, eigenstates of $H(t)$, $|{\sf n}(t)\ra$, and $H_0(t)$, $|{ n}(t)\ra$, differ in general, although in most  processes 
$H_{CD}=0$ is imposed  at $t=0$ and $t_f$,  so that the initial and final eigenvalues and eigenvectors of $H_0$ and $H$ coincide. 
\textcite{Funo2017} redefine $\la W_{ad}(t)\ra=\sum [{{E}}_n(t)-{{E}}_n(0)] p_n^0$ in terms of $H_0$ eigenvalues (this would agree with the previous definition at time $t_f$) and consider the work distributions
\beqa
P[W(t)]&=&\sum_{k,n}p_n^0p_{n\to k}^t\delta\{W(t)-[{\sf E}_k(t)-E_n(0)]\},
\nonumber\\
P_{ad}[W(t)]&=&\sum_n p_n^0 \delta[W(t)-W_{ad}^{(n)}],
\eeqa
where $W^{(n)}_{ad}(t)=E_n(t)-E_n(0)$.  
Since the initial state density operator is assumed diagonal in $\{|n(0)\ra\}$ it will be diagonal in $\{|n(t)\ra\}$ for all time because of the CD driving.  
As $\la n(t)|H(t)|n(t)\ra=\la n(t)|H_0(t)|n(t)\ra$ for all times, it follows that $\la H(t)\ra=\la H_0(t)\ra$ for all times and  thus  
$\la W(t)\ra=\la W_{ad}(t)\ra$ during the process. As for the distributions, they coincide at boundary times $t=0$ and $t_f$. 
\textcite{Funo2017} also consider the work fluctuation with respect to the adiabatic trajectory, 
$\delta(\Delta W)^2=\sum_{m,n} p_n^0 p_{n\to m}^t [{\sf{E}}_m(t)-E_n(t)|^2$, and find that
\beq\label{Funfun}
t_f\geq \frac{\hbar{\cal L}[\rho(0),\rho(t_f)]}{\overline{\delta\Delta W}},
\eeq
where now ${\cal L}$ is the Bures length between initial and final (mixed) states.  
\textcite{Funo2017} identify $\overline{\delta\Delta W}$ as the thermodynamic cost to implement the CD driving.
The inequality (\ref{Funfun}) is tighter than the Aharonov-Anandan  relation, which involves instead the fluctuation $\la H^2(t)\ra -\la H(t)\ra^2$, although the latter is of broader applicability, 
since it is not restricted to CD driving (nevertheless \textcite{Bukov2019} conjecture, and validate for some models, 
that the quantum speed limit for all protocols is bounded by the quantum speed limit for CD protocols). 
An experimental demonstration was carried out with an Xmon qubit \cite{Zhang2018_085001}. 
These results were also extended to classical systems \cite{Bravetti2017}. \textcite{Funo2019_013006} study a quantum speed limit for open quantum systems described by the Lindblad master equation. They find a ``velocity term''  that when the thermal relaxation is dominant compared to the unitary dynamics of the system,  is approximated by the energy fluctuation of the counter-diabatic Hamiltonian.

A related  trade-off relation between time, entropy and state distance is worked out by \textcite{Takahashi2017_115007} 
considering  a canonical equilibrium state in density operator form $\rho(0)$, corresponding to a  Boltzmann distribution with temperature $T$ as the initial state.
Removing the contact with the thermal bath, this evolves unitarily by $H(t)=H_0(t)+H_{CD}(t)$ into 
$\rho(0\to t)$.  Let $\rho(t)$ be the (instantaneous) equilibrium state corresponding to  $H(t)$, and $\rho_0(t)$ the equilibrium state for $H_0$, both at the same temperature as the initial state.  
Using the Kullback-Leibler divergence, or relative entropy,  from $\rho_1$ to $\rho_2$,  $D_{KL}=tr \rho_1 \ln \rho_1 -tr \rho_1 \ln \rho_2$, 
\textcite{Takahashi2017_115007} finds the ``Pythagorean relation'' 
\beqa
&&D_{KL}(\rho(0\to t)||\rho(t))
\nonumber\\
&&=D_{KL}(\rho(0\to t)||\rho_0(t))+D_{KL}(\rho_0||\rho(t)),
\eeqa
which is  interpreted as a decomposition of entropy production.  A trade-off relation mentioned above follows from setting a lower bound to it.

\begin{figure}[t]
\begin{center}
\includegraphics[height=5.3cm,angle=0]{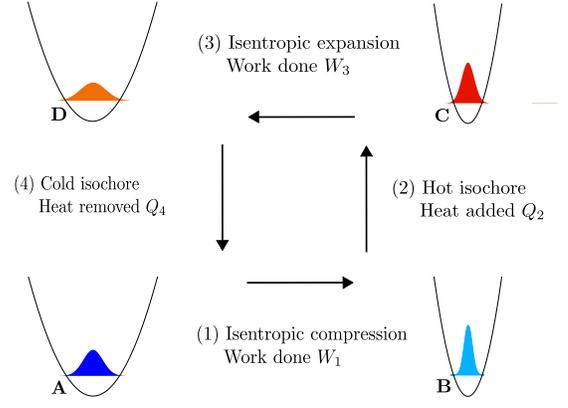}
\end{center}
\caption{\label{Otto}
(Color online) Quantum Otto cycle for a particle in a harmonic oscillator, adapted from  
\textcite{Abah2018_032121}. The colors are intended to represent the temperature ordering as red (C) $>$ orange (D) $>$ light blue (B) $>$ deep blue (A).}
\end{figure}

\textcite{Abah2017} consider STA expansions and compressions of a harmonic trap holding a particle
in the context of microscopic quantum engines and refrigerators, see Fig. \ref{Otto},
making use only of different time-dependences for $\omega(t)$.  
A reference Hamiltonian $H_0(t)$ is complemented by some modification term $H_{STA}(t)$ (found by unitarily equivalence from a CD term
as explained e.g. in  \textcite{Ibanez2012_100403}, see  Sec. \ref{utr} and Eq. (\ref{omegap})), so that the resulting total $H(t)$ 
drives an STA process. The authors associate the energy cost with the time-averaged $\overline{\la H_{STA}\ra}$. 
They also propose to add these terms (one for the compression step, $\overline{\la H^1_{STA}\ra}$, 
and one for the 
expansion, $\overline{\la H^3_{STA}\ra}$)  in the denominator of the efficiency of an Otto cycle 
to the  heat added,
\beq
\eta=\frac{\rm work\, output}{{\rm heat\, added}+\overline{\la H^1_{STA}\ra}+\overline{\la H^3_{STA}\ra}},
\eeq
as an extra energy input.  
Thus this approach ignores the possible role of $H_0(t)$ in actual energy costs and leads to some paradoxical results. 
In particular, choosing 
$H(t)=H_0(t)+H_{STA}(t)$ as the new reference $H_0'(t)=H(t)$, with frequency $\omega'(t)$ given by Eq. (\ref{omegap}), no extra term is needed since $H_0'(t)$ is already a shortcut. 
The efficiency, being of a differential nature,  would then have different values for the same process depending on how the Hamiltonian is partitioned (in one, $H_0'=H$,  
or two pieces, $H_0+H_{STA}$). 

If the unitary transformation on $H_{CD}$ is not performed \cite{Abah2018_032121,Abah2018_00580}, the same type of modified efficiency is proposed with the time averaged $\overline{\la H_{CD}\ra}$ playing the role of $\overline{\la H_{STA}\ra}$. However, $\la H_{CD}(t)\ra$ is zero by construction at  all times for states diagonal in the eigenbasis of $H_0(t)$, see 
Eq. (\ref{h1}). Accordingly, all processes would have the same zero cost independently of the $H_0(t)$ chosen.

Several papers analyze the effect of a control system (CS) (also termed driving or auxiliary system) 
coupled to the primary system (PS) of interest to set the STA driving
and its influence on the energy cost. 
\textcite{Calzetta2018} points out that the time-dependent driving Hamiltonians in STA 
processes are typically semiclassical and thus approximate. 
For a simplified model with a particle in a harmonic oscillator whose frequency depends on a coordinate of a driving system and is subjected to 
quantum fluctuations, Calzetta estimates the excitation 
when the STA process is implemented on average. It grows inversely with process time, but also vanishes  
as the mass of the driving system increases.   In a different vein, \textcite{Horowitz2015}, 
for a model in which the mesoscopic PS system of interest  is coupled to dissipative Markovian noise processes and to an auxiliary control system  
connected to a thermal bath, argue that the minimal work to drive the system through a specified path of states is due to the need for the controller to compensate for 
the dissipation that tries to take the system away from the path. 
The exact formula depends on the Hamiltonian and state of the primary system for weak coupling but 
on the full Hamiltonian and state (including CS+PS) for strong coupling.          

\textcite{Torrontegui2017,Tobalina2018,Tobalina2019} also stress the importance of the control system to find out true energy consumptions. The control system  here is 
the equipment necessary to set the values of the control parameters of the primary system. This equipment is usually  macroscopic and behaves classically.  
Even without direct dissipation of the primary system, implementing the driving will require energy consumption due to an external force, invested to change the 
energy of the global system (PS+CS) and to combat friction of the control system, see Sec. \ref{cranes} and Fig. \ref{grua}. 
Since STA processes are by definition fast, the arguments to justify minimal energy consumptions neglecting friction in the limit of slow motion (see e.g. \textcite{Landauer1961}) are not of much value.\footnote{For a recent analysis of the additional work required by shortcuts to erase a bit  in finite time 
beyond Landauer's bound see \textcite{Boyd2019}. The model used is a Brownian particle in a double well-potential. Instead of a tradeoff between information and energy, as in Landauer's work, more complex tradeoffs are found that depend on  information and its robustness,  energy, statistical bit-bias difference, size of the memory states, and speed.}   
In a scenario with macroscopic control system and microscopic primary system, driving the control system  along a predetermined path $x(t)$ is likely to be the dominant source of consumption, 
with  the energy and backaction of the primary system  being negligible in comparison, e.g. because of a large CS mass $M$ compared to the PS mass $m$. 
These and other effects are indeed shown explicitly with models for transporting a classical load \cite{Torrontegui2017} or an ion in a multisegmented Paul trap \cite{Tobalina2018,Tobalina2019}. 
A further relevant observation in these works is that in practice both positive and negative powers of the external force typically 
imply consumption, i.e., the energy given away by the system in ``braking'' time-segments with a negative power 
is not stored and recovered at will in positive-power segments, although a phenomenological parameter is
introduced to account for the possibility to perform, at least partially, energy-efficient regenerative-braking.
The models also show the importance of the control system to find the experimental gauge to determine the PS energy and corresponding power. 
For example, in a simple harmonic transport of a particle with Hamiltonian  
$
p^2/(2m)+m\omega^2[x-x_0(t)]^2/2+g(t), 
$
driven by the control function $x_0(t)$, the gauge function $g(t)$ does not affect the dynamics and so it is frequently ignored, but it
may affect strongly the PS energy and corresponding power \cite{Campisi2011,Tobalina2019}. A similar effect is found in expansions and compressions of an ion in a Paul trap, where the gauge term implies the opposite behavior to what could be naively expected, namely, an increase of energy during the expansion, and a decrease during the compression.     
%
It remains to be seen if smart engineering and design can equate  the power of the external force to the PS power  (this  implies unrealistic assumptions in the worked out models, such as $M=0$ 
and no friction), or at least make them proportional, as approximately realized in  some model examples \cite{Tobalina2018,Tobalina2019}. 
In any case further  analysis of energy consumption, both fundamental and for STA processes in different systems \cite{Impens2018} is needed.

\subsection{Engines and refrigerators}
A fundamental problem to design heat engine cycles  is that maximal thermal efficiency (output work divided by the heat input from the hot bath) 
is achieved with slow processes that minimize losses but also imply negligible power output (output work divided by cycle time). Shortcuts appear at first sight to solve this problem as the adiabatic evolution can be mimicked in short times. STA-driven engines or refrigerators have been considered mostly for harmonic oscillators performing Otto cycles, see Fig. \ref{Otto},  with two 
isochoric (constant frequency) branches for contact with the hot and cold baths and  thermally isolated compression/expansion branches where STA driving, as developed e.g. 
in \textcite{Salamon2009,Chen2010_123003}, is applied.\footnote{A version of the Otto cycle using single and two-spin-1/2 systems is proposed in \textcite{Cakmak2018}. An STA-enhanced Otto refrigerator based on a superconducting qubit with continuous coupling to two resonant circuits is analyzed in \textcite{Funo2019_03480}.}  For a recent review of the quantum Otto cycle for engines or refrigerators see \textcite{Kosloff2017}. 

%
\begin{figure}[t]
\begin{center}
\includegraphics[height=5.4cm,angle=0]{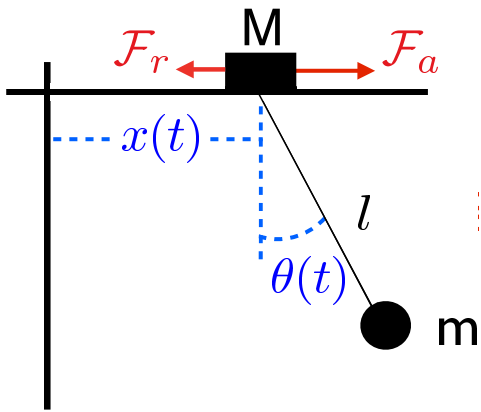}
\end{center}
\caption{\label{grua}
(Color online) Overhead crane composed of a load of mass $m$ (the primary system PS) and a trolley of mass $M$ (the control system CS) connected through a rope of constant length $l$.
 The red solid arrows represent the active force ${\cal F}_a$ and the friction force ${\cal F}_r$ acting on a rightward-moving trolley. Adapted from 
\textcite{Torrontegui2017}.}
\end{figure}

Before the term ``shortcuts to adiabaticity''  existed, Ronnie Kosloff's group had already worked out  ``frictionless'' bang-bang protocols for the isochoric strokes 
in which the adiabatic state was recovered in the final time, see e.g. \textcite{Salamon2009,Rezek2009}.  After 2010,   
two early studies on shortcuts  applied to cycles were \textcite{Torrontegui2013_032103},  which examined the performance of a refrigerator subject to noise, and  
\textcite{Deng2013}, which addressed classical and quantum systems to boost work characteristics and overall heat
engine performance. 

Many other works followed and studied potential advantages or optimization of STA-driven processes \cite{delCampo2014,Tu2014,Xiao2014,Abah2016,Abah2017}. Some have tried to enhance the power output by considering not just STA but also many-body systems \cite{Jaramillo2016,Beau2016,Chotorlishvili2016,Deng2018_eaar5909,Deng2018_013628}, 
and nonlinear BECs \cite{Li2017_015005} driven by STA found with variational methods \cite{Li2016_38258}.   
As well,  \textcite{Babajanova2018} find the equations of state for FF dynamics in expanding cavities that contain an ideal Fermi gas,    
and \textcite{Villazon2019} propose STA methods (unitarily transformed counterdiabatic driving) to manipulate both trap frequency and the coupling to the 
environment to realize fast approximate Otto engines operating near Carnot efficiency.                

The role and design of STA methods in quantum engines and refrigerators is very much an open field where rather fundamental questions are still
under scrutiny and debate. (Further aspects are discussed in Secs. \ref{3ppl} and \ref{sec:open}.) 
For example, most studies systematically focus on the primary system only so that efficiencies and power
computed with usual PS-based formulae 
are idealized limits that ignore CS effects discussed in the previous subsection. 
Some tentative proposals exist to modify the denominator (energy input) in the efficiency, 
e.g. taking into account time-averaged interactions of the terms added to the reference Hamiltonian to perform the STA \cite{Abah2017}, 
energy dissipated by noise in the controls \cite{Kosloff2017}, 
or energy consumptions to set the CS parameters \cite{Tobalina2019}.       
\textcite{Tobalina2019} point out that STA Otto engines based on a trapped ion are not ordinary engines. In the expansion stroke,   the ``piston'', 
whose role is played by the harmonic trap, is not pushed by a hot ion,  which exerts a negligible effect on the trap, but rather it is externally driven, 
in a fast manner specified by the STA protocol, by a controlled circuit  that consumes and dissipates energy to do so. 
In fact so much energy that the microscopic (PS) work output is far from compensating it.
This may well happen for a broad domain of macroscopically controlled quantum systems. A way out might be that the quality of the microscopic work  is worth the energy expense anyway, e.g. because of its effect on relevant degrees of freedom, 
but there is much to do to substantiate this hope
in practice.

I

\subsection{Third law\label{3ppl}}
The third law of thermodynamics was formulated by Nernst as the impossibility to reduce any system to absolute zero in a finite number
of operations,  see \textcite{Kosloff2013} for review. In the context of a quantum Otto heat pump it may be viewed as the vanishing of the cooling rate  
when the temperature of the cold bath approaches zero \cite{Rezek2009}, 
and be quantified by 
the scaling law that relates the cooling rate and cold-bath temperature.  
The fundamental bottleneck is the time needed for the expansion branch.
For a harmonic oscillator with a time-dependent frequency, the temperature of the equilibrium states connected adiabatically (i.e. slowly) 
or by STA is proportional to the frequency $\hbar \omega\propto kT$ to keep the average occupation number constant.   
Relevant questions are how fast we can lower $\omega$ and also finding  scalings between the wanted times and the resources
needed. 
The answers are not unique and  depend on the method and constraints imposed. 
For expansions limited to a designed protocol for $\omega(t)$ minimal times exist if $\omega(t)$ is real and bounded. 
Simple bang-bang solutions for real  $\omega$ give 
$t_f\sim \omega_c^{-1/2}$ ($\omega_c$ being the extreme, target value of $\omega$ in the expansion) \cite{Rezek2009,Salamon2009,Stefanatos2017_4290}. 
If $\omega(t)$ is not restricted and allowed to be imaginary, the expansion  times can be formally arbitrarily
short for any $\omega_c$ \cite{Chen2010_123003}. However it is unrealistic to assume 
that arbitrarily fast processes are viable, as variously argued in Sec. \ref{ecproper}.
In particular, if the time average of the energy is supposed to be bounded, since the trap depth cannot be arbitrarily high, 
the same type of scaling 
arises \cite{Chen2010_053403}.    
In fact more sophisticated bang-bang solutions, allowing for imaginary frequencies 
\cite{Hoffmann2011}, or an arbitrary number of switches \cite{Stefanatos2017_4290,Stefanatos2017_042103} lead to faster processes. 

%
%

%
%
\section{Open quantum systems\label{sec:open}}
In closed quantum systems, a slowly changing Hamiltonian can give rise to adiabatic dynamics as already discussed. By contrast, in open quantum systems, a slowly changing system Hamiltonian, or equivalently
a long evolution time, does not necessarily guarantee adiabatic dynamics. In addition, the cumulative effect of dissipation increases with time. Shortcuts to adiabaticity in this context should therefore be defined with care. Interestingly, the control design of open quantum systems paves the way for the use of shortcuts in the emerging field of thermodynamics of quantum systems.
\subsection{Concept of adiabaticity for open systems}
A direct consequence of the coupling of the system with the environment is 
the need to redefine adiabaticity, as new elements and time scales appear compared to closed systems.  
In 2005, Sarandy and Lidar generalized the adiabatic approximation to open systems  for convolutionless master equations \cite{Sarandy2005}, i.e., master equations of the form $\dot{\rho}(t)=\cL(t)\rho(t)$. Unlike closed systems, for which the Hamiltonian can always be diagonalized, the Lindblad superoperator $\cL(t)$ of an open system is not necessarily diagonalizable and in general can only be written in Jordan normal form. Adiabaticity for open systems is subsequently defined as the regime for which the evolution of the state of a system takes place without mixing the various Jordan blocks. Alternatively, the adiabatic approximation in open systems can  be formulated through an effective Hamiltonian approach \cite{Yi2007}. 

Further work about adiabaticity for open systems also includes the derivation of Markovian master equations suited
for studying the time evolution of a system evolving slowly while coupled
weakly to a thermal bath \cite{Thunstrom2005,Albash2012,Albash2015,Pekola2010,Venuti2016,Kiely2017}, and the derivation of a link between the notion of adiabaticity for open system and the theory of noiseless
subsystems \cite{Oreshkov2010}.

Hereafter, we address different strategies to implement STA protocols in the presence of a coupling with the environment. We first discuss techniques requiring a reservoir engineering,  then focus on methods set up to mitigate the effect of a (non manipulated) environment, see Fig. \ref{opensystems}.
We conclude on a more specific class of open systems that can be described by Non-Hermitian Hamiltonians.

\begin{figure}[t]
\begin{center}
\includegraphics[height=5.0cm,angle=0]{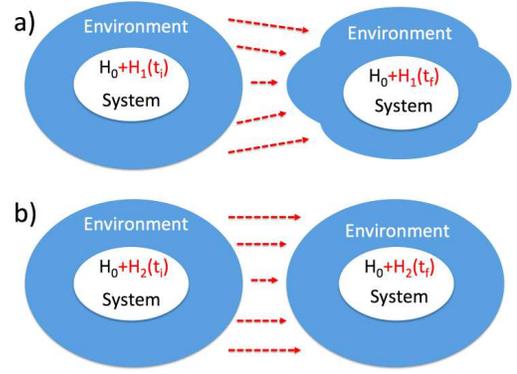}
\end{center}
\caption{\label{opensystems}
(Color online)
Two kinds of control STA protocols of an open quantum system: (a) those involving a reservoir engineering in  addition to Hamiltonian engineering, and those relying only on Hamiltonian engineering (b).
}
\end{figure}

\subsection{Engineering the environment}

Building on the work in \textcite{Sarandy2005}, the concept of a CD  driving was generalized to open systems  by working out the additional term to be superimposed to the Hamiltonian  to guarantee an independent evolution of the different Jordan blocks of the Lindblad operator \cite{Vacanti2014}. While this additional driving is unitary in some cases, in general it does not even provide a completely positive map. 

A natural framework to extend the shortcuts  to open systems is the explicit use of decoherence-free subspaces (DFSs) \cite{Wu2017_042104}. Consider that the coupling to the environment is accounted for by a Lindblad form  $\cL(t)=-\frac{i}{\hbar}\left[H,\rho\right]+\sum_{k} X_{k}^{\mathstrut} \rho X_{k}^{\dagger}-\frac{1}{2}\left\{X_{k}^{\dagger}X_{k}^{\mathstrut},\rho\right\}$, where the Lindblad operators may be time-dependent. Similarly to closed systems, the dynamical invariants of an open system provide us with both an intuitive physical framework and a set of tools to engineer
quantum states \cite{Ma2017}. In this perspective, it is useful to establish the modified equation for the invariant, which reads in this context
\begin{equation}
\partial_{t} I+\frac{i}{\hbar}\left[H,I\right]+\sum_{k} X_{k}^{\dagger} I X_{k}^{\mathstrut}-\frac{1}{2}\left\{X_{k}^{\dagger}X_{k}^{\mathstrut},I\right\}=0. \label{I_Lind}
\end{equation}
In contrast to closed systems, although $\la I \ra$ is still constant in time, its eigenvalues $\lambda_{n}$ are no longer necessarily time independent. More precisely, all $\lambda_{n}$ are time-independent if $I$ and $X_{k}$ have a common basis of eigenvectors for all $k$.  

However, it is possible to design a type of dynamical invariants for open quantum systems, in which a
part of the eigenvalues is constant in time. By construction, the corresponding eigenstates are in the time-dependent DFSs.
The interest of such subspace is that a quantum state evolves unitarily in it. By definition, the vectors of the DFS $\{ \ket{\Phi_{1}}, \ket{\Phi_{2}}, \ldots ,\ket{\Phi_{D}} \}$ are degenerate eigenstates of the Lindblad operators that obey the relation $X_{k}(t)\ket{\Phi_{j}(t)}=c_{k}(t)\ket{\Phi_{j}(t)}$ \cite{Wu2015_062122}. A counterdiabatic driving Hamiltonian in the time-dependent DFS can be shown to be of the form \cite{Wu2017_042104}
\begin{equation}
\widetilde{H}= H(t)+\frac{i}{2}\sum_{k} \left( c_{k}^{*}X_{k}^{\mathstrut}-c_{k}X_{k}^{\dagger} \right ).
\end{equation}
The driving based on invariants in the DFS can be equivalently worked out \cite{Ma2017}. 
An important condition that needs to be fulfilled to control in time  such quantum systems is the dynamical stability of the time-dependent DFS  \cite{Wu2017_042104}. The application to few-spin systems is explicitly worked out in  \textcite{Ma2017,Wu2017_042104}.

We note that another approach to setup the DFS consists in defining the invariant as a superoperator $\cal{I}$ that fulfills the relation \cite{Sarandy2007}%
\begin{equation}
\frac{\partial \cal{I}}{\partial t}=[\cL,\cal{I}].
\end{equation}
$\cal{I}$ is in general non-hermitian. Decoherence free evolution can further be constructed when the commutator of the Lindbladian and the superoperator, $[\cL,\cal{I}]$, is independent of the noise parameters \cite{Sarandy2007}.

Reservoir engineering to shortcut the thermalization process has also been discussed in the context of Non Adiabatic Markov Equations \cite{Dann2018_052129,Dann2018_08821}. Such an approach, well-suited for fast driving within the Markovian approximation, properly accounts for the coupling between population and coherence, and explains the emergence of coherence associated with dissipation.
Systems coupled to a non-Markovian bath have also been investigated. In \textcite{Villazon2019}, the driving of such system is detailed using a protocol that controls 
in time both the system parameters and the coupling strength to the bath. This protocol has been further exploited in an Otto-like engine operating at high power. 

\subsection{Mitigating the effect of environment}

Quite often, the coupling with the environment and the environment itself cannot be designed. 
For instance, the environment may induce noise 
in some parameters \cite{Ruschhaupt2012,Kiely2017}. In such a scenario STA techniques may be designed to 
mitigate its effects. 
In \textcite{Sun2016}, different variants of CD driving for a finite-time Landau-Zener process are investigated  in the presence of a bath. Other strategies exploit the freedom on the phases $\xi_n(t)$ for the 
evolution operator $U(t)=\sum_n e^{i\xi_n(t)}|n(t)\rangle \langle n(0)|$ to minimize the effect of the coupling with the environment \cite{Santos2018_025301} in CD driving.  In the same spirit, STA based on invariants can be readily adapted along the lines set in \textcite{Ruschhaupt2012}, see Sec \ref{ssec:robustness}.
In \textcite{Levy2017}, the effect of noise is minimized by reducing at best the commutators between invariants and the Lindblad noise operators, $X_{k}$, specifically 
by minimizing the quantities $\mathcal{A}_{k} \sim\int_{0}^{t_{f}}ds  \left\lVert\left[X_{k}(s),I(s)\right] \right\rVert$ over the process duration. Designing an invariant $I$ that commutes with the noise operators $X_{k}$ ensures that populations do not decay in the invariant eigenbasis and that decay of coherences is reduced. It is not always possible to minimize all $\mathcal{A}_{k}$ simultaneously, so a weighted average is minimized instead.
Similarly, starting from an inverse engineering protocol, the parameters can be shaped in time to enforce the robustness against stochastic fluctuations in the Hamiltonian for a wide class of noise types \cite{Jing2013}.

\subsection{Non-Hermitian Hamiltonians\label{nhh}}
There is a  class of quantum systems for which the environment can be modeled using  
Non-Hermitian Hamiltonians. Such Hamiltonians usually describe subsystems
of a larger system \cite{Muga2004}. Complex energies imply that ``adiabaticity theorems'' only apply to weak non-hermiticity regimes or to the least dissipative state \cite{Nenciu1992}. Moreover, since    
right and left eigenvectors are normalized in a biorthogonal sense, the normalization factors are ambiguous 
and care must be exercised to extend  the concept of ``population'' to define an adiabaticity criterion \cite{Ibanez2014}.  

Concerning the standard STA techniques, the counterdiabatic driving has been  generalized to weak non-Hermitian Hamiltonians \cite{Ibanez2011_023415,Ibanez2012_019901,Song2016_21674,Chen2016_052109,Li2017_14,Li2017_30135,Chen2018_1700247}. \textcite{Ibanez2011_023415} in particular applied the formalism to control a decaying two-level system.
Later \textcite {Torosov2013} demonstrated  that  auxiliary gain and loss imaginary terms added in the diagonal of the Hamiltonian  of (Hermitian) two-level models, which are feasible in waveguide optics, 
can be chosen to cancel nonadiabatic transitions and  perform fast population transfers. 
The results were  also generalized  to three-level systems \cite{Torosov2014,Li2017_14,Wu2016_22847} and applied to a two-level system coupled to a dissipative 
spin-chain \cite{Diffo2017}.  
In \textcite{Impens2018}, the extra driving field one has to superimpose to the original one  to compensate for the distortion in the spin-$1/2$ {direction} on the Bloch sphere due to a dissipative  non-Hermitian term, is explicitly worked out. 

As for Lewis-Riesenfeld invariants for Non-Hermitian Hamiltonians, they can be generalized in two different forms 
\cite{Simon2018} corresponding to \cite{Gao1992, Khantoul2017,Maamache2017}
\beqa
&&\frac{\partial I}{\partial t}+\frac{i}{\hbar}\left[H,I\right]=0,
\nonumber\\
&&\frac{d}{dt}\la\widehat{\psi}(t)|I(t)|\psi(t)\ra=0, 
\eeqa
where $\widehat{\psi}(t)$ evolves with $H^\dagger(t)$, 
or to
\beqa
&&\frac{\partial I'}{\partial t}+\frac{i}{\hbar}\left[H^\dagger(t) I'(t)-I'(t)H(t)\right]=0,
\nonumber\\
&&\frac{d}{dt}\la {\psi}(t)|I'(t)|\psi(t)\ra=0. 
\eeqa
Combined with inverse engineering,  the first option has been considered e.g. in  \textcite{Ibanez2011_023415,Luo2015}, 
and the potential of the second option is yet to be explored.

\begin{figure}[b]
\begin{center}
\includegraphics[height=5.4cm,angle=0]{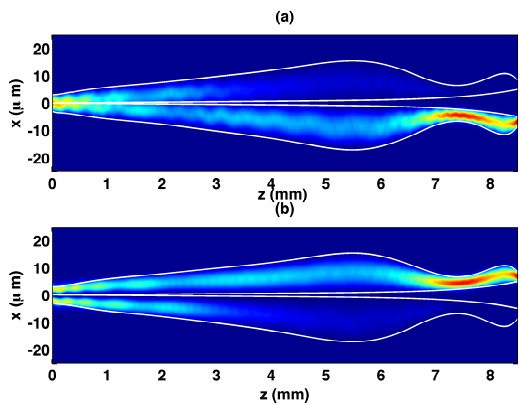}
\end{center}
\caption{\label{Yjunction}
(Color online) Mode-sorting operation of the invariant based Y-junction. Input (a) fundamental mode (b) second mode. From 
\textcite{Martinez-Garaot2014_2306}. Copyright : Optical Society of America.}
\end{figure}
\section{Optical devices\label{od}}
There are some proposals to make use of time-dependent parameters to control 
light waves propagating in linear media via STA approaches  \cite{Lakehal2016}, but the bulk of STA applications in optics substitutes time by a spatial coordinate. 
Thus shortcuts to adiabaticity in optics can lead to more compact waveguide devices, which favors device integration. After numerous theoretical works, the stage where actual STA-enhanced devices are constructed has just been reached.  

STA applications in optics began exploiting the 
analogies \cite{Longhi2009} between the electromagnetic wave propagation inside a waveguide in the paraxial approximation 
and the propagation of a quantum wavefunction inside a well. 
Consider a stationary scalar field ${E}(\textbf{r})$ that satisfies the Helmholtz equation, a time-independent form of the wave equation,
\beq
\label{HE}
\nabla^2 {E}(\textbf{r})+n^2(\textbf{r})k^2 {E}(\textbf{r})=0,
\eeq
where $n(\textbf{r})$ is the position-dependent refractive index inside the waveguide and $k=2\pi/\lambda$, with $\lambda$ the light wavelength in vacuum. 
Two common approximations are the small angle or paraxial approximation, 
which assumes the form   
\beq
\label{PA}
{E}(\textbf{r}) = \mathcal{E} (\textbf{r}) e^{i k_z z}, 
\eeq
and a slowly varying envelope approximation (SVEA) in the scale 
of $2\pi/k_z$ to neglect  $\partial^2\mathcal{E}/\partial z^2$.  
$k_z=k n_0$ plays the role of an ``optical mass'' and $n_0$ is the outer (bulk) refractive index. 
These approximations lead to a simplified form of Eq. (\ref{HE}) with the same form as the time-dependent Schr\"odinger equation, and the longitudinal space $z$ coordinate playing the role of time, 
\beqa
\label{AHE2}
i \frac{\partial \mathcal{E}}{\partial z}&=&\left [-\frac{1}{2kn_0}\nabla_{\bot}^2 +V(\textbf{r}) \right ]\mathcal{E},
\\
V(\textbf{r})&=&\frac{k}{2n_0}\left [ n_0^2-n^2(\textbf{r})\right].
\eeqa

{\it Coupled-mode theory.} Beam dynamics in  coupled waveguides is usually addressed with 
``coupled-mode theory''. 
Among the different formulations
we shall 
focus here on the most used one in STA applications.  
Assume  $N$-coupled waveguides where the refractive index and geometry of the waveguides are allowed to vary along the propagation direction $z$. 
Under the scalar and paraxial approximations and the assumption of weak coupling, the variations of the guided-mode amplitudes of individual waveguides, $\textbf{A}=[a_1,a_2,\dotsc,a_N]^T$, with propagation distance are described by the coupled-mode equations 
\beq
\label{S_cm}
i \frac{d\textbf{A}}{dz}=H(z)\textbf{A}. 
\eeq
Once again, replacing the spatial variation $z$ with the temporal variation $t$, the above equation is equivalent to the time-dependent Schr\"odinger equation ($\hbar \equiv 1$) describing the dynamics of an $N$-state system, with $H(z)$ playing the role of the Hamiltonian.  

\textcite{Lin2012} put forward the use of STA protocols to stabilize and reduce the size of coupled waveguide systems, applying the CD driving approach to the mode conversion in a multimode waveguide with three coupled modes. 
The results using coupled-mode theory were checked using the so called wide-angle beam propagation numerical method, which solves the Helmholtz equation with  SVEA, but no
paraxial approximation. 

Numerous works followed this line. Many of the methods presented in Sec. \ref{methods} were used to improve mode converters \cite{Tseng2012,Tseng2013,Yeih2014} and also to design and optimize other devices such as mode-division multiplexing systems \cite{Martinez-Garaot2014_2306}, see Fig. \ref{Yjunction}, directional couplers \cite{Tseng2014_18849,Tseng2014_6600}, or polarization rotators \cite{Chen2014_24169}.

\textcite{Stefanatos2014_023811} studied how to implement an STA in an optical semi-infinite system. 
A semi-infinite photonic lattice was designed to drive the input light to a controlled location at the output using the invariant-based inverse engineering approach, interpreting the photonic lattice as a quantum harmonic oscillator with time-dependent mass. 
One year later, \textcite{Pan2015} applied the theory in \textcite{Tseng2014_18849} to design silicon-based platforms with high refractive index contrast;    
\textcite{Paul2015} used the CD approach to improve the design of a directional coupler;  and \textcite{Ho2015} optimized the adiabaticity of coupled-waveguides devices using invariant-based inverse engineering. 

\textcite{DellaValle2016} developed an STA for ultra-compact waveguide junctions inspired by the streamlined version of the fast-forward approach \cite{Torrontegui2012_013601}. In this way they  went beyond the coupled-mode equation formalism by extending the optical STA to full-wave problems for the Helmholtz equation, i.e., to an infinite-dimensional system.

In \textcite{Chen2016_18322} the invariants  were used to design stable directional couplers against errors in input wavelength and coupling coefficient simultaneously. Later \textcite{Martinez-Garaot2017} adapted the FAQUAD approach to the optical devices to design mode-sorting asymmetric Y-junctions. This approach is quite useful in this context because the information it requires is 
accessible, and the simplicity of changing only one control parameter is ideal for device fabrication.    
Besides, \textcite{Chung2017} designed a short and broadband silicon asymmetric Y-junction two-mode (de)multiplexer using the theory in \textcite{Martinez-Garaot2017}, \textcite{Wu2017_21084} put forward the generation of  3D entanglement using dressed states, and \textcite{Huang2017} applied the counterdiabatic approach to design shorter and robust two- and three- waveguide couplers. Finally,  \textcite{Chen2018_045804} used the Lie-transform theory in \textcite{Martinez-Garaot2014_053408} to design compact beam splitters. 

In some recent works dealing with silicon waveguides with high index contrast, the scalar and paraxial approach is not accurate, so commercial software is used  to solve the dynamics. Specifically, in \textcite{Martinez-Garaot2017} the shortcut was done using the scalar Helmholtz equation without the paraxial approximation and in \textcite{Chung2017,Chung2018} the FAQUAD approach was applied without approximations,  
since it can incorporate the vectorial fields into the calculations without resorting to the coupled-mode equations.  
Moreover, in \textcite{DellaValle2018} shortcuts beyond the paraxial limit were used to achieve efficient rejection of higher order modes in a broad wavelength range for any two-dimensional multimode optical waveguide. 
Non-Hermitian systems with gain and loss were also  considered \cite{Torosov2013,Torosov2014,Longhi2017}.

The first experimental implementation of STA in optical devices was worked out  by \textcite{Guo2017_9160} demonstrating broadband silicon mode (de)multiplexers with optimized tapers using the method in \textcite{Ho2015}, and 
recently,  the group of S.-Y. Tseng  produced devices using the FAQUAD approach \cite{Hung2019}. 
\section{Extension to classical and statistical physics\label{sec:cms}}
In this section, we review the different STA methods which have been extended to classical mechanics and statistical physics, along with the presentation of a few proof-of-principle experiments that have been carried out to demonstrate their effectiveness.

\subsection{Counterdiabatic methods in classical mechanics} \label{counterclas}
As already discussed in Sec.~\ref{ssec:cd}, the counterdiabatic term that we add to a time-dependent Hamiltonian in quantum mechanics inhibits any excitation in the system.
A natural question is how we can transpose these ideas to a classical system, and to what extent they are related to their quantum counterpart. 

We shall first address this question in the context of one-body classical mechanics. The essence of the method can be readily explained with a 1D integrable system described by a Hamiltonian $H_0(p,x,\lambda)$, where $\lambda$ can be a multicomponent vector (for simplicity, we use a single-component). Using a canonical transformation, there exist angle-action coordinates $(\theta,I)$ such that  $H_0(p,x,\lambda)=\tilde H_0(I,\lambda)$, i.e. the Hamiltonian expressed in terms of the action-angle variables is independent of the angle $\theta$ as a result of its integrability. Assuming that the parameter is now time-dependent, the new Hamiltonian resulting from the canonical transformation in action-angle variables takes the form \cite{Goldstein}
\begin{equation}
H(I,\theta,t)=\tilde H_0(I,\lambda)+\partial_t F=\tilde H_0(I,\lambda)+\dot \lambda\partial_\lambda F,
\label{eqmech1}
\end{equation}
where $F$ is a time-dependent generator whose ``type'' depends on its variables \cite{Goldstein,Deng2013,Deffner2014,Kolodrubetz2017}. For instance, $F=F_1(x,\theta,t)$ for type-I. From Eq.~(\ref{eqmech1}), it is clear that the counterdiabatic Hamiltonian that one may  superimpose to keep $I$ constant with fast 
changes in $\lambda$  is $H_{CD}=-\partial_t F$ \cite{Deng2013}.  In the adiabatic limit ($\dot \lambda \to 0$), the Hamiltonian $H$ boils down to $H_0$.
The counterdiabatic Hamiltonian, $H_{CD}$, added to the original Hamiltonian ensures that the volume of the phase space enclosed by a given energy shell remains constant despite the variation in time of the parameters \cite{Jarzynski2013}.

We now present some illustrative examples. Consider a 1D harmonic oscillator whose angular frequency, $\omega(t)$, is time-dependent: 
\begin{equation}
H_0(p,x,t)=\frac{p^2}{2m}+\frac{1}{2}m\omega^2(t)x^2.
\end{equation}
For such a Hamiltonian, the adiabatic criterion is
\begin{equation}
\frac{d\omega}{dt} \ll \omega^2.
\label{adcrit}
\end{equation}
Using the type-I generating function, the action-angle canonical transformation yields $p=\partial_xF_1$, $I=-\partial_\theta F_1$ and $H(I,\theta,t)=\tilde H_0(I,\lambda)+\dot \omega \partial_\omega F_1$  \cite{Goldstein}.
To get an expression of $H_0$ independent of $\theta$, a natural choice is to search $x$ and $p$ in the form $p=f(I)\cos(\theta)$ and $x=[f(I) \sin \theta]/m\omega$ where the function $f(I)$ needs to be determined. We readily find $f(I)=(2Im\omega)^{1/2}$, $F_1(x,\theta,t)=(m\omega x^2 {\rm cot}\theta)/2=px/2$ and $H_0(I,t)=I\omega$ and  deduce  $H_{CD}(p,x,t)=-\dot \omega F_1/\omega=-\dot \omega px/(2\omega)$. This result 
coincides perfectly with the classical limit of the quantum results (see Sec.~\ref{examplesIBIE}) \cite{Muga2010}. The calculation performed here on a harmonic potential can be readily generalized to even-power-law potentials and to a particle in a one-dimensional box \cite{Jarzynski2013}.

We now consider the example of the transport of a particle by moving a harmonic trap\footnote{The same formalism still holds for an arbitrary transport potential.}:
\begin{equation}
H_0(p,x,t)=\frac{p^2}{2m}+\frac{1}{2}m\omega^2[x-x_0(t)]^2.
\end{equation}
We can again work out the canonical transformation to recast the problem in terms of action-angle variables by replacing $x$ by $x-x_0(t)$ in the previous calculation. The transformed Hamiltonian now reads $H=\tilde H_0(I,\lambda)+\dot x_0 \partial_{x_0} F_1=\tilde H_0(I,\lambda)-p\dot x_0$. The counterdiabatic Hamiltonian is therefore $H_{CD}(p,x,t)=p\dot x_0$ \cite{Sels2017}, as in the quantum result \cite{Torrontegui2011}. The quantum unitary transformation \cite{Ibanez2012_100403}
to find an alternative local interaction corresponds classically to an additional canonical or gauge transformation. This procedure is referred to as the local counterdiabatic driving  \cite{Deffner2014,Ibanez2012_100403,Sels2017}.

For a 1D Hamiltonian, a gauge transformation involves only a scalar potential $\chi(x,t)$. The momentum $p$ is transformed as $\tilde p = p+\partial_x\chi$ for the Hamiltonian $\tilde H = H+\partial_t\chi$. For our first example, the gauge function $\chi(x,t)=\dot \omega mx^2/(4 \omega)$, yields a gauged transformed Hamiltonian $\tilde H$ which is that of a harmonic oscillator of effective time-dependent angular frequency $\omega_{\rm eff}=[\omega^2 + \ddot \omega/(2\omega)-3\dot \omega^2/(4 \omega^2)]^{1/2}$, which coincides with its quantum counterpart, 
see Eq. (\ref{omegap}) \cite{Ibanez2012_100403}. For the second example,  the gauge function $\chi(x,t)=-m\dot x_0 x$, yields $\tilde H(p,x,t)=p^2/(2m)+(1/2)m\omega^2x^2-m\ddot x_0 x$. The extra force $F=m\ddot x_0$ superimposed to the static Hamiltonian plays the role of an effective gravitational field whose amplitude is proportional to the acceleration $\ddot x_0$, and compensates for the inertial force in the frame attached to the potential.

Beyond the two specific examples detailed previously, the formalism can be  generalized to scale-invariant systems, i.e. Hamiltonians of the form \cite{Deffner2014}
\begin{equation}
H_0(p,x,t)=\frac{p^2}{2m}+\frac{1}{\gamma^2(t)}U\left( \frac{x-x_0(t)}{\gamma(t)}\right),
\end{equation}
where $\gamma(t)$ and $x_0(t)$ are real functions depending on time. Shortcuts  are not restricted to scale invariant systems as explicitly shown in \textcite{Patra2017_3403}  on a simple example. The counterdiabatic Hamiltonian is also not unique. For instance, we have worked out its form linear in $p$ for transport, but other solutions, e.g.  cubic in $p$, are also possible; they are related to the dispersionless Korteweg-de Vries hierarchy (see Sec.~\ref{sec:LaxPairs} and \textcite{Okuyama2016,Okuyama2017}). 

Alternatively, one can construct a local dynamical invariant, or equivalently an extra potential to be added to the original Hamiltonian, to preserve the classical action for a fast time variation of the Hamiltonian parameters. Such an approach is reminiscent of the fast-forward method and can be solved implicitly as detailed in \textcite{Jarzynski2017}.

Finally, the counterdiabatic approach  has been also applied to classical spin dynamics \cite{Hatomura2018_00229}. 
The construction makes use of the CD term for each single spin and is much easier to implement than in the corresponding quantum system. 
Moreover it   
does not need knowledge of the instantaneous stationary states. Starting from 
a stationary state of the initial Hamiltonian, it results in a stationary state of the final Hamiltonian if there is no criticality. The method can be used to solve combinatorial optimization problems.

\subsection{Mechanical Engineering}
We discuss  hereafter the interest of STA techniques to control a crane, then  design  robust solutions, and conclude with  the link between STA and flatness based control in mathematics \cite{Fliess1995}.

\subsubsection{Cranes\label{cranes}}

The objective of mechanical cranes is to move loads fast avoiding final pendulations and large 
sway angles on route that could compromise safety.  Since a slow,
adiabatic operation avoids excitations but it takes an impractical long time, 
cranes are a natural domain for STA \cite{Gonzalez-Resines2017}. 
Crane control is an important engineering field,  
see \textcite{Sun2012,Kuo2014}  and references therein. Methods 
and ideas abound and many can be translated or adapted to other STA-driven processes in very different systems.  
For example, the work on closed-loop methods (in which  measurements are performed on route to determine 
control operations) is an inspiring source to develop feedback-based STA in the microscopic realm.     
In reverse, existing STA methods may have quite an impact on crane operation routines. 
Overhead cranes usually operate under a small-oscillations regime so that  
simple operations such as horizontal transport and hoisting/lowering of the load are 
modeled by the same basic (mass independent) equations that apply to 
the transport or compression/expansion of an ion in a time-dependent harmonic trap. 
\textcite{Gonzalez-Resines2017} provide invariant-based STA protocols  for the motion of  the trolley in a transport operation or for hoisting that guarantee final adiabatic energies for the load. Furthermore these energies are shown to be minimal when averaging over a microcanonical ensemble of initial conditions,  consistently with the minimal work principle \cite{Allahverdyan2005,Allahverdyan2007}.  
Indeed the possibility to design robust operations with respect to different perturbations or errors (such as dispersion in the initial conditions, 
or in cable lengths) and  STA\&OCT combinations to limit, for example, on-route pendulations, offer a great potential.  The different techniques to enhance robustness, 
such as the Fourier method (\cite{Guery-Odelin2014_063425}, see the next subsection) may be used to design 
trolley trajectories which are robust with respect to different errors, e.g. in the cable length. 
The bridge to perform analogous inverse engineering in quantum microscopic and classical macroscopic systems is much facilitated by the fact that 
the Lewis-Leach family of potentials implies the same classical \cite{Lewis1982} and quantum formulations \cite{Dhara1984} for the invariants and auxiliary equations.

A crane model that treats the trolley position as a dynamical variable,  instead of as a control function,
subjected to inertia, the engine pulling force ${\cal F}_a$, dissipation, and the backaction of the load, is a neat, explicit testbed to study energy consumptions and  
the implications of shortcuts to adiabaticity  from the point of view of the necessary external controls \cite{Torrontegui2017}.  
In particular the power produced by the engine force is  due to the change of mechanical energy ${\cal H}_0$  of the whole system (load and trolley) 
plus the power needed to compensate the effect of friction, ${\cal P}={\cal F}_a\dot{x}=\frac{d {\cal H}_0}{dt}+\gamma \dot{x}^2$, where $x$ is the trolley's position and 
$\gamma$ the friction coefficient. In the harmonic approximation this becomes
\beq
{\cal P}=(M \ddot{x}-mq\omega^2+\gamma\dot{x})\dot{x},
\eeq
where $M$ is the trolley's mass and $q$ the load horizontal displacement with respect to the  trolley, see Fig. \ref{grua}. The second term is exactly the 
power $P$ defined as the derivative of the mechanical energy of the load.      
${\cal P}=P$ only under rather extreme, and even undesirable conditions.

First of all note that the shortcuts are by definition fast processes, so that 
the friction can hardly be avoided by the trick of slowing down the dynamics applied for ideal reversible operations. 
Only the typically unrealistic $\gamma=0$ scenario would cancel the dissipation term. As for the $M$-dependent term,    
the limit $M=0$ is again rather unrealistic, and in fact does not simplify matters, because  the action of the engine would have to 
depend strongly on the initial conditions of the load; contrast this to ideal state-independent STA operations that require instead a large $M/m$ ratio.  
Another interesting aspect of the model is the analysis of possible  negative values of ${\cal P}$ corresponding to braking. 
Different  scenarios (implying energy consumption or rather partial regenerative braking) are depicted in \textcite{Torrontegui2017} and treated phenomenologically to examine the total energy consumption.

We have focused on cranes but clearly other mechanical machines and robots with moving parts 
can  benefit from shortcuts to adiabaticity  
\cite{Stefanatos2018_17}, as well as other areas of engineering.  
For example \textcite{Faure2018} have introduced inverse engineering to drive an RC circuit.

\subsubsection{Robustness issues\label{Fourier}}

In  Sec.~\ref{ria}, we have shown how the equation of motion of a particle in a moving harmonic potential of angular frequency $\omega_0$ can be reversed. We propose to extend this technique using a Fourier method \cite{Guery-Odelin2014_063425}. From the Newton equation, one can directly relate the excess energy after the transport $\Delta E (t_f)=m|{\cal F}(\omega_0;\{x_0(t)\})|^2$, to the motion of the trap $x_0(t)$ with 
\begin{equation} 
{\cal F}(\omega;\{x_0(t)\})=\int_0^{t_f}\ddot x_0(t')e^{-i\omega t'}dt'.
\end{equation}
An optimal trajectory $\{ \tilde x_0(t) \}$ for the transport shall therefore fulfill the relation ${\cal F}(\omega_0;\{\tilde x_0(t)\})=0$. A systematic way to generate such trajectories is to  define the acceleration $\ddot x_0(t)$ through an auxiliary time-dependent function $g(t)$ as  $\ddot x_0(t)=\ddot g(t) + \omega_0^2 g(t)$, where $g(t)$ obeys the boundary conditions $g(0)=g(t_f)=\dot g(0)=\dot g(t_f)=0$
and the relations
\begin{equation}
\int_0^{t_f}g(t) dt=0\;\; {\rm and}\;\;\int_0^{t_f}dt' \int_0^{t'}g(t'') dt''=\frac{d}{\omega_0^2}.
\end{equation}
These conditions ensure that $x_0(0)=0$, $\dot x_0(0)=0$, $x_0(t_f)=d$ and $\dot x_0(t_f)=0$. We then find 
\begin{equation} 
{\cal F}(\omega;\{x_0(t)\})=(\omega_0^2-\omega^2)\int_0^{t_f}g(t')e^{-i\omega t'}dt',
\end{equation} 
which vanishes for $\omega=\omega_0$, as expected for an optimal transport. Interestingly, using a fourth order differential equation to relate $\ddot x_0$ and $g(t)$, and with appropriate boundary conditions for $g(t)$, one can factorize a polynomial in $\omega^2$ in front of the Fourier transform of $g(t)$ of the form $(\omega_1^2-\omega^2)(\omega_2^2-\omega^2)$ \cite{Guery-Odelin2014_063425}. With such a solution, the same trajectory of the trap would be optimal for two different angular frequencies, $\omega_1$ and $\omega_2$. This would apply  for instance for two different atoms transported by the same moving optical tweezers \cite{Couvert2008}. Alternatively, the protocol repeated for higher order polynomial with the same root $\omega_0$ provides a generic method to enforce robustness against the exact value of the angular frequency experienced by the atoms \cite{Guery-Odelin2014_063425}. 
For an application in ion transport see \textcite{An2016}.\\ 

{\it Flatness based control theory.}
From a more mathematical point of view, the previous solution can also be recovered using flatness based control theory commonly used for steering a system from one state to another \cite{Rouchon2005}. This formalism applies to differential systems of the form 
\begin{equation}
dx/dt=f(x,u),
\label{eqflat}
\end{equation}
where the vector $u$ contains the control variables. The control problem is readily solved when there exists a so-called flat-output map $h$, $y(t)=h\left( x, u, \dot u, ..., u^{(\alpha)}\right)$, such that $x=h_1 \left( y, \dot y, ... , y^{(\beta)}   \right)$ and $u=h_2 \left( y, \dot y, ... , y^{(\beta+1)}   \right)$, where $\alpha$ and $\beta$ are some finite numbers, and $h_1$ and $h_2$ some smooth functions. For a given system described by Eqs.~(\ref{eqflat}) there is no algorithm to determine if a flat-output map $h$ exists. However, many examples of engineering interest turn out to be flat including  transport \cite{Rouchon2005} and cranes \cite{Fliess1995}.

\subsection{STA for isolated dilute gases}
So far we have essentially considered one-body problems. 
One may wonder to what extent the results can be generalized to an assembly of interacting atoms. We shall first consider a dilute gas trapped by a 3D isotropic harmonic potential of angular frequency $\omega_0$. The notion of adiabaticity shall be revisited in this new context. Indeed, the gas has a relaxation time $\tau$ that is related to the collision rate $\gamma_c=n\sigma \bar v$ where $n$ is the mean atomic density, $\sigma$ the total cross section and $\bar v \propto (k_BT/m)^{1/2}$ the mean thermal velocity for a temperature $T$. The relation between $\tau$ and $\gamma_c$ depends on the collision regime: for $\gamma_c \ll \omega_0$, i.e. when there are few collisions per oscillation period, $\tau \propto \gamma_c^{-1}$, while $\gamma_c \gg \omega_0$ in the hydrodynamic limit, $\tau \propto \gamma_c/\omega_0^2$. The thermodynamical criterion for adiabaticity associated with a slow change of the angular frequency $\omega(t)$ now reads
\begin{equation}  
\frac{d\omega}{dt} \ll \frac{\omega}{\tau}.
\end{equation}
Once this criterion is fulfilled, the quantity $T(t)/\omega(t)$ remains constant. The physical interpretation of this conserved quantity is clear, it ensures that the populations $\pi_n \propto \exp(-\hbar n \omega/ k_BT )$ of the eigenstates are conserved during a slow change of confining strength. Otherwise stated, the transformation corresponds to a work and is not accompanied by heat (modifications of the populations).

\subsubsection{Boltzmann equation}
The search for a shortcut on such transformations requires a rigorous mathematical modeling of the out-of-equilibrium dynamics. Under the approximation of diluteness, the evolution of the phase space distribution, $f({\bf{r}}\,,{\bf{v}}\,,t)$,  of the gas  is well-described by the Boltzmann equation where collisions are accounted for through the two-body collisional integral. In \textcite{Guery-Odelin2014_180602}, an exact solution of this equation for a time-dependent angular frequency $\omega(t)$ was worked out, 
\begin{equation}
f({\bf{r}}\,,{\bf{v}}\,,t)=\frac{(\alpha\beta-\gamma^2/4)^{3/2}}{\pi^3}e^{-\alpha r^2 - \beta v^2 - \gamma {\bf{r}} \cdot {\bf{v}}},
\end{equation} 
where $\alpha(t)$, $\beta(t)$ and $\gamma(t)$ are functions that depend only on time. They are related through a linear set of differential equations
which can be recast as a single third order differential equation on the quantity $\beta$ which plays the role of the inverse of an effective temperature, 
\begin{equation}
\dddot \beta + 4 \omega^2 \dot \beta + 4 \omega \dot \omega \beta =0.
\label{eqbeta}
\end{equation}
For a slow transformation, the third order derivative can be neglected and the quantity $T(t)/\omega(t) \propto (\beta(t)\omega(t))^{-1}$ is conserved. Fast transformations can be  designed by inverse engineering, i.e., by fixing the boundary conditions on $\beta$ and its derivatives, interpolating the $\beta$ function accordingly,  and inferring $\omega(t)$ from the equation obeyed by $\beta$. For very fast decompression, we find as in quantum mechanics intervals of time over which the sign of the curvature of the potential is reversed. So far, we have kept the trap isotropic. The extension of STA protocols to anisotropic 2D Bose gas and 3D unitary Fermi gases including in the presence of topological defects such as soliton or vortices is discussed in \textcite{Papoular2015}.
\subsubsection{Extension to Navier-Stokes equation}
The solution outlined above hints at a related solution for hydrodynamics. Indeed, as originally demonstrated by Chapman and Enskog \cite{Chapman}, the hydrodynamic equations can be derived from the Boltzmann equation. These equations relate the velocity field ${\bf v}({\bf r},t)$, the temperature field, $T({\bf r},t)$ and the density $n({\bf r},t)$, 
\begin{eqnarray}
  \frac{\partial n}{\partial t}+\mbox{\boldmath $\nabla$}\cdot(n {\bf v}) & = & 0,
 \nonumber \\
 mn\left(\frac{\partial}{\partial t}+{\bf v}\cdot\mbox{\boldmath $\nabla$}\right){\bf
u} & = &-n\mbox{\boldmath $\nabla$}U-\mbox{\boldmath $\nabla$}\bigg(P-\frac{\eta}{3}
(\mbox{\boldmath $\nabla$}\cdot{\bf v})\bigg)
\nonumber\\
&+&\frac{\eta}{\rho}\nabla^2{\bf v},
\nonumber\\
\left(\frac{\partial}{\partial t}+{\bf v}\cdot\mbox{\boldmath
$\nabla$}\right)T & = & -\frac{T}{c_V}(\mbox{\boldmath $\nabla$}\cdot{\bf v})+
\frac{\kappa\nabla^2T}{mn c_V},
\end{eqnarray}
where $\eta$ is the viscosity, $c_V$ the specific heat and $\kappa$ the thermal conductivity. The exact solution for a 3D isotropic harmonic trap with time-dependent angular frequency $\omega(t)$ is found through the search of an exact scaling solution of the form $n({\bf r},t)  =  b^{-3}n_0\left( {\bf r}/b \right)$, ${\bf v}({\bf r},t)  = \dot b{\bf r}/b$, and $T({\bf r},t) = \beta^{-1}(t)$. Remarkably, we find that the inverse of the effective temperature parameter $\beta$ and $\omega(t)$ are connected once again by Eq. (\ref{eqbeta}). The same STA strategy can therefore be  applied in this context.

\subsection{Shortcuts for systems in contact with a thermostat\label{thermostat}}
So far, we have considered only isolated systems. In this section, we answer the question of how a classical system in contact with a thermal bath can benefit from an accelerated equilibration protocol. Indeed, the control in the presence of a thermostat is of general interest with applications ranging from nano-oscillators, nanothermal engines, to the driving of mesoscopic chemical or biological processes.

The one-body equation to be considered is therefore a stochastic differential equation, the Langevin equation, whose noise is related to the temperature of the bath. In its most general form, it is given by
\begin{equation}
m\ddot x = -\partial_xU(x,t)-m\gamma \dot x+\xi(t),
\label{Langevin}
\end{equation}
where we will consider the noise $\xi(t)$ as a white noise delta correlated in time, $\langle \xi(t) \xi(t')\rangle=2m\gamma k_BT \delta(t-t')$. The inverse engineering method cannot be readily applied directly to a stochastic equation. Instead, the equation for the probability density $\rho(x,t)$ function associated with such a Brownian motion is used. 

\subsubsection{The overdamped regime}

First, we consider the overdamped regime for which inertial effects become negligible. This amounts to vanishing the mass in Eq.~(\ref{Langevin}). The probability density then obeys the Fokker-Planck equation
\begin{equation}
\partial_t \rho(x,t)=\gamma^{-1}\partial_x[\rho(x,t)U(x,t)]+D\partial^2_{xx}\rho,
\label{fpeq}
\end{equation}
with the diffusion constant $D=k_BT/m\gamma$. Two standard STA methods can be applied to accelerate the equilibration: the transposition of the counterdiabatic ideas and the inverse engineering. 

For the first method, we consider an equilibrium solution of Eq.~(\ref{fpeq}), $\rho_0(x,\{ \lambda \})$ that corresponds to the potential energy $U_0(x,\{ \lambda \})$, where $\{ \lambda \}$ refers to the control parameters of the potential. $U_1$ is the extra potential that we have to add to the original potential $U_0$ to compensate for the time variation of the parameters $\{ \lambda \}$. This potential is the solution of the equation
\begin{equation}
\partial_t \rho_0(x,\{ \lambda(t) \})=\gamma^{-1}\partial_x [ \rho_0(x,\{ \lambda(t) \})U_1(x,t)].
\end{equation}
For instance, with a potential $U_0(x,t)=m\omega^2(t)(x-x_0(t))^2/2$, we find
\begin{equation}
U_1(x,t)=-m\gamma \dot x_0(x-x_0(t))+ \frac{m\gamma \dot \omega (x-x_0(t))^2}{2\omega}.
\end{equation}
The potential $U_1(x,t)$ is nothing but the classical counterdiabatic Hamiltonian determined in Sec.\ref{counterclas} \cite{Li2017_012144}.

The second strategy is to  apply inverse engineering to the Fokker-Planck equation.  For the compression/decompression of a harmonic oscillator from an angular frequency $\omega_i$ to $\omega_f$, we use an exact scaling solution of the overdamped Fokker-Planck equation \cite{Martinez2016},
\begin{equation}
\rho(x,t)=\sqrt{\frac{\alpha(t)}{\pi}} \exp[- \alpha(t) x^2]
\end{equation}
with
\begin{equation}
\frac{\dot \alpha}{\alpha}=\frac{2\omega^2(t)}{\gamma} -\frac{4k_BT \alpha}{\gamma}.
\label{eqalpha}
\end{equation}
For our purpose, we impose boundary conditions on the $\alpha$ parameter ($\alpha(0)=m\omega_i^2/2k_BT$, $\alpha(t_f)=m\omega_f^2/2k_BT$, $\dot \alpha (0)=0$ and $\dot \alpha(t_f)=0$), interpolate the $\alpha(t)$ functions, and infer from Eq.~(\ref{eqalpha}) how one shall shape $\omega(t)$. This solution has been successfully implemented experimentally for a compression using a 1 $\mu$m size microsphere trapped by an optical tweezer and immersed in a thermalized fluid chamber  \cite{Martinez2016}. The strength of the confinement was simply increased by designing in time the intensity of the trap beam. In this manner, the system has reached equilibrium 100 times faster than the natural equilibration rate. Alternatively, the minimum time to perform transitions between thermal equilibrium states has also been studied using Optimal Control Theory (see Sec.~\ref{sOCT}) under different constraints on the domain of variation of the time-dependent angular frequency $\omega(t)$ \cite{Stefanatos2017_4290,Plata2019}.

As for the system described by the Boltzmann equation or for the quantum counterpart, the protocol requires for very fast decompression the transient use of a repulsive potential. In practice, this is arduous if not impossible depending on the system. However, and contrary to the quantum case, an extra parameter can be tuned in statistical physics, namely the temperature. Indeed the temperature can be related to noise as explicitly written in the Langevin equation \cite{Martinez2013}. By a proper shaping of the noise, it is therefore possible to accelerate dramatically the decompression keeping the trap attractive as recently demonstrated experimentally \cite{Chupeau2018_010104}.

\subsubsection{Connection with free energy and irreversible work}

A natural question arises: what is the work that can be extracted from a given transformation? In the context of an overdamped dynamics, this question has been theoretically addressed in \textcite{Sekimoto1997_3326}. The mean work $W$ done by the systems reads \cite{Tu2014,Acconcia2015,Martinez2016,Li2017_012144}:
\begin{equation}
W=\Delta F + W_{\rm irr},
\label{work}
\end{equation}
where $\Delta F$ refers to the increment of the Helmholtz free energy associated with the transformation and $W_{\rm irr}>0$ to the irreversible work. In the limit of slow variation of the control parameter the work done boils down to the variation of the Helmholtz free energy which becomes independent of the path used for the transformation. If $t_f$ refers to the interval over which the transformation is performed, one can show that the product $W_{\rm irr} t_f$ is bounded in the limit $t_f \to \infty$. For a compression/decompression, we find $\Delta F=k_BT \log (\omega_f/\omega_i)$ and $W_{\rm irr} =\eta k_BT (\omega^2_f/\omega_i^2)(\tau_{\rm{relax}}/t_f)$ where the numerical factor $\eta$ depends upon the chosen protocol and  $\tau_{\rm{relax}}=\gamma/m\omega^2_f$. 

\subsubsection{Extensions}

The extension of the previous approach to the underdamped regime has been worked out in \textcite{LeCunuder2017,Li2017_012144,Chupeau2018_10512}.
The inverse engineering approach can be readily generalized to manipulate  the phase space distribution $\rho(x,v,t)$ \cite{Chupeau2018_10512}. This technique has been used experimentally to accelerate the equilibration of a micro mechanical oscillator \cite{LeCunuder2017}. However, the transposition of the counterdiabatic method provides an auxiliary potential linear in $p$ that is not relevant from an experimental point of view \cite{Li2017_012144}.

%
%
%
%
%
%
%
\section{Outlook, open questions\label{sec:oq}}
%
%
%
%
%

Since the birth of the term in 2010 till today, ``shortcuts to adiabaticity'' have experienced a phenomenal 
growth.\footnote{ 
A Web-of-Science search including the main keywords gives 7 citations in 2010,  359 in 2014, and 1543 in 2018; 
with  an $h$ index $\sim 40$.}  
We attribute this expansion to a double appeal, both practical and fundamental:     
      
The practical side is rooted in that  adiabatic invariance is ubiquitous as a phenomenon and as a route for state preparation, 
in quantum physics and beyond.
Shortcuts overcome adiabatic protocols,  which imply long times  and the
concomitant accumulation of perturbations from the environment or the control system.  
Moreover a rich network of different pure or hybrid STA approaches provide a flexible toolbox that can be applied and adapted to many systems
and operations.  Only in the quantum arena proposals exist for STA-mediated cooling, interferometers, photon production, enantiomer separation,  quantum gates, or information transfer. Methodological progress in one area can be translated to others allowing for  synergies.
Quantum physics has been indeed the main field to develop STA methodology and applications so far. 
As the control of microscopic systems improves,  we get closer to realize new quantum technologies.    
Yet, decoherence remains a stumbling block to go beyond  proof-of-principle results.  Shortcuts contribute to 
fight decoherence via shorter process times and robust protocols.
%
An ideal quantum device is expected to operate fast and accurately despite a noisy environment or perturbations, and with minimal 
consumption of resources. These are all goals that fit into the agenda and capabilities of STA methods.  

The fundamental side rests on the fact that basic concepts and physical quantities and phenomena such as   robustness, timing, energy and work, information, entropy,  needed resources, controllability,  environment effects, or classical/quantum borders and connections,  all play a role in shortcut design, so shortcuts motivate and contribute to the quest for their interrelations. 
Early work on the time-dependent  harmonic oscillator made clear that the process time and time-averaged  energies 
involved in the STA-driven  process implied nontrivial inequalities  and speed limits that helped to quantify cooling speeds.   
The scrutiny of energy-time relations has grown in different directions, e.g. to analyze energy costs and their scaling with process times. The fast nature of STA processes, makes some conventional estimates based on slow reversible processes invalid. As well, fundamental questions on the  meaning and appropriate definitions  of work, heat, or efficiencies arise.   
The counterdiabatic Hamiltonian bridges the gap between actual and ideal transitionless dynamics and so it 
enters in a number of fundamental inequalities to set  speed limits, see  Sec. \ref{ec}, but also as an aid to reach 
maximal precision limits as discussed in Sec. \ref{metrology}.  Recent  work demonstrates  energy and time relations may be
quite rich for  open systems bringing to the fore further elements such as entropy production, robustness, and/or  information erasure \cite{Takahashi2017_115007,Funo2019_013006,Boyd2019}.   
 

%



In this review we have seen a number of problem- or field-specific challenges. Here we want to underline a few 
open questions we consider to be important along broader conceptual or methodological lines: 
%

- We have discussed examples for which the full spectral information of the original Hamiltonian is not necessary to perform the shortcuts. As well, approximate schemes 
are being developed. Improving approximations and spectral-information independence  are of upmost importance for complex quantum systems such as many-body systems, multiple levels, or for adiabatic computing. The question of the minimum required information for a given
transformation that needs to be accelerated has not been investigated systematically so far. 

- We have discussed  the key role played by dynamical invariants and their link with other STA techniques.  Open questions are (a) to find further families of Hamiltonian-invariant pairs  beyond the Lewis-Leach family,  e.g. via Lax pairs; (b)  in $n$-dimensional systems, to find and implement effective  schemes 
when the  dynamical 
 normal modes need a generalized transformation (involving coordinates and momenta); (c) to explore the use for STA design of different generalizations of ``invariant'' 
 operators for  non-Hermitian systems.
 
- Steering a dynamical quantum system from an initial to a final state in the presence of an environment poses a challenge for quantum control on: controllability,  i.e. to which extent the target state is reachable; and control design,  possibly including reservoir engineering. A framework to draw  the frontiers of controllability is highly desirable for quantum control in open systems  \cite{Glaser2015,Koch2016},  but note that
shortcuts may be applied even if the system is not fully controllable \cite{Petiziol2018}.

- We have  provided examples of hybrid control approaches, for instance  hybridation of shortcuts  and Optimal Control Theory that allows for optimal 
protocol selection. From a control perspective, shortcuts to adiabaticity have contributed to open-loop (with no-feedback) design.  
One could also envision the hybridation of such feed-forward techniques with feedback oriented techniques. Indeed, STA could help to approach the target very fast and the final convergence could be ensured by a feedback procedure. Such a strategy would benefit from the advantages of both techniques: short time processing and strong robustness.

- We need to clarify the energetic and resource cost of STA approaches, and the associated trade off relations. This is of general interest both at fundamental level and for 
specific experiments, for example  to determine the actual performance of STA-enhanced microscopic or mesoscopic engines and refrigerators. 

- STA protocols can a priori be adapted to a large class of other dynamical/differential equations e.g. in engineering, plasma physics, optics, soft condensed matter or biology.

\section*{Acknowledgments}
We thank P. Claeys, S. Deffner, C. Jarzynski, R. Kosloff, E. Sherman, M. Sarandy, D. Sugny, K. Takahashi, E. Trizac, and S.-Y. Tseng   for clarifying comments, or critical reading of the manuscript or sections of it. 

Many colleagues and collaborators, too numerous to mention individually,  have contributed through the last ten years to our work on shortcuts. We are deeply indebted to all of them. 

This work was supported by 
the Basque Country Government (Grant No.
IT986-16); 
PGC2018-101355-B-100 (MCIU/AEI/FEDER, UE);  MINECO/FEDER, UE FIS2015-70856-P;  CAM/FEDER Project No. S2018/TCS-4342 (QUITEMAD-CM); and by Programme Investissements d'Avenir under the Grant ANR-11-IDEX-0002-02, reference ANR-10-LABX-0037-NEXT, as well as the Grant ANR-18-CE30-0013%

\newpage

\begin{table*}[h!]
\caption{\label{tabletransport}List of papers on STA-mediated transport.\\ CD: Counterdiabatic driving;  IE: inverse-engineering;  OCT: Optimal control theory;  FF: Fast-forward; Exp.: Experiment}. 
\begin{ruledtabular}
\begin{tabular}{lll}
Reference&Method&System
\\\hline
\textcite{Couvert2008}&Fourier-based IE (Exp)&Cold atoms in moving optical tweezers
\\
\textcite{Masuda2010}&FF&1 particle in arbitrary trap
\\
\textcite{Torrontegui2011}&Invariants, Compensating-force, Bang-bang&1 particle 
\\
\textcite{Chen2011_043415}&Invariants+OCT&1 particle harmonic transport 
\\
\textcite{Torrontegui2012_013031}& Invariants+OCT &BEC 
\\
\textcite{Sun2012}&Bang-bang (Exp)& Load  in a 2-D planar overhead mechanical crane  
\\
\textcite{Bowler2012}&Fourier transform (Exp)& 1, 2 or 9 ions in Paul trap   
\\ 
\textcite{Walther2012}&Optimized drivings (Exp)& 1 or 2 ions in Paul trap 
\\
\textcite{Ibanez2012_100403}&Unitary transformations &1 particle in harmonic trap 
\\
\textcite{Palmero2013}&Invariants &2 ions  in anharmonic traps 
\\
\textcite{Stefanatos2014_733}&OCT&1 particle 
\\
\textcite{Lu2014_063414}&Invariants+Perturbation theory+OCT & 1 ion 
\\
\textcite{Furst2014}&OCT+Compensating force&1 ion  in Paul trap 
\\
\textcite{Palmero2014}&Invariants&Mixed-species ion chains in Paul trap
\\
\textcite{Guery-Odelin2014_063425}&Fourier method&1 particle or BEC
\\
\textcite{Pedregosa-Gutierrez2015}&Numerical simulations&Large ion clouds
\\
\textcite{Lu2015}&Invariants&2 ions of different mass
\\
\textcite{Zhang2015_043410}&Inverse engineering&1 particle in anharmonic trap
\\
\textcite{Kamsap2015}&Numerical simulations& Large ions clouds 
\\
\textcite{Martinez-Garaot2015_053406}&FAQUAD, Compensating-force& 1 particle
\\
\textcite{Alonso2016}&Bang-bang (Exp)&1 ion 
\\
\textcite{Zhang2016}&Invariants+OCT&Cold atoms
\\
\textcite{Okuyama2016}&Lax pairs+local CD& Soliton-like potentials 
\\
\textcite{An2016}&CD and CD+unitary transformation (Exp)& Simulated transport of 1 ion 
\\
\textcite{Funo2017}&CD&1 ion  in  Paul trap 
\\
\textcite{Tobalina2017}&Invariants&1 ion  in a nonrigid trap 
\\
\textcite{Li2017_3272}&Trigonometric protocols&Cold atoms  in anharmonic traps
\\
\textcite{Torrontegui2017}&Invariants&Load  in  mechanical crane
\\
\textcite{Dowdall2017}&Pauli blocking&Ultracold Fermi gases
\\
\textcite{Gonzalez-Resines2017}&Invariants&Load  in  mechanical crane 
\\
\textcite{Corgier2018}&Inverse engineering&BECs  in atom chips
\\
\textcite{Kaufmann2018}&Invariants (Exp)&1 ion  in  Paul trap
\\
\textcite{Lu2018}&Invariants&1 ion
\\
\textcite{Tobalina2018}&OCT+Compensating-force&1 ion  in Paul trap
\\
\textcite{Ness2018}&Invariants (Exp)& Cold atoms in optical dipole trap
\\
\textcite{Chen2018_013631}&Inverse engineering&Spin-orbit-coupled BECs 
\\
\textcite{Li2018_113029}&Inverse engineering&Qubit in double quantum dots 
\\
\textcite{Amri2018}&OCT \& STA&BECs in atom chips

\end{tabular}
\end{ruledtabular}
\end{table*}

%
\begin{turnpage}
\begin{table*}
\caption{\label{tableexp}Experiments using STA methods. \\ CD: Counterdiabatic driving;  IE: inverse-engineering;  OL: optical lattice; STIRAP: Stimulated Raman adiabatic passage; NV: Nitrogen vacancy; FAQUAD: Fast quasi adiabatic dynamics.}
\begin{ruledtabular}
\begin{tabular}{llll}
Reference&System&Operation&Method
\\\hline
\textcite{Couvert2008}&Ultracold atoms&Transport&Fourier-based IE
\\
\textcite{Bason2012}&2-level Landau-Zener system (BEC in OL)&Population inversion&CD+unitary transformation
\\
\textcite{Bowler2012}&Trapped ions (1, 2)&Transport&Fourier transform 
\\
                            &Trapped ion chain (9 ions)&Separation&FAQUAD
\\
\textcite{Sun2012}&Overhead crane&Transport&Bang-bang
\\
\textcite{Walther2012}&1 and 2 ions&Transport&Optimized drivings
\\
\textcite{Richerme2013}&14 spins in linear trap (Ising model)&Adiabatic quantum simulation&Local adiabatic                   
\\
\textcite{Zhang2013_240501}&NV center in diamond&Assisted adiabatic passage&CD and accelerated CD
\\
\textcite{Ruster2014}&2 ions&Separation&Implement a function for equilibrium distance
\\
\textcite{Kamsap2015}&Large ion clouds&Transport&Numerical simulations
\\
\textcite{Rohringer2015}&1D BEC&Expansion/Compression&Scaling
\\
\textcite{Alonso2016}&Trapped ion&Transport&Bang-bang
\\
\textcite{Du2016}&3-level Rb&STA-STIRAP&CD+unitary transform
\\
\textcite{Martinez2016}&Brownian particle in  optical potential&Equilibration&Engineered swift equilibration
\\
\textcite{An2016}&Trapped ion&Simulated transport&CD and unitarily equivalent protocols
\\
\textcite{Navez2016}&Atoms in state-dependent ``moving buckets''&Interferometer&Choose convenient timing
\\
\textcite{Guo2017_9160}&Optical waveguides&Demultiplexing&Invariant-based IE
\\
\textcite{Zhang2017_042345}&Superconducting phase qubit&Measure Berry phase&CD
\\
\textcite{Chupeau2018_10512}&Brownian particle in quadratic potential&Control of temperature&Inverse engineering 
\\
\textcite{Cohn2018}&2D array of 70 ions in Penning trap&Create highly entangled state&Bang-bang
\\
\textcite{Deng2018_013628}&Anisotropic Fermi gas&Expansion/Compression&Scaling
\\
\textcite{Kaufmann2018}&Trapped ion&Transport&Invariant-based IE
\\
\textcite{Ness2018}&Ultracold K atoms in optical lattice&Transport&Invariants
\\
\textcite{Smith2018}&2-level in $^171$Yb$^+$&Qubit rotations&CD
\\ 
\textcite{Wang2018_065003}&Superconducting Xmon qubit&Quantum gates&CD
\\
\textcite{Hu2018}&Trapped $^{171}$Yb$^+$ ion&Speed up Landau-Zener&Optimized-phase CD   
\\
\textcite{Zhang2018_085001}&Superconducting Xmon qubit&Measure work statistics&CD
\\
\textcite{Zhou2018}&BEC&Prepare BEC in  bands of OL&Optimized pulse sequences
\\
\textcite{Faure2018}&RC circuit&Change stationary regime&Inverse engineering
\\
\textcite{Boyers2018}&NV center&qubit population inversion&Floquet engineering of Hamiltonian components
\\
\textcite{Kolbl2019}&NV center&Control of dressed states&CD
\\
\textcite{Yan2019_080501}&Xmon superconducting qutrit&Holonomic gates&CD
\\
\textcite{Vepsalainen2019}&Transmon&STA-STIRAP& CD
\\
\textcite{Santos2019}&NMR of a Cloroform molecule&Single qubit gates&CD driving
\end{tabular}
\end{ruledtabular}
\end{table*}
\end{turnpage}

\newpage

\appendix
\section{List of acronyms}
\noindent 1D, 2D, ... = one dimensional, two dimensional, ...
\\
BEC = Bose Einstein condensate
\\
CD = Counterdiabatic
\\
CNOT = Controlled not
\\
CS = Control system
\\
DC = Direct current
\\
DFS = Decoherence free subspace
\\
DRAG = Derivative removal of adiabatic gate
\\
FAQUAD = Fast quasiadiabatic dynamics
\\
FF = fast-forward
\\
GHZ = Greenberger-Horne-Zeilinger
\\
IE = Inverse engineering
\\
IP = Interaction picture
\\ 
NV = Nitrogen vacancy 
\\
OCT = Optimal control theory
\\
PDE = Partial differential equation
\\
PS = Primary system
\\
QED = Quantum electrodynamics 
\\
QZD = Quantum Zeno dynamics 
\\
STA = Shorcut to adiabaticity
\\
STIRAP = Stimulated Raman adiabatic passage 
\\
SVEA = Slowly varying envelope approximation 
\\
WAHWAH =Weak anharmonicity with
average Hamiltonian

\section{Example of Lie transform\label{elt}}
As an example to illustrate the use of ``physical'' unitary transformations to generate alternative shortcuts 
we consider a two-level system in which we apply a Lie transform to get rid of a $\sigma_y$ term in the 
Hamiltonian found with the CD approach.  The reference Hamiltonian $H_0$ is given in Eq. (\ref{H0_Lie})
and $H_{CD}$ in Eq. (\ref{counter}).  
%
%
%
%
The generators of the dynamical algebra are the Pauli matrices,
\beq
G_1=\left ( \begin{array}{cc}
0 & 1\\
1 & 0 
\end{array}\right ), \, \, \, 
G_2=\left ( \begin{array}{cc}
0 & -i\\
i & 0 
\end{array}\right ), \, \, \, 
G_3=\left ( \begin{array}{cc}
1 & 0\\
0 & -1 
\end{array}\right ),
\eeq
which satisfy the commutation relations $[G_a,G_b]=2i \epsilon_{abc}G_c$.  
The total Hamiltonian $H(t)=H_0(t)+H_{CD}(t)$ in terms of the algebra generators can be written as 
\beq
H(t)=\frac{\hbar}{2} [\Omega_R(t) G_1-\Delta(t) G_3]+\frac{\hbar}{2} \Omega_a(t) G_2. 
\eeq
Suppose that the generator $G_2$ is difficult or inconvenient to implement,  
see for example \textcite{Bason2012}. Setting $G=G_3$ in Eq. (\ref{Lie_transformation}), and substituting into 
Eq. (\ref{HI_series}), the series of repeated commutators may be summed up. $H'$ becomes
\beqa
\label{HI_new}
H'(t)&=&\frac{\hbar}{2} \left \{\left [ \Omega_R(t) \cos{(2g(t))} + \Omega_a(t) \sin{(2g(t))} \right ] G_1  \right.  \nonumber \\
&-&\left [ \Omega_R(t) \sin{(2g(t))} - \Omega_a(t) \cos{(2g(t))} \right ] G_2 \nonumber \\
&-&\left. \left [ \Delta(t) + 2 \dot g(t)  \right ] G_3 \right \} . 
\eeqa
To cancel the $G_2$ term, we choose
\beq
\label{Lie_g_condition}
g(t)=\frac{1}{2} \arctan{\left[\frac{\Omega_a(t)}{\Omega_R(t)}\right ]}. 
\eeq
Substituting Eq. (\ref{Lie_g_condition}) into Eq. (\ref{HI_new}) we have finally
\beqa
H'(t)&=& \frac{\hbar}{2} \left \{ \left [ \sqrt{1+\frac{\Omega_a^2(t)}{\Omega_R^2(t)}} \Omega_R(t) \right ] G_1 \right. \nonumber \\
&-& \left.  \left [ \Delta+\frac{\Omega_R(t) \dot\Omega_a(t)-\dot\Omega_R(t) \Omega_a(t)}{\Omega_R^2(t)+\Omega_a^2(t)} \right ]\!G_3 \right \}\!, 
\eeqa
which has the same structure (generators) as the reference Hamiltonian (\ref{H0_Lie}) but different time-dependent coefficients. 
A similar result can be found without starting from the reference shortcut (\ref{counter}), but following the 
bottom-up approach of \textcite{Torrontegui2014}, see Sec. \ref{lri}.
%
%
%
%
%
\section{Counterdiabatic Hamiltonian for a 2-level Hamiltonian with complex coupling\label{2gen}}
%
%
%
%
We find here $H_{CD}$ for a two-level Hamiltonian $H_0$ with complex-valued coupling $\Omega$,
\begin{eqnarray}
H_0(t)=\frac{\hbar}{2}\left(\begin{array}{cc}
-\Delta(t) & \fabs{\Omega(t)} e^{i \alpha(t)}\\
\fabs{\Omega(t)} e^{-i \alpha(t)} & \Delta(t)
\end{array}\right)\!
\label{eq:app_cd1}
\end{eqnarray}
where $\fabs{\Omega(t)}$ is the modulus and $\alpha(t)$ the argument of the coupling.
The instantaneous eigenvectors of this Hamiltonian are
\beqa
|\lambda_{-}(t)\rangle &=& - \sin[\theta(t)/2] e^{i \alpha(t)/2} |1 \rangle +
\cos[\theta(t)/2] e^{-i \alpha(t)/2} |2 \rangle, \nonumber 
\\
|\lambda_{+}(t)\rangle &=& \cos[\theta(t)/2] e^{i \alpha(t)/2} |1 \rangle 
+ \sin[\theta(t)/2] e^{-i \alpha(t)/2} |2 \rangle, \nonumber
\\
\eeqa
with the mixing angle $\theta (t)\equiv \arccos [-\Delta(t)/\widetilde\Omega(t)]$
and  eigenvalues $E_{\mp}(t)= \mp \hbar \widetilde \Omega /2$,
where $\widetilde\Omega = \sqrt{\Delta^2 (t) + \fabsq{\Omega (t)}}$.
The counterdiabatic Hamiltonian is given by $H_{CD} = H_{CD}^{(1)} + H_{CD}^{(2)}$
where 
$H_{CD}^{(1)} =   i \hbar \sum_n  |\dot{n}(t)  \rangle \langle n(t) |$
and  
$H_{CD}^{(2)} =  - i \hbar \sum_n  \langle n(t) | \dot{n}(t)  \rangle | n(t) \rangle \langle n(t) |$.
For the Hamiltonian $H_0 (t)$ in Eq. \eqref{eq:app_cd1}, we get
\beqa
H_{CD}^{({1})} (t)
= \frac{\hbar}{2} \left(\begin{array}{cc} -\dot{\alpha} (t) & -i e^{i \alpha(t)} \dot{\theta}(t)
\\
i e^{-i \alpha(t)} \dot{\theta}(t) & \dot{\alpha} (t)
\end{array}\right)
\eeqa
and
\beqa
H_{CD}^{({2})} (t)
&=& \frac{\hbar}{2} \cos \theta(t) \, \dot{\alpha} (t)\nonumber\\
& & \times \left(\begin{array}{cc}
\cos \theta (t) & e^{i \alpha(t)} \sin \theta(t)
\\
e^{-i \alpha(t)} \sin \theta(t) & -\cos \theta(t)
\end{array}\right).
\eeqa
If $\alpha(t) = 0$, we have that $H_{CD}^{(2)} = 0$, and the expression of
$H_{CD}^{({1})} (t)$ simplifies to the one given e.g. in Eq. (\ref{counter}) or  \textcite{Chen2010_123003}.

\bibliographystyle{apsrmp4-1} 
\bibliography{Bibliography_review}  

\end{document}